\documentclass[aps,preprint,onecolumn,tightenlines,superscriptaddress,amsmath,amssymb,longbibliography,showkeys,floatfix]{revtex4-2}
\usepackage{graphicx}

\usepackage{booktabs}
\usepackage[british]{babel}
\usepackage[caption=false]{subfig}
\usepackage{hyperref}
\usepackage{tikz}
\usetikzlibrary{decorations.pathmorphing,arrows.meta}
\providecommand{\dd}{\mathop{}\!\mathrm{d}}
\providecommand{\ii}{\mathrm{i}}
\providecommand{\CD}[2]{{}^{\mathrm C}\!D_{#1}^{#2}}
\providecommand{\sech}{\mathop{\mathrm{sech}}\nolimits}
\providecommand{\Mh}{M_H}

\begin{document}

\title{A Scaling Framework for Mechanical Memristance: Dimensionless Metrics and Material Design Maps}

\author{Abdulla Alhembar}
\thanks{ORCID: \url{https://orcid.org/0000-0002-5045-9827}.}
\affiliation{Department of Mechanical and Nuclear Engineering, Khalifa University of Science and Technology, Abu Dhabi 127788, United Arab Emirates}
\affiliation{Bristol Composites Institute, School of Civil, Aerospace and Design Engineering (CADE), University of Bristol, BS8 1TR Bristol, UK}

\author{Fabrizio Scarpa}
\thanks{ORCID: \url{https://orcid.org/0000-0002-5470-4834}}
\email{f.scarpa@bristol.ac.uk}
\affiliation{Bristol Composites Institute, School of Civil, Aerospace and Design Engineering (CADE), University of Bristol, BS8 1TR Bristol, UK}

\author{Chrystel D. L. Remillat}
\thanks{ORCID: \url{https://orcid.org/0009-0005-1742-0268}.}
\affiliation{Bristol Composites Institute, School of Civil, Aerospace and Design Engineering (CADE), University of Bristol, BS8 1TR Bristol, UK}

\author{Rodrigo J. da Silva}
\thanks{ORCID: \url{https://orcid.org/0000-0001-8016-3165}.}
\affiliation{CERN, The European Organization for Nuclear Research, R\&D Programme EP-DT, Geneva, Switzerland}

\author{Ross Anderson}
\thanks{ORCID: \url{https://orcid.org/0000-0002-6796-0482}.}
\affiliation{School of Biochemistry and Biomedical Sciences, University of Bristol, BS8 1TD Bristol, UK}

\author{Ludovico Cademartiri}
\thanks{ORCID: \url{https://orcid.org/0000-0001-8805-9434}.}
\affiliation{Department of Chemistry, Life Sciences, and Environmental Sustainability, University of Parma, Parco Area delle Scienze 11/a, 43124, Parma, Italy}

\author{Abderrezak Bezazi}
\thanks{ORCID: \url{https://orcid.org/0000-0002-4461-6689}.}
\affiliation{Laboratory of Applied Mechanics of New Materials (LMANM), University 8 May 1945, Guelma, Algeria}

\author{James P. K. Armstrong }
\thanks{ORCID: \url{https://orcid.org/0000-0002-0599-0643}.}
\affiliation{Bristol Medical School, University of Bristol, BS1 3NY BRISTOL, UK}

\author{Adam W. Perriman}
\thanks{ORCID: \url{https://orcid.org/0000-0003-2205-9364}.}
\affiliation{School of Biochemistry and Biomedical Sciences, University of Bristol, BS8 1TD Bristol, UK}
\affiliation{Research School of Chemistry and John Curtin School of Medical Research, Australian National University, Canberra, ACT 2601, Australia}

\begin{abstract}
Mechanical memristors are systems whose dissipative response depends on the history of previous loading through an evolving internal state. History-dependent forces and dissipation occur in a wide range of materials and devices, including viscoelastic polymers, shape-memory materials, piezoelectrics, granular media and field-responsive fluids. Determining which of these responses admits a mechanical-memristor representation requires a constitutive test, as well as a comparison of scales. In this work, a fractional-order mechanical memristor model is developed and cast into a nondimensional form to identify the governing parameters controlling memory-dependent dissipation. The formulation leads to a set of dimensionless groups that characterise dissipation magnitude, memory-state scale and memory transfer. These quantities are combined into an effective mechanical memristance screening index \(\Mh = \beta\gamma |\mathcal H_\alpha(\Omega)|\), which provides a conditional measure of local damping modulation at matched response amplitude, constitutive slope and reference scales. Illustrative parameter scenarios are then constructed for material classes including shape-memory polymers, shape-memory alloys, hydrogels, nanocellulose, lignin-rich materials, natural fibres, piezoelectric polymers, piezoelectric ceramics, electrorheological fluids, magnetorheological fluids and granular dampers. The framework establishes a common basis for comparing memory-dependent damping within the adopted constitutive description and identifies the calibration required for its application to candidate materials and devices.
\end{abstract}


\maketitle
\clearpage

\section{Introduction}

The ideal memristor, originally introduced by Chua as the fourth fundamental
passive circuit element \cite{1083337}, is characterised by a
constitutive relation between electrical charge and flux linkage. The
generalised memristive system extends this description by allowing the
resistance to depend on evolving internal states \cite{ChuaKang1976}. Since the first experimental demonstrations of
resistive-switching devices, memristors have attracted considerable
attention for applications in non-volatile memory, neuromorphic
computing, in-memory processing, flexible electronics and intelligent
sensing systems
\cite{WOS:000348204900002,WOS:000581738300050,
WOS:001384802000001,WOS:001460474400001}.
Recent developments have further established memristive devices as
multifunctional platforms capable of integrating sensing, memory and
information processing within a common architecture, enabling
technologies such as tactile sensors, artificial nociceptors, wearable
computing systems and neuromorphic electronic skins
\cite{WOS:000711790600083,WOS:001278072600001,
WOS:001453720000001,WOS:001768750700001}.

Inspired by these developments, mechanical memristance has emerged as
the mechanical counterpart of electrical memristance. In mechanical
memristive systems, forces depend not only on the current mechanical
state but also on the history of previous loading, giving rise to
path-dependent dynamics and memory-dependent dissipation. For a mechanical memristive dissipator, the relevant pinching is in the
dissipative force--velocity plane \cite{WOS:000352695100037}. A pinched
loop alone is not sufficient to establish
memristive behaviour, the force--velocity law and internal-state evolution
must also be identified. This perspective has
motivated the development of memory-enabled mechanical elements,
including the memdamper, mem-dashpot and mem-inerter, which extend the
classical concepts of springs, dampers and inertial elements by
introducing state-dependent memory effects
\cite{WOS:000380571000049,WOS:000521624000001}. More recently, pinched hysteresis loops have also been reported in MEMS
resonators \cite{WOS:001518837600001}.

From a mathematical perspective, the connection between electrical and
mechanical memristors follows naturally from the force--voltage
(impedance) analogy. In the impedance representation, the
power-conjugate variables and their time integrals are mapped
according to

\begin{equation}
  V_{\mathrm e}\longleftrightarrow F,
  \qquad
  I_{\mathrm e}\longleftrightarrow \dot u,
  \qquad
  Q_{\mathrm e}=\int I_{\mathrm e}\,\dd t\longleftrightarrow u,
  \qquad
  \Phi_{\mathrm e}=\int V_{\mathrm e}\,\dd t
  \longleftrightarrow p_{\mathrm I}=\int F\,\dd t.
  \label{eq:electromechanical_equivalence}
\end{equation}

Here $V_{\mathrm e}$ and $I_{\mathrm e}$ are electrical voltage and current,
$Q_{\mathrm e}$ is charge, and $\Phi_{\mathrm e}$ is flux linkage. Their
mechanical counterparts are force $F$, velocity $\dot u$, displacement $u$
and force impulse $p_{\mathrm I}$, respectively. Time is denoted by $t$, and an overdot
denotes its derivative. The integrated quantities are measured relative to
a common reference state. The correspondence preserves instantaneous power:
$V_{\mathrm e}I_{\mathrm e}\longleftrightarrow F\dot u$.

Accordingly, an electronic current-controlled memristive system
\(V_{\mathrm e}=R(z_{\mathrm e},I_{\mathrm e},t)I_{\mathrm e}\), with resistance
$R$ and electronic internal state $z_{\mathrm e}$ evolving according to
\(\dot z_{\mathrm e}=g_{\mathrm e}(z_{\mathrm e},I_{\mathrm e},t)\),
maps to a mechanical memristive dissipator

\begin{equation}
F_{\mathrm d}
=
c_{\mathrm m}(q,\dot u,t)\dot u,
\end{equation}

with memory-state evolution

\begin{equation}
\dot q
=
g_{\mathrm m}(q,u,\dot u,t).
\end{equation}

Here $F_{\mathrm d}$ is the dissipative force, $c_{\mathrm m}$ is the
memory-dependent damping coefficient, and $q$ is a mechanical internal
state that records the effect of earlier loading. The functions $g_{\mathrm e}$
and $g_{\mathrm m}$ specify the rates at which the respective internal states
change. A displacement-like representation of $q$ is specified in
Section~\ref{sec:dimensional}.

Electrical resistance therefore corresponds to a damping coefficient,
Joule heating corresponds to mechanical dissipation and the electronic
internal state maps to a mechanical memory coordinate. The analogy is
established at the level of constitutive equations and state-space
representation, rather than through identical physical mechanisms
\cite{1083337,WOS:000352695100037,
10.1121/1.1915605,Jeltsema14052010}.

The significance of mechanical memristors extends beyond their role as
mechanical analogues of electrical devices. Memory-dependent behaviour
provides a route towards adaptive structures capable of integrating
vibration control, sensing, energy management and learning within a
single physical system. Similar ideas are currently being explored in
adaptive metamaterials, smart structures, soft robotics and
neuromorphic mechanical devices
\cite{WOS:000382548000014,WOS:001040804900001,
WOS:001562294200001}.

Among the many manifestations of mechanical memristance, damping is one
of the most important and experimentally accessible. Classical viscous
damping is memoryless and depends only on instantaneous velocity,
whereas memristive damping evolves through an internal state variable
that stores information about previous excitation. Significant progress
has been achieved through the study of viscoelastic materials,
fractional-order constitutive models and shape-memory materials, where
distributed relaxation processes naturally generate memory-dependent
energy dissipation and hysteresis
\cite{WOS:000226389800004,WOS:000335708500008,
WOS:000239158700003,WOS:000385246400024,
WOS:000430031700004,WOS:000436101200005}.
At the system level, memory-dependent damping concepts have been
incorporated into port-Hamiltonian formulations and memdamper
frameworks, enabling adaptive vibration attenuation without active
control
\cite{WOS:000329274500004,WOS:000380571000049,
WOS:000554423300001}.

In the present work, the term mechanical memristance is employed in the broader sense of the memristive-systems framework introduced by Chua and Kang, in which the constitutive response depends upon the evolution of internal state variables. It should therefore be distinguished from a strict mechanical analogue of the ideal memristor, whose constitutive relation is defined directly between conjugate state variables.

Despite the growing body of work on mechanical memristors, a fundamental question remains
largely unresolved: how can the memory-dependent responses of
different materials and devices be compared within a common framework?
Existing studies have focused primarily on constitutive modelling,
device implementation and hysteresis characterisation, whereas
comparatively little attention has been devoted to scaling laws and
dimensionless performance measures capable of linking memory,
dissipation and structural dynamics across multiple classes of
mechanical memristors. As a result, there is currently no widely
accepted framework for comparing systems operating across different
length scales, timescales and material platforms.

A significant limitation of the existing literature is that mechanical memristive systems are generally studied within material- or device-specific constitutive descriptions. Although hysteresis, memory retention and dissipation have been widely investigated, there is currently no established dimensionless framework for comparing memory-dependent dissipative systems across different materials, geometries and operating conditions. Nor is there a commonly adopted measure that relates the magnitude of memory-state evolution to the resulting modulation of mechanical dissipation.

The present study addresses this challenge through a nondimensional and
scaling-based formulation of mechanical memristance. By identifying the dimensionless parameters governing memory-dependent dissipation, hysteresis and dynamic response, this framework enables direct comparison of systems whose dissipative behaviour is represented by the state-dependent dashpot considered here. A further contribution is the use of a specific bounded, strictly positive function $\mathcal G(z)$ as a common constitutive basis for the scaling analysis. Building on established mechanical mem-dashpot formulations \cite{PeiEtAl2015}, it combines symmetric saturation with controllable smoothing and couples the damping response to a separate relaxation state, including fractional memory. This choice makes it possible to connect local changes in damping to the dimensionless parameters and material design maps.

The principal contribution of the present work, however, is the introduction of the effective mechanical memristance screening index:

\begin{equation}
\Mh=\beta\gamma|\mathcal H_\alpha(\Omega)|.
\label{eq:MH_screening_definition}
\end{equation}

where $\beta$ measures damping relative to stiffness on a chosen time
scale, $\gamma$ compares the displacement scale with the memory-state scale,
and $|\mathcal H_\alpha(\Omega)|$ is the ratio of memory-state amplitude to
displacement amplitude under harmonic motion. Here $\Omega$ is the
nondimensional forcing frequency and $\alpha$ is the order of the memory
law. These quantities are defined quantitatively in
Section~\ref{sec:nondimensional}. The index combines these effects for
screening under comparable loading and scaling conventions, its relation
to actual damping modulation is qualified in
Section~\ref{sec:conditional_effective_memristance}. Unlike equivalent damping or hysteresis measures, $M_H$ is designed to capture how strongly an evolving memory state can modulate the dissipative coefficient under comparable scaling and loading conditions. The framework is subsequently illustrated through a series of Ashby-type design maps motivated by shape-memory
polymers, shape-memory alloys, hydrogels, lignocellulosic materials,
piezoelectrics, electrorheological fluids,
magnetorheological fluids and granular dampers and constructed from representative parameter scenarios. These maps serve to demonstrate the use of the scaling framework and do not constitute experimentally identified material rankings. 

The present work makes three principal contributions. First, a fractional-order mechanical memristor model is recast into a dimensionless form that identifies the governing groups controlling memory-dependent dissipation. Second, the resulting scaling relations are used to derive measures of state transfer, damping modulation and cycle dissipation. Third, these quantities are combined into the effective mechanical memristance screening index $M_H$, which provides a common basis for locating candidate systems within a unified design space. The objective of the present work is therefore not to establish material rankings, but to provide a common scaling framework within which memory-dependent dissipative systems can be represented, compared and subsequently calibrated.

The paper first develops a fractional-order mechanical memristor model,
then derives the corresponding nondimensional governing parameters.
Scaling laws are then established and used to define the effective
mechanical memristance index. Design maps of the model response are constructed using representative parameters from a broad range of candidate material systems. The maps identify promising regions for constitutive calibration and experimental evaluation. Details of the
mathematical derivations and consistency analyses are provided in
Appendices~\ref{app:derivations} and \ref{app:Consistency}.

\section{Constitutive model}

\begin{figure}[htbp]
\centering
    \includegraphics[width=\linewidth]{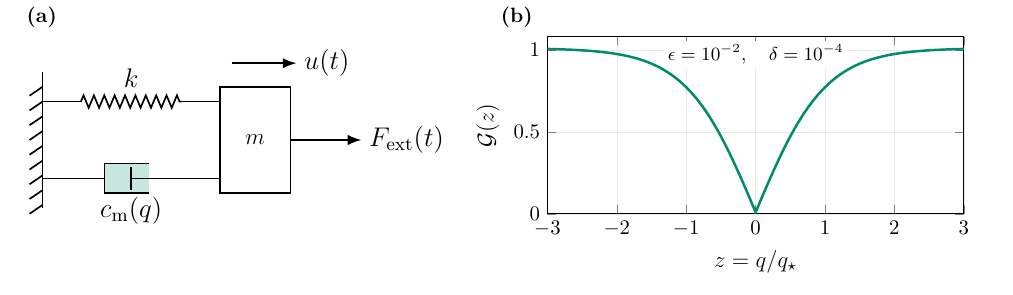}
    \caption{(a) Single-degree-of-freedom spring--mem-dashpot system. The spring and shaded mem-dashpot act in parallel and share the displacement $u(t)$. The internal memory coordinate $q(t)$ governs the damping coefficient $c_{\mathrm m}(q)$. (b) Example of the state-dependent function $\mathcal G(z)$ in Eq.~\eqref{eq:smooth_positive_law}, shown for $\epsilon=10^{-2}$ and $\delta=10^{-4}$, with $z=q/q_{\star}$.}
\label{fig:canonical_oscillator}
\end{figure}

Consider the single-degree-of-freedom system in
Fig.~\ref{fig:canonical_oscillator}a, with displacement $u(t)$ and
memory coordinate $q(t)$. A linear spring and a mem-dashpot act in parallel so that
\begin{equation}
  F_{\mathrm r}(t)=k u(t)+F_{\mathrm d}(t),
  \label{eq:restoring_force}
\end{equation}
where $F_{\mathrm r}$ is the total resisting force, $ku$ is the spring force,
and $k>0$ is the spring stiffness. With mass $m\ge0$, the structural balance is
\begin{equation}
  m\ddot u(t)+F_{\mathrm r}(t)=F_{\mathrm ext}(t),
  \label{eq:general_balance}
\end{equation}
where the applied force $F_{\mathrm ext}$ consists of a sinusoidal load and an optional constant bias, which allows oscillations about a preloaded state. Purely sinusoidal forcing is recovered by setting $F_{\mathrm b}=0$:
\begin{equation}
  F_{\mathrm ext}(t)=F_{\mathrm b}+F_{\mathrm a}\sin(\omega t).
  \label{eq:forcing_dimensional}
\end{equation}

Here $F_{\mathrm b}$ is the constant bias force, $F_{\mathrm a}\ge0$ is the
oscillatory force amplitude and $\omega>0$ is its angular frequency. Thus, the
force oscillates about $F_{\mathrm b}$ with period $2\pi/\omega$. The force
decomposition and associated energy balance are given in
Eqns.~\eqref{eq:appendix_parallel_forces} and
\eqref{eq:appendix_power_balance}.
An inertia-free model naturally follows by setting $m=0$.  The solution remains time dependent, because both the viscous force and the memory coordinate
continue to evolve. When model parameters are fitted to measured forces and displacements and inertia is significant, the force
assigned to the spring--mem-dashpot assembly is
$F_{\mathrm r}=F_{\mathrm ext}-m\ddot u$.

\section{Canonical dimensional model}
\label{sec:dimensional}

The general memristive description considers a damping coefficient
that depends on a single internal state:
\begin{equation}
  F_{\mathrm d}(t)=c_{\mathrm m}\bigl(q(t)\bigr)\dot u(t),
  \label{eq:memdashpot_force}
\end{equation}
The coefficient $c_{\mathrm m}(q)$ has units of force times time per
displacement. This force--velocity form follows the mechanical counterparts
of state-dependent resistive systems \cite{ChuaKang1976,PeiEtAl2015}. The
particular coefficient and relaxation law adopted below are constitutive
choices for the present model. The damping coefficient is assumed to evolve as follows:
\begin{equation}
  c_{\mathrm m}(q)=c_{\star}\mathcal G\!\left(\frac{q}{q_{\star}}\right),
  \qquad c_{\star}>0,\quad q_{\star}>0.
  \label{eq:general_memristance}
\end{equation}

Here $c_{\star}$ is a reference damping coefficient, $q_{\star}$ is the
state-amplitude scale over which damping changes and $\mathcal G$ is a
dimensionless function describing that change. The components $q$ and $q_{\star}$ are assigned
displacement units so that $q$ can be compared with $u$ in the relaxation
law. The conditions \(c_{\star}>0\) and \(q_{\star}>0\) define positive reference scales. Expressing the memory state on a displacement scale renders the relaxation law dimensionally consistent and places different physical state variables on a common basis within the same modelling framework. For a material-specific interpretation, let $\eta_{\mathrm m}$ denotes an underlying
physical state, such as, recoverable strain and polymer-network configuration \cite{Liu2006}, martensitic fraction \cite{Lagoudas2008}, polarisation (electric dipole moment per unit volume) \cite{Damjanovic2006}, particle-chain
morphology \cite{Winslow1949}, or force-chain connectivity (the network of contacts
carrying load in a granular assembly) \cite{Jaeger1996}. Their connection to
the canonical coordinate (the common displacement-scale variable used in this reduced model) is represented by:
\begin{equation} q = q_\star h(\eta_{\mathrm m}), \label{eq:general_state_mapping}
\end{equation}
where $h$ converts the physical state into a dimensionless measure of its
effect on damping. This mapping requires material-specific calibration. Recoverable energy associated with the physical state, and any coupling to thermal or electrical fields, are not captured by the present formulation and require separate constitutive description. With the normalised state $z$, the coefficient is:
\begin{equation}
  z=\frac{q}{q_{\star}}=h(\eta_{\mathrm m}),
  \qquad c_{\mathrm m}(q)=c_{\star}\mathcal G(z).
  \label{eq:qstar_state_scale}
\end{equation}

For the present study, a bounded coefficient is chosen such that it gives the same
damping for positive and negative states of equal magnitude. The hyperbolic
tangent describes saturation as $|z|$ grows. For $\delta>0$, the sharp corner at $z=0$ is replaced by a smooth minimum, so that the slope is defined there. Compared with the displacement-dependent power-law examples discussed by Pei et al.~\cite{PeiEtAl2015}, this specific function remains bounded at large state amplitudes and has no corner at the origin when $\delta>0$. Its positive floor ensures dissipation, while saturation limits the damping coefficient and smoothing permits local differentiation. These features motivate the common constitutive choice used in the present scaling analysis, which still requires material-specific calibration. The function in Eq.~\eqref{eq:general_memristance} is:
\begin{equation}
  \mathcal G(z)
  =\epsilon+\sqrt{\tanh^2 z+\delta^2}-\delta,
  \qquad \epsilon>0,\quad \delta\ge0.
  \label{eq:smooth_positive_law}
\end{equation}
 The dimensionless parameter $\epsilon>0$ sets a strictly positive minimum
damping. The dimensionless parameter $\delta\ge0$ sets how gradually the curve changes slope near its minimum: $\delta>0$ gives a continuously differentiable coefficient,
whereas $\delta=0$ gives $\epsilon+|\tanh z|$, which has a corner at $z=0$. Figure \ref{fig:canonical_oscillator}b shows an example of the function $\mathcal{G}(z)$ for a specific combination of $\epsilon$ and $\delta$ parametres.
For every allowed value of $\delta$, the coefficient $\epsilon$ satisfies
\begin{equation}
  \epsilon\le \mathcal G(z)
  \le \epsilon+\sqrt{1+\delta^2}-\delta.
  \label{eq:positive_law_bounds}
\end{equation}

The bounds in Eq.~\eqref{eq:positive_law_bounds} imply a finite, positive
damping coefficient at every state. Consequently, the instantaneous power
absorbed by the mem-dashpot is non-negative:
\begin{equation}
  F_{\mathrm d}\dot u
  =c_{\mathrm m}(q)\dot u^2\ge0,
  \label{eq:passivity}
\end{equation}

Equation~\eqref{eq:passivity} is the passivity condition which states that the
mem-dashpot dissipates mechanical energy and cannot supply energy through
its force--velocity interaction \cite{ChuaKang1976,JeltsemaVanDerSchaft2010}.
Its contribution to the system energy balance is given in
Eq.~\eqref{eq:appendix_power_balance}. The role of \(\delta\) in determining the width and shape of this smooth minimum is derived in Eqns.~\eqref{eq:appendix_delta_transition_scale} and \eqref{eq:appendix_delta_origin}.

The mem-dashpot is a dissipative element embedded within an energy-storing system. In the present model, recoverable energy arises solely from the elastic term \(ku^2/2\) and the kinetic term \(m\dot{u}^2/2\). It should be noted that no recoverable energy is associated with \(q\).

Combining Eqns.~\eqref{eq:restoring_force}--\eqref{eq:memdashpot_force} gives
the common dimensional mechanical equation:
\begin{equation}
  m\ddot u+k u+c_{\mathrm m}(q)\dot u
  =F_{\mathrm b}+F_{\mathrm a}\sin(\omega t).
  \label{eq:common_mechanics_dimensional}
\end{equation}

The simplest relaxation law within this state-variable framework lets $q$
approach the imposed displacement on one characteristic time scale. The first-order law can be written as:
\begin{equation}
  \tau_{\mathrm m}\dot q+q=u,
  \qquad \tau_{\mathrm m}>0,
  \label{eq:first_order_memory_dimensional}
\end{equation}

where $\tau_{\mathrm m}$ is the memory relaxation time. Its positivity ensures
that, at fixed $u$, the difference $q-u$ decays rather than grows. After a
time $\tau_{\mathrm m}$, this difference is reduced by a factor $e^{-1}$.
The complete system is:

\begin{subequations}
\label{eq:first_order_dimensional_system}
\begin{align}
  m\ddot u+k u+c_{\star}\mathcal G
  \!\left(\frac{q}{q_{\star}}\right)\dot u
  &=F_{\mathrm b}+F_{\mathrm a}\sin(\omega t),
  \label{eq:first_order_dimensional_mech}\\
  \tau_{\mathrm m}\dot q+q&=u.
  \label{eq:first_order_dimensional_mem}
\end{align}
\end{subequations}

To quantify how rapidly the state can follow an oscillatory displacement,
consider a sinusoidal component after the initial transient has decayed.
Writing its displacement and state as the real parts of
$\widehat u\exp(\ii\omega t)$ and $\widehat q\exp(\ii\omega t)$, respectively,
gives:
\begin{equation}
  \frac{\widehat q}{\widehat u}
  =\frac{1}{1+\ii\omega\tau_{\mathrm m}}.
  \label{eq:first_order_memory_transfer_dimensional}
\end{equation}

Here the complex amplitudes $\widehat u$ and $\widehat q$
encode both the amplitude and phase of each sinusoid. Their ratio measures
attenuation and phase delay of the memory state, the same substitution is
shown in Eq.~\eqref{eq:appendix_first_order_harmonic}.

A single exponential relaxation may not adequately represent systems in which multiple interacting processes relax over different time scales. Fractional relaxation provides
a compact description of such distributed relaxation
\cite{Mainardi2010}. Therefore, the state law is extended using the Caputo
derivative \cite{Caputo1967,Podlubny1999}:
\begin{equation}
  \CD{t}{\alpha}q(t)
  =\frac{1}{\Gamma(1-\alpha)}
   \int_0^t\frac{\dot q(\zeta)}{(t-\zeta)^{\alpha}}\,\dd\zeta,
  \qquad 0<\alpha<1,
  \label{eq:caputo_definition}
\end{equation}

Here \(\alpha\) is the fractional order, \(\Gamma\) is the gamma function, and \(\zeta\) is the past-time integration variable. The integral incorporates the complete history of changes in \(q\) since \(t=0\), with the kernel \((t-\zeta)^{-\alpha}\) assigning greater weight to recent changes and progressively less weight to older ones. The corresponding state law is

\begin{equation}
  \tau_{\alpha}^{\alpha}\CD{t}{\alpha}q+q=u,
  \qquad 0<\alpha\le1.
  \label{eq:fractional_memory_dimensional}
\end{equation}

The parameter $\tau_{\alpha}>0$ has units of time and sets the characteristic
scale of fractional relaxation. The multiplier $\tau_{\alpha}^{\alpha}$
has units of time to the power $\alpha$, so both terms on the left have
displacement units. At $\alpha=1$, the model is reduced to the first-order approximation, choosing $\tau_{\alpha}=\tau_{\mathrm m}$ then recovers
Eq.~\eqref{eq:first_order_memory_dimensional}. For $0<\alpha<1$, relaxation
is generally non-exponential. The corresponding dimensional system is:
\begin{subequations}
\label{eq:fractional_dimensional_system}
\begin{align}
  m\ddot u+k u+c_{\star}\mathcal G
  \!\left(\frac{q}{q_{\star}}\right)\dot u
  &=F_{\mathrm b}+F_{\mathrm a}\sin(\omega t),
  \label{eq:fractional_dimensional_mech}\\
  \tau_{\alpha}^{\alpha}\CD{t}{\alpha}q+q&=u.
  \label{eq:fractional_dimensional_mem}
\end{align}
\end{subequations}
After the transient associated with the initial state has decayed, the
fractional law gives the following amplitude and phase relation for a
harmonic displacement:
\begin{equation}
  \frac{\widehat q}{\widehat u}
  =\frac{1}{1+(\ii\omega\tau_{\alpha})^{\alpha}},
  \label{eq:fractional_memory_transfer_dimensional}
\end{equation}
For positive \(\omega\), the complex power is evaluated as \[ (\ii\omega\tau_{\alpha})^{\alpha} = (\omega\tau_{\alpha})^{\alpha}\exp(\ii\pi\alpha/2). \] This relation specifies the phase and establishes the branch convention used in Eq.~\eqref{eq:appendix_fractional_branch}. The initial-state
contribution, which must be retained for finite-time calculations, is shown
in Eq.~\eqref{eq:appendix_fractional_transform_initial}.

The calculations and design maps presented below employ only the symmetric saturating law in Eq.~\eqref{eq:smooth_positive_law}, together with the first-order or fractional memory equation. Throughout, the damping coefficient is determined by the internal state through the common mapping \(\mathcal G\).

\section{Nondimensional formulation}
\label{sec:nondimensional}

Nondimensionalisation separates the effects of inertia, damping and memory from the units and characteristic scale of a particular device. A positive reference force \(F_{\mathrm c}\) is adopted and held fixed across all loading cases under consideration. The applied oscillation amplitude \(F_{\mathrm a}\) may then vary independently of the reference scales. The associated reference displacement is the static
spring displacement under $F_{\mathrm c}$, and $t_{\mathrm c}$ is a positive
reference time:

\begin{equation}
  u_{\mathrm c}=\frac{F_{\mathrm c}}{k},
  \qquad t_{\mathrm c}>0,
  \label{eq:scales}
\end{equation}

The dimensionless displacement \(x=u/u_{\mathrm c}\), memory coordinate \(y=q/u_{\mathrm c}\), time \(\tau=t/t_{\mathrm c}\), and force \(f=F/F_{\mathrm c}\) are introduced. Here \(F\) denotes either the applied force \(F_{\mathrm ext}\) or the resisting force \(F_{\mathrm r}\); in each case, the force is scaled by the common reference force \(F_{\mathrm c}\) to obtain its dimensionless form. In particular,
$z=\gamma y$. Substitution of these variables into the dimensional balance
gives the dimensionless groups:
\begin{equation}
  \mu=\frac{m}{k t_{\mathrm c}^{2}},
  \qquad
  \beta=\frac{c_{\star}}{k t_{\mathrm c}},
  \qquad
  \gamma=\frac{u_{\mathrm c}}{q_{\star}},
  \qquad
  \Omega=\omega t_{\mathrm c},
  \label{eq:nondim_groups_main}
\end{equation}

and the dimensionless bias force and oscillatory force amplitude, respectively:

\begin{equation}
  f_{\mathrm b}=\frac{F_{\mathrm b}}{F_{\mathrm c}},
  \qquad
  f_{0}=\frac{F_{\mathrm a}}{F_{\mathrm c}}.
  \label{eq:nondim_force_levels}
\end{equation}

The natural angular frequency of the undamped mass--spring system is \(\sqrt{k/m}\) for \(m>0\), and \(\sqrt{m/k}\) therefore defines its characteristic inertial time scale. The dimensionless group \(\mu=m/(k t_{\mathrm c}^2)\) compares the square of this time scale with \(t_{\mathrm c}^2\) and vanishes when \(m=0\). Similarly, \(\beta\) compares the viscous relaxation time \(c_{\star}/k\) with the reference time scale \(t_{\mathrm c}\). The group $\gamma$ compares the reference
displacement with the memory-state amplitude scale, while $\Omega$ measures
the forcing frequency relative to $1/t_{\mathrm c}$. These interpretations
follow directly from Eq.~\eqref{eq:appendix_group_interpretation}.

A convenient choice is to measure time in units of the memory time:
$t_{\mathrm c}=\tau_{\mathrm m}$ for first-order relaxation, or
$t_{\mathrm c}=\tau_{\alpha}$ for fractional relaxation.
Using $u_{\mathrm c}=F_{\mathrm c}/k$, the definitions then become

\begin{equation}
\begin{aligned}
  t_{\mathrm c}=\tau_{\mathrm m}:&\qquad
  c_{\star}=\beta k\tau_{\mathrm m},
  &q_{\star}=\frac{u_{\mathrm c}}{\gamma}=\frac{F_{\mathrm c}}{k\gamma}
  &&\text{(first order)},\\
  t_{\mathrm c}=\tau_{\alpha}:&\qquad
  c_{\star}=\beta k\tau_{\alpha},
  &q_{\star}=\frac{u_{\mathrm c}}{\gamma}=\frac{F_{\mathrm c}}{k\gamma}
  &&\text{(fractional)}.
\end{aligned}
\label{eq:nondim_groups_model_specific}
\end{equation}

Equation \eqref{eq:nondim_groups_model_specific} gives a direct engineering interpretation of the two parameters. Once the stiffness and memory time have been selected, $\beta$ specifies the reference viscous coefficient $c_{\star}$, for example, $\beta=1$ gives $c_{\star}=k\tau_{\mathrm m}$ in the first-order model and $c_{\star}=k\tau_{\alpha}$ in the fractional model. Likewise, $\gamma$ specifies the memory-state displacement scale, with $\gamma=1$ giving $q_{\star}=u_{\mathrm c}$. The same relation for $q_{\star}$ applies to both memory laws. More details about the nondimensionalisation and the derivation of the nondimensional coefficients of the mechanical memristor oscillator are described in Appendix \ref{app:nondimens_eom_details} and \ref{app:fo_scalng_memory_transfer}.

The dimensionless memory-dependent damping coefficient is obtained by
dividing $c_{\mathrm m}$ by $kt_{\mathrm c}$:

\begin{equation}
  c_{\mathrm e}(y)
  =\beta\,\mathcal G(\gamma y)
  =\beta\left[\epsilon+\sqrt{\tanh^2(\gamma y)+\delta^2}-\delta\right].
  \label{eq:nondim_coefficient}
\end{equation}

For the first-order model, $\lambda_1=\tau_{\mathrm m}/t_{\mathrm c}$ is the
ratio of the memory time to the reference time. It follows from the change
of variables, as derived in Appendix~\ref{app:fo_scalng_memory_transfer}.
The nondimensional system is

\begin{subequations}
\label{eq:first_order_nondim_general}
\begin{align}
  \mu x''+x+c_{\mathrm e}(y)x'&=f_{\mathrm b}+f_0\sin(\Omega\tau),
  \label{eq:first_order_nondim_mech}\\
  \lambda_1 y'+y&=x.
  \label{eq:first_order_nondim_mem}
\end{align}
\end{subequations}

Here primes denote differentiation with respect to $\tau$. The numerical
comparisons below use the inertia-free case $\mu=0$ and a sinusoidal load
with zero mean, $f_{\mathrm b}=0$. Setting $f_0=1$ makes the force amplitude
equal to the reference force, while $t_{\mathrm c}=\tau_{\mathrm m}$ gives
$\lambda_1=1$. These specified loading and scaling choices isolate the
coupling between damping and memory, giving

\begin{subequations}
\label{eq:first_order_nondim_attached}
\begin{align}
  c_{\mathrm e}(y)x'+x&=\sin(\Omega\tau),\\
  y'+y&=x.
\end{align}
\end{subequations}

For the fractional nondimensional model, the scaling property of the Caputo derivative is considered \cite{Podlubny1999,Mainardi2010}:

\begin{equation}
  \CD{t}{\alpha}q
  =u_{\mathrm c}t_{\mathrm c}^{-\alpha}\CD{\tau}{\alpha}y.
  \label{eq:caputo_scaling_main}
\end{equation}

Substitution of Eq.~\eqref{eq:caputo_scaling_main} into the dimensional
state law gives the dimensionless coefficient
$\lambda_{\alpha}=(\tau_{\alpha}/t_{\mathrm c})^{\alpha}$.
This factor follows from the fractional derivative's time scaling. Consequently, the system becomes

\begin{subequations}
\label{eq:fractional_nondim_general}
\begin{align}
  \mu x''+x+c_{\mathrm e}(y)x'&=f_{\mathrm b}+f_0\sin(\Omega\tau),
  \label{eq:fractional_nondim_mech}\\
  \lambda_{\alpha}\CD{\tau}{\alpha}y+y&=x,
  \qquad 0<\alpha\le1.
  \label{eq:fractional_nondim_mem}
\end{align}
\end{subequations}

Choosing $t_{\mathrm c}=\tau_{\alpha}$ sets $\lambda_{\alpha}=1$.
To analyse oscillatory loading, the response after the initial
transient has decayed is considered.
A response with the forcing period can be expressed as a sum of harmonics:
\begin{equation}
  x(\tau)=\sum_{n\in\mathbb Z}\widehat x_n\exp(\ii n\Omega\tau).
\end{equation}
Here $n$ is the integer harmonic number, $\widehat x_n$ is the corresponding
complex displacement coefficient, and $n=0$ gives the mean displacement.
Because the state equation is linear in $x$ and $y$, each harmonic of $x$
produces a state harmonic at the same frequency:
\begin{equation}
  \widehat y_n
  =\frac{\widehat x_n}
  {1+\lambda_{\alpha}(\ii n\Omega)^{\alpha}}.
  \label{eq:fractional_harmonic_transfer_nondim}
\end{equation}
The coefficients for negative $n$ are the complex conjugates of those for
positive $n$, ensuring real displacement and state histories. This relation
is derived from Eq.~\eqref{eq:appendix_fractional_transform_initial}.

\subsection{Harmonic state transfer and memory-state amplitude}
\label{sec:harmonic_transfer}

Damping modulation depends on the extent to which the memory state traverses the curve \(c_{\mathrm m}=c_\star\mathcal G(z)\), which defines the damping coefficient associated with each memory state \(z=q/q_\star\). Its component at the forcing frequency, called the
first harmonic ($n=1$), is related to the corresponding displacement
component by the transfer function
\begin{equation}
  \mathcal H_{\alpha}(\Omega)
  =\frac{\widehat y}{\widehat x}
  =\frac{1}{1+\lambda_{\alpha}(\ii\Omega)^{\alpha}}.
  \label{eq:Halpha}
\end{equation}
For $\Omega>0$, the convention
$(\ii\Omega)^{\alpha}=\Omega^{\alpha}
[\cos(\pi\alpha/2)+\ii\sin(\pi\alpha/2)]$
from Eq.~\eqref{eq:appendix_fractional_branch} gives its magnitude:
\begin{equation}
  \left|\mathcal H_{\alpha}(\Omega)\right|
  =\left[
    1+2\lambda_{\alpha}\Omega^{\alpha}
      \cos\!\left(\frac{\pi\alpha}{2}\right)
    +\lambda_{\alpha}^{2}\Omega^{2\alpha}
  \right]^{-1/2},
  \label{eq:Halpha_magnitude}
\end{equation}

The phase delay $\phi_{\alpha}$, measured in radians, specifies how far
the memory-state sinusoid lags behind the displacement sinusoid:
\begin{equation}
  \phi_{\alpha}(\Omega)
  =\operatorname{atan2}\!\left(
    \lambda_{\alpha}\Omega^{\alpha}
      \sin\frac{\pi\alpha}{2},
    1+\lambda_{\alpha}\Omega^{\alpha}
      \cos\frac{\pi\alpha}{2}
  \right).
  \label{eq:Halpha_phase}
\end{equation}
Here $\operatorname{atan2}(a,b)$ denotes the angle of the complex number
$b+\ii a$. Equations~\eqref{eq:Halpha_magnitude} and \eqref{eq:Halpha_phase}
follow by taking the modulus and negative argument of
Eq.~\eqref{eq:Halpha}. The magnitude describes how closely the state follows an
ongoing sinusoidal displacement.

For the amplitude estimates that follow, an unbiased response
with negligible mean and a dominant first harmonic is assumed. Higher displacement
harmonics are neglected in this approximation, although they are retained
in the numerical solution. If \(A\) denotes the dimensionless peak displacement amplitude, the time origin is chosen such that \(\tau=0\) corresponds to a displacement maximum, giving
\begin{equation}
  x=A\cos(\Omega\tau),
  \qquad
  y=A\left|\mathcal H_{\alpha}\right|
  \cos(\Omega\tau-\phi_{\alpha}),
  \label{eq:harmonic_state_response}
\end{equation}

The peak magnitude of the normalised state $z=\gamma y$ is therefore

\begin{equation}
  \chi(\Omega,A)
  =\gamma A\left|\mathcal H_{\alpha}(\Omega)\right|.
  \label{eq:state_excursion_parameter}
\end{equation}
Thus $\chi$ is the largest value of $|q|/q_{\star}$ reached during the
approximately sinusoidal cycle. It identifies the part of $\mathcal G$ that
the response explores. For $\delta>0$ and
$\chi\ll\min(1,\delta)$, the state remains near the rounded minimum and
the damping variation is quadratic in the state amplitude, as shown in
Eq.~\eqref{eq:appendix_delta_inner}. If $\delta\ll\chi\ll1$, the response
extends beyond that rounded region while remaining below saturation, the
state-dependent contribution then varies approximately with $|z|$,
according to Eq.~\eqref{eq:appendix_delta_outer}. Values $\chi=O(1)$ explore
the nonlinear transition, while for $\chi\gg1$ the coefficient is close to
its saturation value over much of the cycle.

The dimensional displacement amplitude is $Au_{\mathrm c}$, so
$\gamma A=Au_{\mathrm c}/q_{\star}$ and $\chi$ is independent of the
arbitrary reference displacement. For a biased response or one with
substantial higher harmonics, the full state history must instead be used
to determine the range of $z$. The connection between state amplitude and
the screening index is developed in
Section~\ref{sec:conditional_effective_memristance}.

\section{Linearised models}
\label{sec:linearisation}

\subsection{Small-amplitude structure of the adopted passive law}

Linearisation identifies the response to small changes about an operating
state. At first, the coefficient near the zero state is examined, then the perturbation equations about a general reference motion are derived.

In this section, $C(y)=c_{\mathrm e}(y)$
is a shorthand for the nondimensional damping coefficient in
Eq.~\eqref{eq:nondim_coefficient}. For $\delta>0$ and
$|\gamma y|\ll\min(1,\delta)$, expanding that complete coefficient gives:
\begin{equation}
  C(y)
  =\beta\epsilon
  +\frac{\beta\gamma^2}{2\delta}y^2+O(y^4).
  \label{eq:positive_small_argument}
\end{equation}

Differentiating Eq.~\eqref{eq:positive_small_argument} at $y=0$ gives
the coefficient value, slope and curvature, respectively:

\begin{equation}
  C(0)=\beta\epsilon,
  \qquad C_y(0)=0,
  \qquad C_{yy}(0)=\frac{\beta\gamma^2}{\delta}.
  \label{eq:coefficient_origin_derivatives}
\end{equation}
Here $C_y$ and $C_{yy}$ denote the first and second derivatives with respect
to the memory state $y$. Near $y=0$, the damping curve has zero slope. A linear approximation therefore sees only the constant damping $\beta\epsilon$ and misses the change caused by the memory state. The first such change in the damping force is proportional to $y^2x'$: if both the state and velocity amplitudes double, this contribution increases by a factor of eight. Memory-dependent damping near the undeformed state must therefore be examined beyond the linear approximation. This expansion requires $\delta>0$; for $\delta=0$, the curve has a sharp corner at $y=0$ and cannot be expanded in this way.

\subsection{Small perturbations about an operating response}

To assess small changes about a biased or moving response, let
$(x_{\mathrm b}(\tau),y_{\mathrm b}(\tau))$ denote a solution of the full
nonlinear equations under the nondimensional force $f_{\mathrm b}^{\ast}(\tau)$.
This selected response is the reference solution, and
$V_{\mathrm b}=x_{\mathrm b}'$ is its velocity. Introduce small changes in
its displacement, memory state and applied force as follows:
\begin{equation}
  x=x_{\mathrm b}+\varepsilon\xi,
  \qquad
  y=y_{\mathrm b}+\varepsilon\eta,
  \qquad
  f=f_{\mathrm b}^{\ast}+\varepsilon p,
  \qquad 0<\varepsilon\ll1.
  \label{eq:perturbations}
\end{equation}
The functions $\xi$, $\eta$ and $p$ describe these changes, and
$\varepsilon$ controls their small magnitude. Expanding $C(y)x'$ and retaining terms proportional to $\varepsilon$ gives
the mechanical perturbation equation (Appendix~\ref{app:Tangent_linearisation}):
\begin{equation}
  \mu\xi''+\xi+C_{\mathrm b}\xi'+G_{\mathrm b}\eta=p,
  \label{eq:linearised_mechanics}
\end{equation}

where:

\begin{equation}
  C_{\mathrm b}=C(y_{\mathrm b}),
  \qquad
  G_{\mathrm b}=C_y(y_{\mathrm b})V_{\mathrm b}.
  \label{eq:tangent_coefficients}
\end{equation}
The coefficient $C_{\mathrm b}$ is the nondimensional damping at the selected
memory state.  It sets the change in force produced by a small change in
velocity.  The coefficient $G_{\mathrm b}$ combines the local change of the
damping law with the velocity of the reference solution, and therefore
describes how a small state perturbation enters the mechanical balance while
that reference solution is moving.

For $\delta>0$, differentiating Eq.~\eqref{eq:nondim_coefficient} gives
\begin{equation}
  C_y(y)
  =\beta\gamma
  \frac{\tanh(\gamma y)\operatorname{sech}^2(\gamma y)}
       {\sqrt{\tanh^2(\gamma y)+\delta^2}}.
  \label{eq:coefficient_derivative}
\end{equation}
The first-order memory perturbation satisfies:
\begin{equation}
  \lambda_1\eta'+\eta=\xi,
  \label{eq:linearised_first_order_memory}
\end{equation}
whereas, the fractional memory perturbation satisfies:
\begin{equation}
  \lambda_{\alpha}\CD{\tau}{\alpha}\eta+\eta=\xi.
  \label{eq:linearised_fractional_memory}
\end{equation}
Equation~\eqref{eq:linearised_mechanics}, together with either
Eq.~\eqref{eq:linearised_first_order_memory} or
Eq.~\eqref{eq:linearised_fractional_memory}, describes small changes about
the selected response. The approximation is first order in perturbation
amplitude for both memory laws. If the reference response is periodic,
$C_{\mathrm b}$ and $G_{\mathrm b}$ vary through the cycle, because the
reference state and velocity vary.

A local comparison can be obtained by evaluating both coefficients at a single response point and assuming them constant. This approximation is valid only if their variation is negligible over the perturbation timescale.. Under this assumption, the
Laplace transform converts the differential equations into algebraic
relations (Appendix~\ref{app:Frequency_domain_models}). With zero initial
perturbations, let $\Xi(s)$ and $P(s)$ be the transforms of $\xi$ and $p$,
and let $s$ denote the Laplace variable. The displacement change per unit
force change for the first-order memory law is:
\begin{equation}
  \frac{\Xi(s)}{P(s)}
  =\left[
    \mu s^2+1+C_{\mathrm b}s
    +\frac{G_{\mathrm b}}{1+\lambda_1s}
  \right]^{-1},
  \label{eq:first_order_linear_transfer}
\end{equation}
and the fractional transfer becomes:
\begin{equation}
  \frac{\Xi(s)}{P(s)}
  =\left[
    \mu s^2+1+C_{\mathrm b}s
    +\frac{G_{\mathrm b}}
      {1+\lambda_{\alpha}s^{\alpha}}
  \right]^{-1}.
  \label{eq:fractional_linear_transfer}
\end{equation}
For harmonic perturbations, the same complex-exponential representation
used in Eq.~\eqref{eq:Halpha} gives $s=\ii\Omega$. This standard frequency
description represents amplitude and phase together. The resulting
nondimensional dynamic stiffness, defined as the force amplitude divided
by the displacement amplitude, is:
\begin{subequations}
\label{eq:linear_complex_stiffnesses}
\begin{align}
  K_1^{\ast}(\Omega)
  &=1-\mu\Omega^2+\ii\Omega C_{\mathrm b}
    +\frac{G_{\mathrm b}}{1+\ii\lambda_1\Omega},
  \label{eq:first_order_complex_stiffness}\\
  K_{\alpha}^{\ast}(\Omega)
  &=1-\mu\Omega^2+\ii\Omega C_{\mathrm b}
    +\frac{G_{\mathrm b}}
      {1+\lambda_{\alpha}(\ii\Omega)^{\alpha}}.
  \label{eq:fractional_complex_stiffness}
\end{align}
\end{subequations}
Here, $K_1^{\ast}$ and $K_{\alpha}^{\ast}$ correspond to the first-order and
fractional memory laws, respectively, the superscript $\ast$ denotes a
complex-valued quantity.  The term $1-\mu\Omega^2$
combines spring and inertial effects, $\ii\Omega C_{\mathrm b}$ represents
local damping, and the final term describes the force change caused by a
memory-state perturbation.

The role of motion in the memory coupling follows directly from
Eq.~\eqref{eq:tangent_coefficients}. At a static equilibrium the reference
velocity is zero, so $G_{\mathrm b}=0$: a small state change cannot alter
the damping force to first order when the reference velocity vanishes.
Both memory laws therefore give:
\begin{equation}
  \mu\xi''+C_{\mathrm b}\xi'+\xi=p.
  \label{eq:static_tangent}
\end{equation}
The memory-state change still follows
Eq.~\eqref{eq:linearised_first_order_memory} or
Eq.~\eqref{eq:linearised_fractional_memory}. Consequently, small mechanical
measurements about a static equilibrium identify $C_{\mathrm b}$ but cannot
distinguish the two memory laws at first order. Their mechanical effects
require terms beyond this approximation or a moving reference response.
For the inertia-free calculations considered below, setting $\mu=0$ in
Eq.~\eqref{eq:static_tangent} gives $C_{\mathrm b}\xi'+\xi=p$.

The full nondimensional models in
Eqs.~\eqref{eq:first_order_nondim_general} and
\eqref{eq:fractional_nondim_general}, together with the harmonic state
transfer in Eq.~\eqref{eq:Halpha}, now provide the basis for comparing
changes in damping and energy loss across operating conditions.

\section{Explicit scaling laws}

\subsection{Conditional effective-memristance measures}
\label{sec:conditional_effective_memristance}

The small-response analysis shows that damping variation depends on both
 the amplitude of the memory-state change and the slope of the constitutive
law at the operating state. Therefore, the screening index $\Mh$
defined in Eq.~\eqref{eq:MH_screening_definition} is used, together with the
state-amplitude parameter $\chi$ in Eq.~\eqref{eq:state_excursion_parameter}.
The index $\Mh$ combines the damping scale $\beta$, the state-scale ratio
$\gamma$ and the fraction of the displacement amplitude transmitted to the
memory state, $|\mathcal H_{\alpha}|$. Together with the response amplitude
and the constitutive slope, it sets the change in damping derived below.
The parameter $\chi$ instead locates the range of $\gamma y$ sampled during
a cycle. Thus the two measures connect state motion to coefficient
variation without assuming that either alone determines dissipated energy.

Consider a small harmonic displacement of amplitude $A$ about a fixed
mean memory state $y_{\mathrm b}$, and write its constitutive argument as
$z_{\mathrm b}=\gamma y_{\mathrm b}$. The resulting state-argument amplitude
is $\chi$ by Eq.~\eqref{eq:state_excursion_parameter}. Expanding
$c_{\mathrm e}=\beta\mathcal G(\gamma y)$ about $z_{\mathrm b}$ gives the
peak departure of the coefficient from its operating value,
$\Delta c_{\mathrm e,\mathrm{amp}}$. To first order in the response amplitude,
\begin{equation}
  \Delta c_{\mathrm e,\mathrm{amp}}
  \simeq
  \beta\left|
    \mathcal G'(z_{\mathrm b})
  \right|\chi
  =A\left|
    \mathcal G'(z_{\mathrm b})
  \right|\Mh.
  \label{eq:local_coefficient_modulation}
\end{equation}

Equation~\eqref{eq:local_coefficient_modulation} applies to both the
first-order and fractional memory models through their respective
$\mathcal H_{\alpha}$. It shows that coefficient modulation is proportional
to $\Mh$ when the displacement amplitude and the non-zero constitutive
slope $|\mathcal G'(z_{\mathrm b})|$ are held fixed. A material comparison
using this index therefore requires consistent specimen geometry,
reference scales, loading conditions, coefficient law and operating bias.
In particular, $\epsilon$ and $\delta$ must be matched because they set the
coefficient's baseline and shape. These limitations make $\Mh$ a
conditional screening measure, its use in material selection is discussed
in Section~\ref{sec:material_mapping}.

As established in Section~\ref{sec:harmonic_transfer}, $\chi$ is unchanged
by the arbitrary choice of displacement scale $u_{\mathrm c}$. It therefore
allows responses with the same coefficient law to be compared by the
range of memory states sampled, even if different displacement units or
reference scales are used. Although $\chi$ has no explicit factor of $\beta$, the response amplitude
$A$ can depend on $\beta$ under force-controlled loading. State amplitude
alone therefore cannot rank damping magnitude or energy loss.

The non-zero slope condition fails for the adopted symmetric law when
the mean state is at the origin, $z_{\mathrm b}=0$ (zero bias).
The relation $\mathcal G'(0)=0$ means that the damping coefficient has zero
slope there. A small state change therefore affects the coefficient only
at second order. For $\delta>0$ and $\chi\ll\min(1,\delta)$,
Eq.~\eqref{eq:positive_small_argument} gives the maximum increase above
$c_{\mathrm e}(0)$ during the cycle:
\begin{equation}
  \Delta c_{\mathrm e}
  \simeq\frac{\beta}{2\delta}\chi^2
  =\frac{\beta\gamma^2A^2}{2\delta}
   \left|\mathcal H_{\alpha}(\Omega)\right|^2.
  \label{eq:zero_bias_modulation_scale}
\end{equation}

Thus the leading damping change at zero bias scales with $\chi^2$,
so a linear ranking by $\Mh$ is not applicable in this regime. Both
measures describe a specified operating response rather than an intrinsic
material constant.

\subsection{Exact periodic energy and coefficient modulation}
\label{sec:cycle_measures}

A damping coefficient can vary strongly while dissipating little energy
if its largest values occur when the velocity is small. To evaluate the
actual damping performance, the work dissipated
over a complete cycle and the constant viscous coefficient are calculated. This comparison follows the
established energy-equivalence approach to nonlinear damping
\cite{Jacobsen1930Damping,Papagiannopoulos2018Equivalent}.

For a closed periodic response, let $T$ be its period in nondimensional
time $\tau$, $T=2\pi/\Omega$ for a response at the forcing period.
Multiplying Eq.~\eqref{eq:first_order_nondim_mech} or
\eqref{eq:fractional_nondim_mech} by $x'$ and integrating over that period
uses the power balance in Eq.~\eqref{eq:appendix_power_balance} to give:
\begin{equation}
  \mathcal E_{\mathrm{cyc}}
  =\oint f\,\dd x
  =\int_0^T c_{\mathrm e}\bigl(y(\tau)\bigr)
    x'(\tau)^2\,\dd\tau
  \ge0.
  \label{eq:exact_periodic_energy}
\end{equation}
Here $\mathcal E_{\mathrm{cyc}}$ is the nondimensional work lost per cycle,
the corresponding dimensional energy is $k u_{\mathrm c}^{2}\mathcal E_{\mathrm{cyc}}$.
The integral $\oint f\,\dd x$ measures the work of the applied force along
the closed displacement cycle. Stored inertial and spring energies return
to their initial values, leaving only dissipation. For a cycle with non-zero
motion, $\int_0^T x'^2\,\dd\tau>0$, the energy-equivalent nondimensional
coefficient is therefore
\begin{equation}
  c_{\mathrm{eq}}
  =\frac{
    \displaystyle\int_0^T c_{\mathrm e}(y)x'^2\,\dd\tau
  }{
    \displaystyle\int_0^T x'^2\,\dd\tau
  }.
  \label{eq:energy_equivalent_coefficient}
\end{equation}
This is a velocity-squared-weighted average of $c_{\mathrm e}$, replacing
the mem-dashpot by a constant coefficient $c_{\mathrm{eq}}$ would dissipate
exactly the same energy along the specified displacement history.
The dimensional coefficient is $k t_{\mathrm c}c_{\mathrm{eq}}$.
Equations~\eqref{eq:exact_periodic_energy} and
\eqref{eq:energy_equivalent_coefficient} are exact cycle measures and
require no sinusoidal-response approximation.

The same cycle energy can arise from nearly constant damping or from
substantial changes in damping as the state evolves. To distinguish these
responses, an index comparing the full range of coefficient
values with the energy-equivalent coefficient is introduced:
\begin{equation}
  \mathcal I_C
  =\frac{c_{\mathrm e,\max}-c_{\mathrm e,\min}}
         {c_{\mathrm{eq}}},
  \label{eq:coefficient_modulation_index}
\end{equation}
where $c_{\mathrm e,\max}$, $c_{\mathrm e,\min}$ and $c_{\mathrm{eq}}$ are all evaluated over the same periodic cycle. The numerator of $\mathcal I_C$ is the full within-cycle range of the instantaneous damping coefficient, and division by $c_{\mathrm{eq}}$ expresses that range relative to the constant coefficient that would dissipate the same energy over that cycle. A constant dashpot therefore gives $\mathcal I_C=0$, whereas a larger value means that the damping changes more strongly as the memory state evolves. A large $\mathcal I_C$ does not by itself imply a large energy loss, because Eq.~\eqref{eq:exact_periodic_energy} weights the coefficient by the squared velocity and therefore also depends on when the larger values occur. Both $c_{\mathrm{eq}}$ and $\mathcal I_C$ remain dependent on amplitude, frequency and operating state. The force--velocity relation $F_{\mathrm d}=c_{\mathrm m}(q)\dot u$
describes the state-dependent damping at each instant, whereas the closed
force--displacement integral measures work over a cycle. These relations
therefore provide complementary measures of memory-dependent damping
and energy dissipation.

\section{Mechanical memristor performance in the nondimensional design space} \label{sec:parametric_space}

\begin{figure}[htbp]
    \centering
    \includegraphics[width=\linewidth]{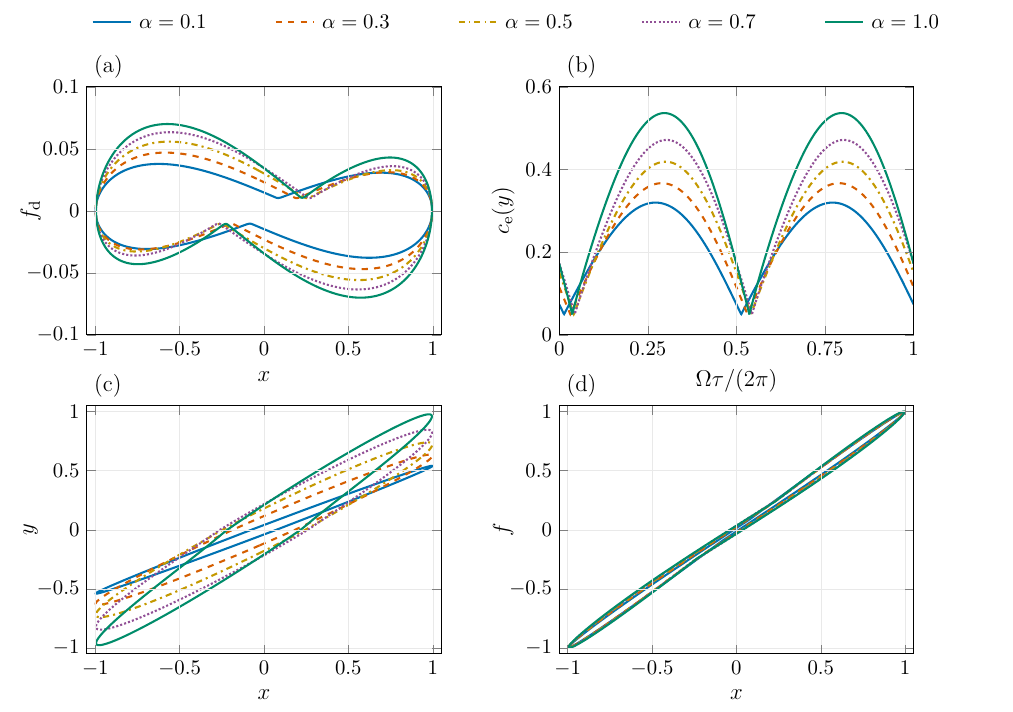}
    \caption{Long-time periodic response for several memory orders: (a) damping force $f_{\mathrm d}=F_{\mathrm d}/F_c$ against displacement $x$; (b) variation of $c_{\mathrm e}$ over one forcing cycle; (c) nondimensional memory state against displacement; and (d) nondimensional force--displacement loops. The first-order memory model corresponds to $\alpha=1.0$.}
    \label{fig:periodic_response}
\end{figure}

The nonlinear formulations introduced above were used to compute the long-time periodic responses discussed below. Details of the numerical procedure are provided in Appendix~\ref{app:numerical}. The role of memory order is first isolated in Fig.~\ref{fig:periodic_response}. Figure~\ref{fig:periodic_response}(a) shows the corresponding damping-force trajectories, while Fig.~\ref{fig:periodic_response}(b) shows the variation of the nondimensional damping coefficient $c_{\mathrm e}$ within each cycle. Increasing \(\alpha\) enlarges the memory-state excursion and modifies its phase relation with the displacement. Through Eq.~\eqref{eq:nondim_coefficient}, this enhances the variation of the instantaneous damping coefficient over the cycle and broadens the force--displacement hysteresis loop. The memory-state \((y,x)\) loop area increases from \(0.123\) at \(\alpha=0.1\) to a maximum of \(0.658\) at \(\alpha=0.7\), before decreasing slightly in the first-order limit. These areas describe the lag of the memory state, not dissipated energies. By contrast, the force--displacement loop area increases monotonically from \(0.104\) to \(0.171\) as \(\alpha\rightarrow1\), indicating increasing cycle dissipation for the parameters considered here.

\begin{figure}[htbp]
    \centering
    \includegraphics[width=\linewidth]{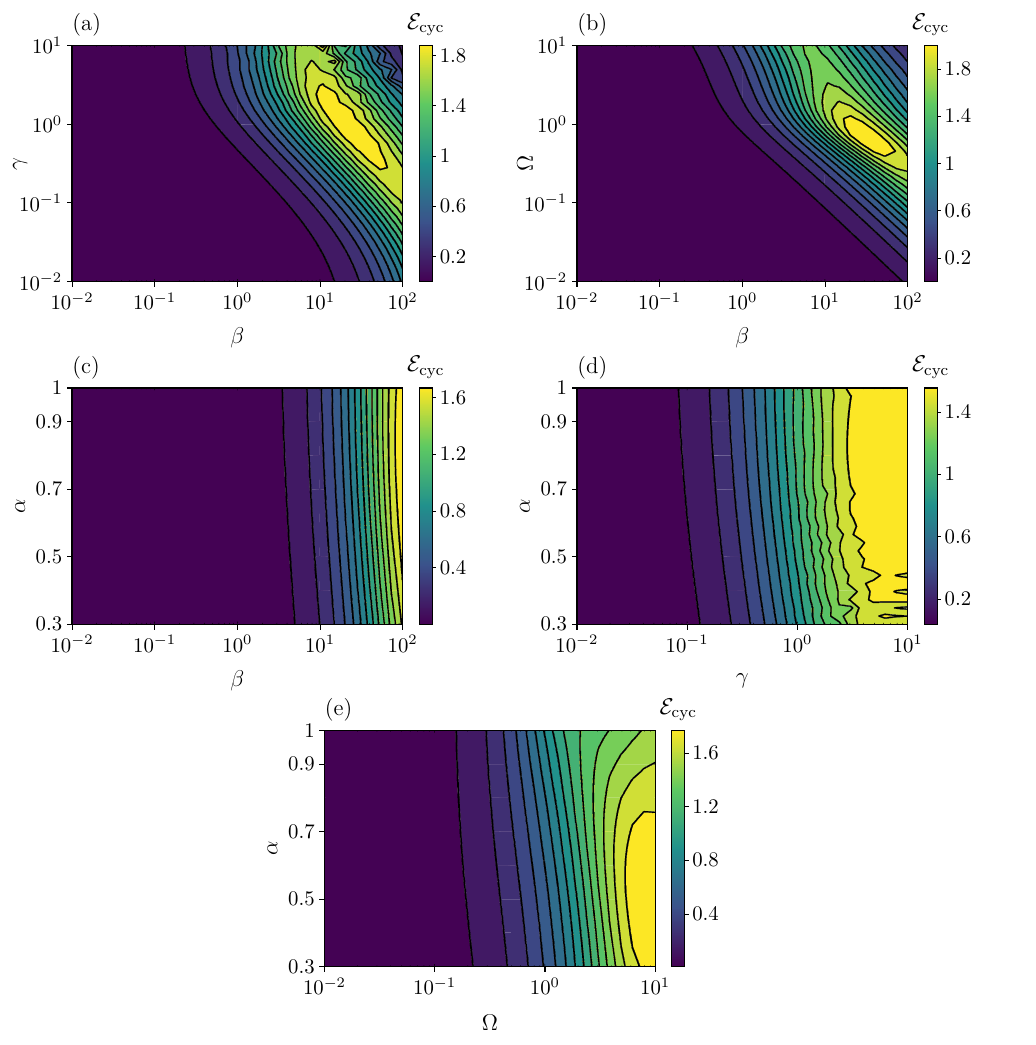}
    \caption{Exact periodic-energy maps. Panels (a) and (b) show the first-order memory model ($\alpha=1.0$): (a) $(\beta,\gamma)$ at $\Omega=0.2$ and (b) $(\beta,\Omega)$ at $\gamma=0.1$. The fractional memory model is shown for (c) $(\beta,\alpha)$ at $(\gamma=0.1,\Omega=0.2)$, (d) $(\gamma,\alpha)$ at $(\beta=5,\Omega=0.2)$ and (e) $(\Omega,\alpha)$ at $(\beta=5,\gamma=0.1)$. The plotted quantity $\mathcal E_{\mathrm{cyc}}$ is defined in Eq.~\eqref{eq:exact_periodic_energy}.}
    \label{fig:exact_periodic_energy_maps}
\end{figure}

The exact cycle energy \(\mathcal E_{\mathrm{cyc}}\), defined in Eq.~\eqref{eq:exact_periodic_energy}, shows a common trend across both the first-order and fractional memory models (Fig.~\ref{fig:exact_periodic_energy_maps}). Small values of either \(\beta\) or \(\gamma\) produce weak dissipation, whereas intermediate combinations maximise cycle energy. At large \(\beta\), the response becomes strongly damped and the displacement amplitude decreases, leading to a reduction in \(\mathcal E_{\mathrm{cyc}}\). The fractional order \(\alpha\) primarily shifts the frequency range over which memory can be effectively retained rather than uniformly increasing or decreasing the cycle energy. Consequently, maximum dissipation is controlled by a balance between damping magnitude, memory-state excursion and excitation frequency. Figures~\ref{fig:exact_periodic_energy_maps}(c) and (d) show how the fractional order $\alpha$ modifies these trends when varied with $\beta$ and $\gamma$, respectively.

\begin{figure}[htbp]
    \centering
    \includegraphics[width=\linewidth]{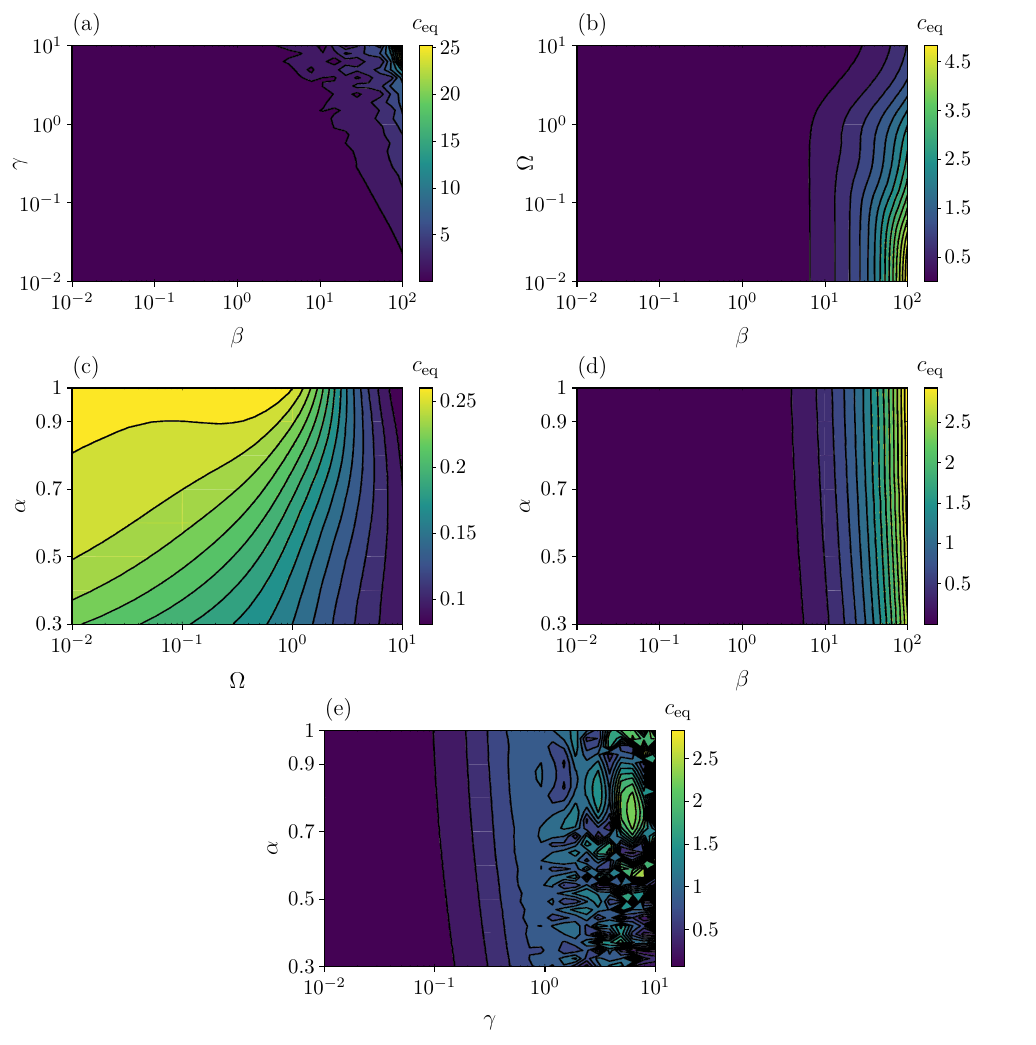}
    \caption{Energy-equivalent coefficient $c_{\mathrm{eq}}$ defined in Eq.~\eqref{eq:energy_equivalent_coefficient}. Panels (a) and (b) show the first-order memory model ($\alpha=1.0$): (a) $(\beta,\gamma)$ at $\Omega=0.2$ and (b) $(\beta,\Omega)$ at $\gamma=0.1$. The fractional memory model is shown for (c) $(\Omega,\alpha)$ at $(\beta=5,\gamma=0.1)$, (d) $(\beta,\alpha)$ at $(\gamma=0.1,\Omega=0.2)$ and (e) $(\gamma,\alpha)$ at $(\beta=5,\Omega=0.2)$.}
    \label{fig:ceq_maps}
\end{figure}

The energy-equivalent coefficient \(c_{\mathrm{eq}}\) converts cycle energy into an equivalent viscous representation (Fig.~\ref{fig:ceq_maps}). In both memory models, \(\beta\) provides the dominant scaling, while \(\gamma\) controls the sensitivity of the equivalent damping to memory-state evolution. Increasing \(\Omega\) generally reduces \(c_{\mathrm{eq}}\) because the memory state has less time to follow the imposed displacement. The role of \(\alpha\) is primarily to shift the frequency range in which memory transfer remains effective. Thus, \(c_{\mathrm{eq}}\) provides a useful global measure of dissipation, but it does not fully capture the local coefficient variations responsible for memristive behaviour. Because the average in Eq.~\eqref{eq:energy_equivalent_coefficient} is weighted by squared velocity, these maps also depend on the response waveform and the timing of the coefficient variation.

\begin{figure}[htbp]
    \centering
    \includegraphics[width=\linewidth]{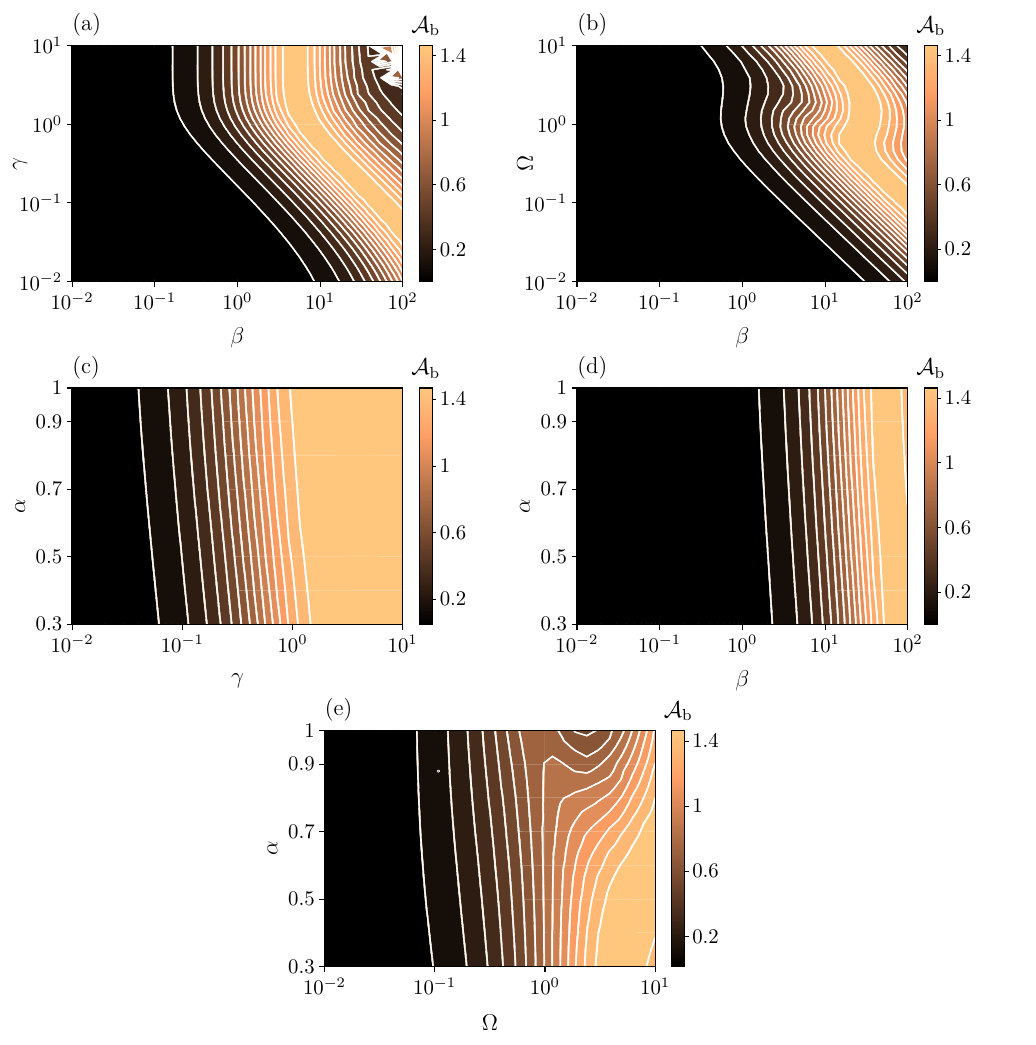}
    \caption{Loop-area maps obtained from the tangent representation based on $C_{\mathrm b}$. Panels (a) and (b) show the first-order memory model ($\alpha=1.0$): (a) $(\beta,\gamma)$ at $\Omega=0.2$ and (b) $(\beta,\Omega)$ at $\gamma=0.1$. The fractional memory model is shown for (c) $(\gamma,\alpha)$ at $(\beta=5,\Omega=0.2)$, (d) $(\beta,\alpha)$ at $(\gamma=0.1,\Omega=0.2)$ and (e) $(\Omega,\alpha)$ at $(\beta=5,\gamma=0.1)$.}
    \label{fig:Cb_maps}
\end{figure}

The tangent coefficient \(C_{\mathrm b}\) obtained from the frozen-state approximation provides a local representation of the periodic response (Fig.~\ref{fig:Cb_maps}). Although \(C_{\mathrm b}\) follows many of the same trends as \(c_{\mathrm{eq}}\), it does not reproduce the extreme values generated by the full nonlinear cycle. Consequently, \(C_{\mathrm b}\) is useful for local linearisation and small-signal analysis but should not be interpreted as a substitute for either \(c_{\mathrm{eq}}\) or \(\mathcal E_{\mathrm{cyc}}\). Figure~\ref{fig:Cb_maps} shows the loop area $\mathcal A_{\mathrm b}=\pi\Omega C_{\mathrm b}/[1+(\Omega C_{\mathrm b})^2]$ obtained for unit harmonic forcing in the inertia-free tangent model, with $C_{\mathrm b}$ defined in Eq.~\eqref{eq:tangent_coefficients}.

\begin{figure}[htbp]
    \centering
    \includegraphics[width=\linewidth]{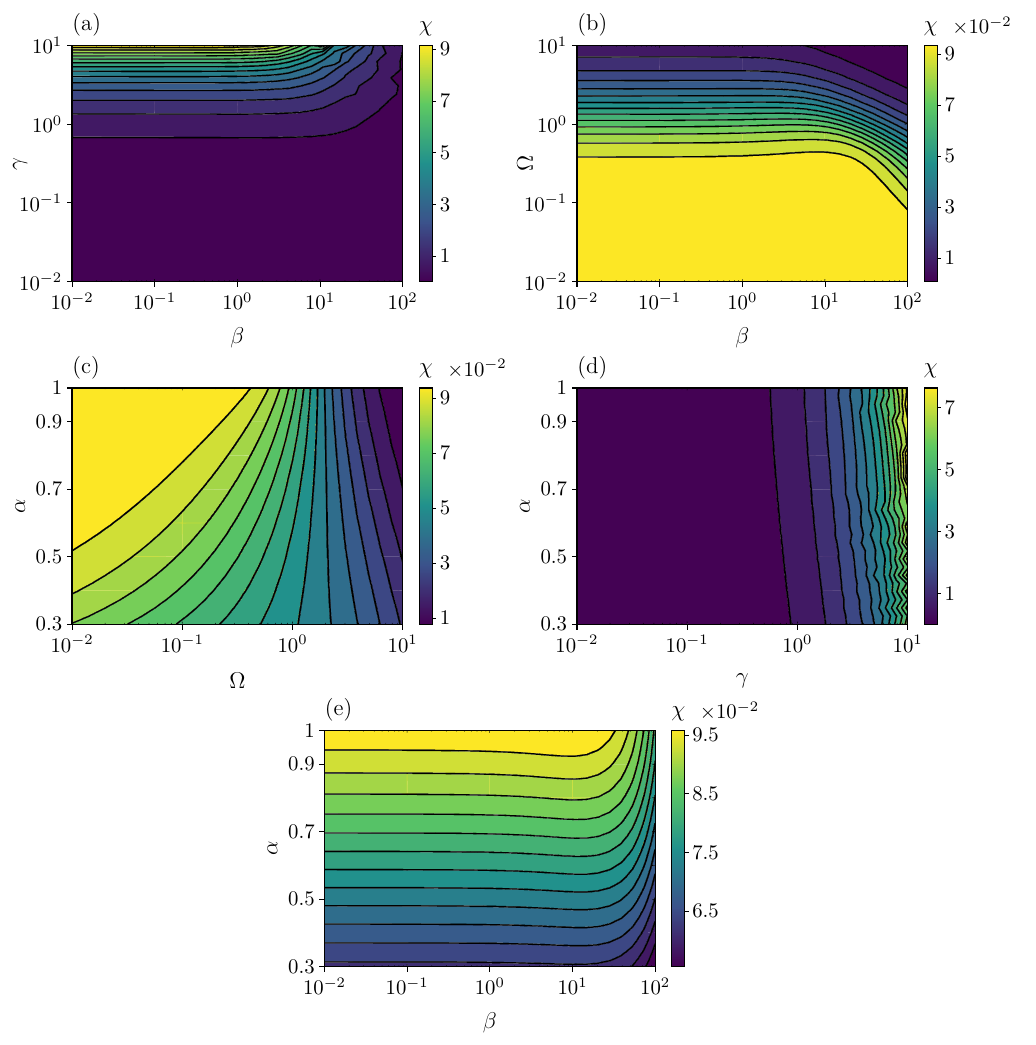}
    \caption{State-excursion parameter $\chi$ defined in Eq.~\eqref{eq:state_excursion_parameter}. Panels (a) and (b) show the first-order memory model ($\alpha=1.0$): (a) $(\beta,\gamma)$ at $\Omega=0.2$ and (b) $(\beta,\Omega)$ at $\gamma=0.1$. The fractional memory model is shown for (c) $(\Omega,\alpha)$ at $(\beta=5,\gamma=0.1)$, (d) $(\gamma,\alpha)$ at $(\beta=5,\Omega=0.2)$ and (e) $(\beta,\alpha)$ at $(\gamma=0.1,\Omega=0.2)$.}
    \label{fig:chi_maps}
\end{figure}

Unlike \(\mathcal E_{\mathrm{cyc}}\) and \(c_{\mathrm{eq}}\), the state-excursion parameter \(\chi\) isolates the extent to which the nonlinear memory map is explored during a cycle (Fig.~\ref{fig:chi_maps}). Because \(\chi\) is independent of the damping scale \(\beta\), it separates memory-state evolution from dissipation magnitude. The maps show that \(\gamma\) primarily controls the amplitude of memory-state excursions, whereas \(\alpha\) shifts the frequency range over which the memory state can effectively follow the imposed displacement. In this sense, \(\chi\) quantifies the degree to which the nonlinear constitutive law is sampled during cyclic operation.

\begin{figure}[htbp]
    \centering
    \includegraphics[width=\linewidth]{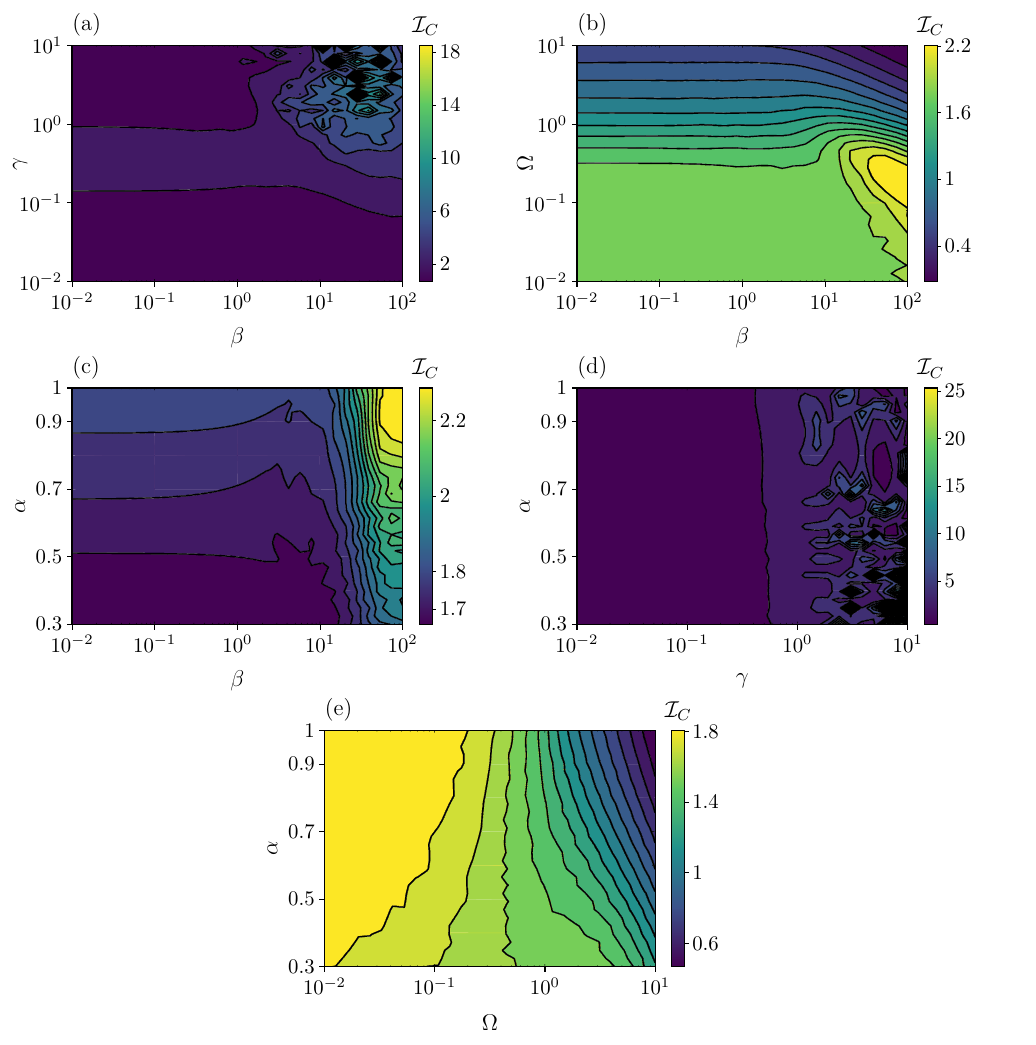}
    \caption{Coefficient-modulation index $\mathcal I_C$ defined in Eq.~\eqref{eq:coefficient_modulation_index}. Panels (a) and (b) show the first-order memory model ($\alpha=1.0$): (a) $(\beta,\gamma)$ at $\Omega=0.2$ and (b) $(\beta,\Omega)$ at $\gamma=0.1$. The fractional memory model is shown for (c) $(\beta,\alpha)$ at $(\gamma=0.1,\Omega=0.2)$, (d) $(\gamma,\alpha)$ at $(\beta=5,\Omega=0.2)$ and (e) $(\Omega,\alpha)$ at $(\beta=5,\gamma=0.1)$.}
    \label{fig:Ic_maps}
\end{figure}

The coefficient-modulation index \(\mathcal I_C\) (Fig.~\ref{fig:Ic_maps}) completes the picture by quantifying the variation of the damping coefficient over a cycle. Unlike \(\chi\), which measures state excursion alone, \(\mathcal I_C\) also depends on the local slope of the constitutive law. Large memory excursions do not necessarily imply strong coefficient modulation. Instead, large values of \(\mathcal I_C\) occur only when substantial state evolution, significant response amplitudes and strong constitutive sensitivity are present simultaneously. The maps therefore confirm that memristive behaviour depends on both the extent of memory-state evolution and the sensitivity of the damping law to that evolution. Near an operating state with nonzero slope, their connection follows from Eq.~\eqref{eq:local_coefficient_modulation}, near zero bias, the quadratic relation in Eq.~\eqref{eq:zero_bias_modulation_scale} applies instead.

As a summary, the mechanical memristor is therefore represented as a state-dependent dashpot in parallel with a linear spring, with a strictly positive coefficient that ensures passive behaviour and an internal displacement-like state governed by either first-order or Caputo fractional relaxation. The harmonic transfer $\mathcal H_{\alpha}$ related displacement to the memory state, while $\chi=\gamma A|\mathcal H_{\alpha}|$ measures the portion of the nonlinear coefficient sampled during each cycle. The periodic results show that $\beta$ primarily sets the damping level, $\gamma$ sets its sensitivity to the memory state and $\alpha$ together with $\Omega$ controls how closely that state follows the displacement. Energy dissipation is greatest at intermediate coupling, where the coefficient varies appreciably without strongly suppressing the response. The complementary measures $c_{\mathrm{eq}}$ and $\mathcal I_C$ then distinguish cycle-averaged damping from within-cycle coefficient modulation. These response-based relations provide the basis for mapping the scaled formulation to candidate material mechanisms and practical devices.

\section{Mapping scaled quantities to structures and materials for designing mechanical memristors}
\label{sec:material_mapping}

\subsection{Interpretation of candidate material classes} \label{sec:material_candidates}

The canonical mem-dashpot introduced here is a reduced state-space representation of memory-dependent dissipation. In practical systems, the memory variable is not necessarily displacement-like and may instead represent a martensitic fraction, polarisation state, moisture content, solvent distribution, particle-chain morphology or force-chain topology. To relate such variables to the canonical formulation, a constitutive mapping is introduced, \begin{equation} q=q_{\star}h(\eta), \label{eq:physical_state_map} \end{equation} where \(\eta\) is the physical memory variable, \(q_\star\) is a characteristic memory scale and \(h\) is a dimensionless mapping function. This mapping places the internal coordinate on a common scale, but does not by itself determine its force coupling or evolution law. The material classes considered below provide guidance on plausible choices of state variables and illustrative parameter sets. Shape-memory polymers, hydrogels, nanocellulose and natural fibres derive memory from distributed relaxation and network reconfiguration; shape-memory alloys from reversible martensitic transformations; piezoelectric and ferroelectric materials from evolving polarisation states; electrorheological and magnetorheological fluids from field-induced particle networks; and granular systems from contact and force-chain evolution. For a given specimen, the canonical representation is admissible only if the measured force and state evolution are consistent with the assumed force--velocity and relaxation laws. State-dependent restoring forces and transformation hysteresis require constitutive elements beyond the present constant spring and bounded mem-dashpot.

\subsection{General mapping of materials}
\label{sec:mapping_materials}

To map a material into the model, we first specify the specimen geometry, the way it is loaded and the measurement procedure.  The dimensional
quantities $m$, $k$, $c_{\star}$, $q_{\star}$,
$\tau_{\mathrm m}$ or $(\tau_{\alpha},\alpha)$, $\epsilon$ and $\delta$
must refer to that same configuration.  Model parameters are identified through direct comparison of the measured force response with Eq.~\eqref{eq:common_mechanics_dimensional}, subject to the adopted memory evolution law. This procedure avoids division by velocity, which approaches zero at turning points and may therefore magnify measurement errors.

For small-strain linear viscoelastic data represented by a generalised
Maxwell spectrum \cite{Christensen1982},
\begin{equation}
  G(t)=G_{\infty}+\sum_{i=1}^{N}G_i\exp(-t/\tau_i),
  \label{eq:gmm_relaxation_modulus}
\end{equation}
the zero-frequency viscosity of the relaxing branches is
\begin{equation}
  \eta_0
  =\int_0^{\infty}\bigl[G(t)-G_{\infty}\bigr],\dd t
  =\sum_{i=1}^{N}G_i\tau_i.
  \label{eq:gmm_zero_frequency_viscosity}
\end{equation}
A device-level reference coefficient can then be written as
\begin{equation}
  c_{\mathrm{ref}}=K_f\eta_0,
  \label{eq:gmm_device_coefficient}
\end{equation}
where $K_f$ contains the geometry, for a uniform axial specimen,
$K_f=A_{\mathrm s}/L$.  Setting $K_f=1$ is meaningful only after an explicit
per-unit-geometry normalisation.  If $c_{\mathrm{ref}}$ is measured at the
state $z_{\mathrm{ref}}$, then
\begin{equation}
  c_{\star}
  =\frac{c_{\mathrm{ref}}}
  {\mathcal G_{\epsilon,\delta}(z_{\mathrm{ref}})}.
  \label{eq:cstar_reference_state}
\end{equation}
In particular, the zero-state coefficient of the present law is
$c_{\mathrm m}(0)=c_{\star}\epsilon$, not $c_{\star}$.

The first two frequency-dependent terms of the low-frequency Maxwell
expansion define a characteristic time of that spectrum,
\begin{equation}
  \tau_{\mathrm m}^{(\mathrm{mom})}
  =\frac{\sum_iG_i\tau_i^2}{\sum_iG_i\tau_i}.
  \label{eq:tau_m_moment_matched}
\end{equation}
This moment characterises the Maxwell spectrum and does not identify the
canonical memory time or the nonlinear dependence of $c_{\mathrm m}$ on $q$.
The relaxing Maxwell branches store energy and can change force at fixed
displacement, whereas the present spring-mem-dashpot retains $F_{\mathrm r}=ku$.  For the
fractional model, $\alpha$ and $\tau_{\alpha}$ must instead be fitted jointly
from the measured magnitude and phase of the state transfer over the intended
frequency interval \cite{Podlubny1999,Mainardi2010}.  The nonlinear
coefficient and the state map in Eq.~\eqref{eq:physical_state_map} must still
be identified from amplitude-dependent force data.

Once the dimensional quantities and the reference scales have been fixed for
the same specimen, the dimensionless groups are calculated using Eq.~\eqref{eq:nondim_groups_main} and the memory-time ratio $\lambda_{\alpha}$ introduced with Eq.~\eqref{eq:fractional_nondim_mem}.

Table~\ref{tab:material_mapping_summary} lists candidate physical memory states and the conditions needed to relate them to the model. These states may involve changes in elastic forces, phase transformations or field coupling that the present single-state spring--mem-dashpot does not capture. The table therefore also identifies limitations of the proposed mapping.

\begin{table}[t]
\centering
\caption{Routes for mapping candidate material classes to
the canonical passive mem-dashpot.}
\label{tab:material_mapping_summary}
\scriptsize
\begin{tabular}{@{}p{0.14\textwidth}p{0.18\textwidth}p{0.27\textwidth}p{0.31\textwidth}@{}}
\hline
\textbf{Class} & \textbf{State} &
\textbf{Supported behaviour} & \textbf{Required qualification}\\
\hline
Geometry-programmed hydraulic dashpot & Terminal or spool position controlling
flow resistance & Tapered dashpots and shaped laminar resistance provide a
direct mechanical analogue \cite{OsterAuslander1972,PeiEtAl2015,HappelBrenner1983}. &
The mapping assumes Newtonian, laminar, incompressible flow with negligible
friction and a calibrated geometry.\\
Viscoelastic polymers, shape-memory polymers and hydrogels & Relaxation
spectrum, frozen strain, network or solvent state & Broad spectra and
fractional viscoelastic descriptions are established
\cite{Bonfanti2020,Liu2006}. & Linear hereditary response alone is not a
state-dependent damping coefficient; the nonlinear coupling must be measured.\\
Shape-memory alloys & Martensitic fraction, transformation strain and
temperature & Transformation history and hysteresis are established
\cite{Lagoudas2008}. & A history-dependent force at held displacement or finite quasistatic
hysteresis requires constitutive terms beyond the present constant spring
and bounded mem-dashpot, together with appropriate thermal conditions.\\
Piezoelectric polymers and ferroelectric ceramics & Polarisation and domain
state & Electromechanical loss and hysteresis are documented
\cite{Vinogradov2004,Damjanovic2006}. & Polarisation is not displacement; a
coupled state model and Eq.~\eqref{eq:physical_state_map} are required.\\
Electro- and magnetorheological devices & Field-dependent particle network and
controller state & Field-induced yield and controllable hysteretic damping are
established \cite{Winslow1949,Spencer1997}. & These devices are normally
semi-active and may include yield forces outside $c(q)\dot u$; field power and
control must be stated.\\
Granular dampers & Packing fraction, contact network, force chains and granular
temperature & Compaction memory and history dependence are observed
\cite{Jaeger1996,Josserand2000}. & Impact and friction are non-smooth,
multi-state and amplitude dependent, so one smooth viscous coefficient may be
insufficient.\\
Lignocellulosic and natural-fibre systems & Moisture, cell-wall relaxation,
fibril orientation and interface state & Hierarchical viscoelasticity and
moisture sensitivity are established \cite{Salmen2004,Thybring2022}. & A
calibrated coupled hygro-viscoelastic law is required; biological origin alone
does not establish a $\tanh$ memristance.\\
\hline
\end{tabular}
\end{table}

The group $\beta\gamma|\mathcal H_{\alpha}|$ is used in this mapping only as a conditional screening index.  Equation~\eqref{eq:local_coefficient_modulation} shows that it is proportional to local first-harmonic coefficient modulation only when the displacement amplitude, reference scales, specimen geometry, loading protocol, coefficient law, $\epsilon$, $\delta$ and the bias $z_{\mathrm b}$ are fixed, the bias has a common non-zero constitutive slope, and higher harmonics are negligible.  These conditions do not hold at zero bias for the adopted even law because $\mathcal G_{\epsilon,\delta}'(0)=0$, the leading rounded-core change then scales with $\beta\gamma^2A^2|\mathcal H_{\alpha}|^2/\delta$.  The parameter $\chi=\gamma A|\mathcal H_{\alpha}|$ remains the most general state-excursion measure, but it is an operating-response quantity, rather than an intrinsic material constant.  

\subsection{Limitations of the material mapping} \label{sec:material_mapping_limitations}

The illustrative scenarios considered below are examined through Eq.~\eqref{eq:local_coefficient_modulation}. Within the canonical model, this relation describes local coefficient modulation only for a fixed response amplitude and a common non-zero constitutive slope. The resulting design map illustrates representative parameter choices within the model. Nevertheless, several limitations should be recognised when interpreting the parameter ranges summarised later in Table~\ref{tab:materialmap}. First, the ranges should not be interpreted as relative measures of material performance, since no common experimental identification procedure has been applied across the cited systems. Second, the mapping is based on a reduced state-variable description. Although the generalised representation $q=q_\star h(\eta)$ provides a common coordinate scale, the underlying state variable may correspond to very different physical quantities, including recoverable strain, polarisation, martensitic fraction, network configuration, particle chain morphology or force-chain topology. Consequently, the state mapping neither establishes constitutive equivalence between material classes nor validates the canonical force law. The numerical ranges are illustrative parameter assignments, informed by the relaxation mechanisms and internal state variables reported in the literature, and should be regarded as assumed scenarios rather than material-specific calibrations. Also, the effective mechanical memristance \(M_H = \beta\gamma |\mathcal H_\alpha(\Omega)|\), is not an intrinsic material property. The quantity depends not only on the material memory mechanism, but also on specimen geometry, characteristic displacement scale, loading frequency and state-variable normalisation. Furthermore, it should be noted that the quantities reported in Table~\ref{tab:materialmap} are contingent upon the assumption that the memory-transfer magnitude satisfies: \begin{equation} 
    |\mathcal H_\alpha(\Omega)| = O(0.1-1) 
\end{equation}

corresponding to operating frequencies comparable to the dominant memory timescales of the material. Changes in loading frequency may alter the effective value of \(M_H\) significantly even when the underlying material remains unchanged. 

Finally, many material classes considered here exhibit multiple interacting memory mechanisms rather than a single dominant memory state. Examples include lignocellulosic materials, hydrogels, piezoelectric systems and field-responsive fluids, in which several relaxation processes may coexist over different spatial and temporal scales.

\subsection{Comparing materials for mechanical memristor configurations at fixed geometry, frequency and amplitudes} \label{sec:materials}

The material classes discussed below should be viewed as candidate physical realisations of the canonical mem-dashpot framework rather than direct embodiments of Eq.~\eqref{eq:smooth_positive_law}. The ranges of \(\alpha\), \(\beta\), \(\gamma\) and \(M_H\) reported in this section are intended as order-of-magnitude engineering estimates that combine published constitutive, rheological and damping data with the scaling relations developed in the present work. In general, published studies most directly support the fractional-memory parameter \(\alpha\), whereas the parameters \(\beta\), \(\gamma\) and the resulting effective mechanical memristance \(M_H\) are inferred through the present framework and should therefore be regarded as evidence-informed screening quantities rather than intrinsic material constants. The materials considered span conventional smart-material systems, including shape-memory polymers (SMPs), shape-memory alloys (SMAs), piezoelectric polymers, piezoelectric ceramics, electrorheological (ER) fluids and magnetorheological (MR) fluids, together with bio-derived systems such as hydrogels, nanocellulose, lignin-rich materials and natural fibres. Illustrative examples of the parameter-identification procedure are given in Appendices~\ref{app:SMPexample}, \ref{app:PVDF_example} and \ref{app:SMA_example}. The resulting estimates should be interpreted as representative locations within the design space developed above and summarised in Section~\ref{sec:unified_material_map} rather than as definitive rankings of the corresponding material classes.

\subsubsection{Shape-Memory Polymers} Available thermomechanical and fractional-order constitutive studies of epoxy and polyurethane SMPs suggest \(0.6\lesssim\alpha\lesssim0.9\), together with estimated values \(1\lesssim\beta\lesssim10\) and \(0.05\lesssim\gamma\lesssim0.5\) \cite{Liu2006,Chen2014,Leng2011,Fang2015,Fang2016}. These values imply \(M_H=O(0.1-5)\), placing SMP-based systems in a moderate-memristance region characterised by large recoverable deformation and relatively long memory retention.

\subsubsection{Shape-Memory Alloys} Fractional-order descriptions and thermomechanical characterisation of NiTi systems suggest \(0.75\lesssim\alpha\lesssim0.95\), with \(5\lesssim\beta\lesssim20\) and \(0.5\lesssim\gamma\lesssim5\) estimated from transformation strains and hysteretic energy losses \cite{Lagoudas2008,PuenteCordova2023,Wang2023RateEffect,Guo2024FractionalSMA}. The resulting estimate \(M_H=O(2.5-100)\) indicates comparatively strong memory-dependent dissipation.

\subsubsection{Hydrogels} Hydrogels derive memory primarily from poroelastic relaxation, solvent transport and polymer-network restructuring. Fractional-rheology studies support \(0.3\lesssim\alpha\lesssim0.8\), while scaling estimates suggest \(0.1\lesssim\beta\lesssim10\) and \(0.2\lesssim\gamma\lesssim2\) \cite{Bonfanti2020,Raffaelli2021,Lenoch2022}. These values imply \(M_H=O(0.1-10)\), positioning hydrogels towards the long-memory region of the design space.

\subsubsection{Nanocellulose} Fractional-rheology studies of cellulose nanofibril networks support \(0.3\lesssim\alpha\lesssim0.8\), while rheological and network-relaxation data suggest \(0.5\lesssim\beta\lesssim10\) and \(0.2\lesssim\gamma\lesssim2\) \cite{Bonfanti2020,MirandaValdez2024,PuenteCordova2025}. The corresponding estimate \(M_H=O(0.1-20)\) places nanocellulose systems between hydrogels and strongly dissipative smart materials.

\subsubsection{Lignin-Rich Materials} Available viscoelastic and damping data for wood, technical lignins and lignin-based composites suggest \(0.2\lesssim\alpha\lesssim0.7\), \(1\lesssim\beta\lesssim20\) and \(0.5\lesssim\gamma\lesssim3\) \cite{Song2018,Bhattacharyya2020,Sternberg2023,Nadanyi2025,Ruwoldt2024}. These estimates correspond to \(M_H=O(0.5-60)\), indicating moderate-to-high memristive potential combined with long-lived memory effects.

\subsubsection{Natural Fibres} Studies of flax, hemp, jute and related lignocellulosic composites suggest \(0.4\lesssim\alpha\lesssim0.9\), \(0.5\lesssim\beta\lesssim15\) and \(0.1\lesssim\gamma\lesssim1\) \cite{Placet2009,Shah2013}. The resulting estimate \(M_H=O(0.05-15)\) places natural-fibre systems in an intermediate region of the design space.

\subsubsection{Piezoelectric Polymers} The candidate polymers include poly(vinylidene fluoride) (PVDF) and its copolymer with trifluoroethylene, P(VDF-TrFE). Published studies of PVDF and PVDF-TrFE support \(0.4\lesssim\alpha\lesssim0.8\), while dielectric, viscoelastic and ferroelectric data suggest \(1\lesssim\beta\lesssim15\) and \(0.2\lesssim\gamma\lesssim3\) \cite{Lovinger1983,Furukawa1989,Martins2014,Mohammadpourfazeli2023PVDFReview,Ahbab2025PVDFReview}. These values yield \(M_H=O(0.2-45)\), reflecting the coexistence of mechanical and polarisation-driven memory mechanisms.

\subsubsection{Piezoelectric Ceramics} Ferroelectric relaxation and domain-wall dynamics reported for PZT and related ceramics suggest \(0.7\lesssim\alpha\lesssim1.0\), \(5\lesssim\beta\lesssim50\) and \(1\lesssim\gamma\lesssim10\) \cite{Jaffe1971,LinesGlass1977,Uchino2010,Damjanovic1998}. The corresponding estimate \(M_H=O(5-500)\) indicates one of the strongest forms of passive mechanical memristance considered here.

\subsubsection{Electrorheological and Magnetorheological Fluids} Field-responsive fluids derive memory from the formation and reorganisation of particle-chain networks. Available rheological data suggest \(0.4\lesssim\alpha\lesssim0.8\), \(1\lesssim\beta\lesssim20\) and \(0.5\lesssim\gamma\lesssim5\) for ER fluids, giving \(M_H=O(0.5-100)\) \cite{Conrad1991,Munteanu2025,Liu2025}. For MR fluids, stronger field-dependent dissipation leads to \(0.5\lesssim\alpha\lesssim0.9\), \(20\lesssim\beta\lesssim100\) and \(1\lesssim\gamma\lesssim10\), yielding \(M_H=O(20-1000)\) \cite{Morillas2020,Maurya2024,Osial2023,Escalante2020}.

\subsubsection{Granular Dampers} Granular systems derive memory from evolving contact networks and force-chain structures. Available experimental and theoretical studies suggest \(0.3\lesssim\alpha\lesssim0.9\), \(5\lesssim\beta\lesssim50\) and \(0.5\lesssim\gamma\lesssim5\) \cite{Saeki2002,Marhadi2005,Lu2011,Jaeger1996,DeGiuli2016}. The resulting estimate \(M_H=O(2.5-250)\) suggests strong dissipation and substantial memory effects associated with collective particle dynamics.

\section{Unified Material Design Map}
\label{sec:unified_material_map}

\begin{figure}
    \centering
    \includegraphics[width=\linewidth]{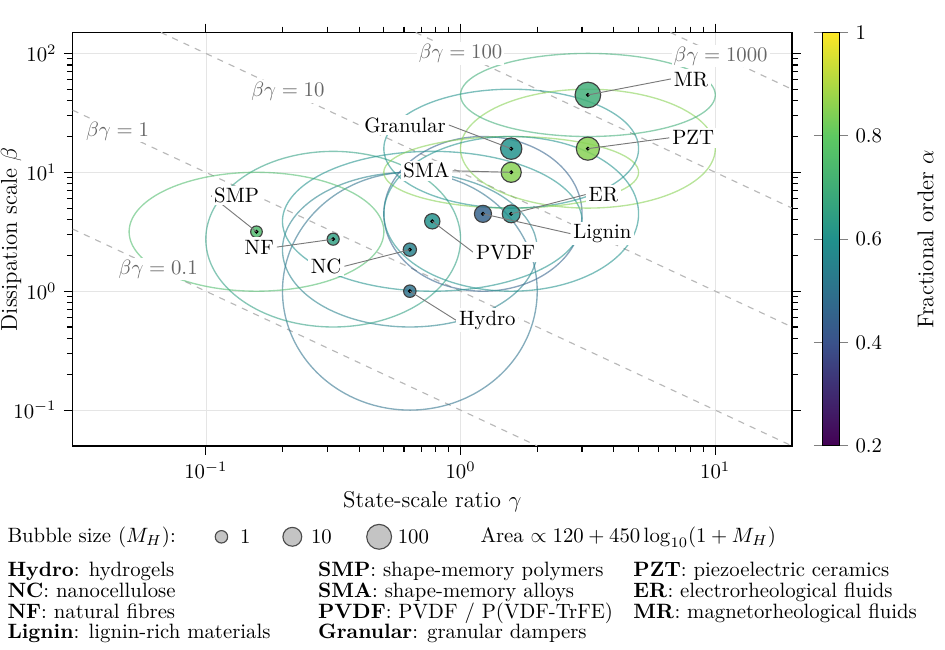}
    \caption{Parameter map for the canonical mechanical memristor in the $(\gamma,\beta)$ plane. The horizontal axis represents the state-scale ratio $\gamma=u_c/q_{\star}$, while the vertical axis represents the dissipation scale $\beta=c_{\star}/(k t_c)$. Dashed lines denote constant values of the effective mechanical memristance screening index $M_H=\beta \gamma |\mathcal H_\alpha(\Omega)|$ in the unit-transfer limit $|\mathcal H_\alpha|\to1$. Assigned $M_H$ scenario is proportional to a logarithmic function of the effective mechanical memristance screening index $M_H$, and bubble colour represents the fractional-memory parameter \(\alpha\). Elliptical envelopes show the assumed scenario ranges in Table~\ref{tab:materialmap}. Bubble locations and sizes represent illustrative scenario assignments within the canonical model and should not be interpreted as comparative measures of material performance.}
    \label{fig:ashby_gamma_beta_alphaMH}
\end{figure}

\begin{figure}
    \centering
    \includegraphics[width=\linewidth]{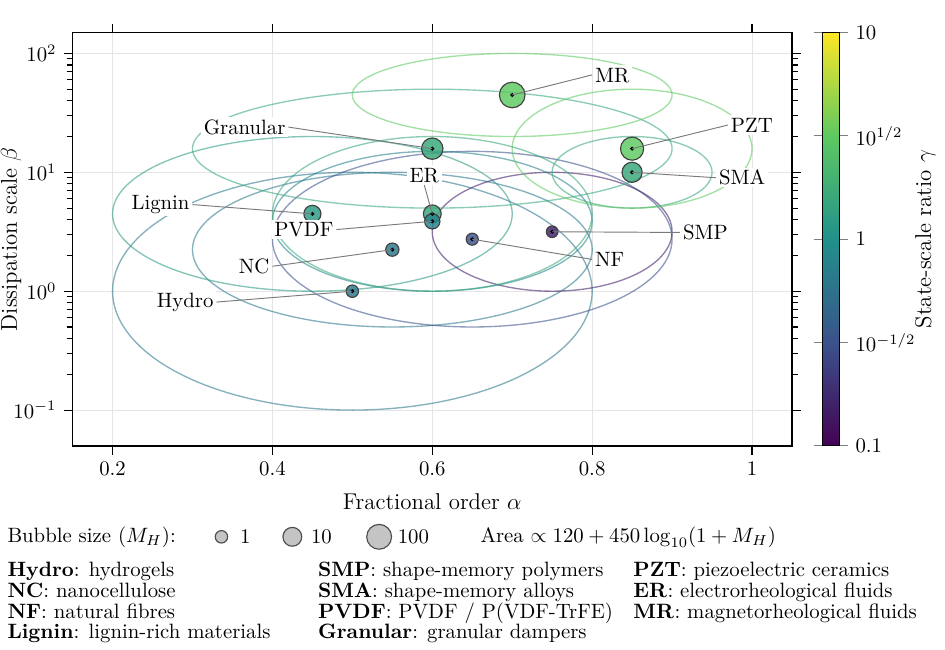}
    \caption{Canonical-model parameter map in the $(\alpha,\beta)$ plane. The parameter $\alpha$ sets the order of the assumed state-relaxation law, while $\beta$ represents the reference damping scale. Bubble area is proportional to a logarithmic function of the effective mechanical memristance screening index $M_H$. Bubble and envelope colours correspond to $\log_{10}(\gamma)$, where $\gamma$ is the state-scale ratio.}
    \label{fig:ashby_alpha_beta_gammaMH}
\end{figure}

The preceding analyses indicate that a wide variety of mechanical memory systems may be represented within a common dimensionless framework. The damping scale $\beta$ and state-scale ratio $\gamma$ are defined in Eq.~\eqref{eq:nondim_groups_main}, and the harmonic memory-transfer magnitude follows from Eq.~\eqref{eq:Halpha_magnitude}. Their combination in the screening index $M_H$, defined in Eq.~\eqref{eq:MH_screening_definition}, shows that mechanical memristive behaviour is governed by three independent mechanisms: \begin{enumerate} \item the magnitude of dissipation (\(\beta\)); \item the relative scale of memory-state excursions (\(\gamma\)); \item the ability of the memory state to follow or retain imposed deformation (\(|\mathcal H_\alpha(\Omega)|\)). \end{enumerate} Consequently, materials and devices may be represented within a common three-dimensional design space defined by \((\alpha,\beta,\gamma)\), while \(M_H\) provides a convenient scalar metric for comparing their expected memristive behaviour under similar operating conditions. The use of \(M_H\) as a screening metric is motivated by Eq.~\eqref{eq:local_coefficient_modulation}, which establishes its leading-order relationship with coefficient modulation. Materials occupying regions of large \(M_H\) are therefore expected to exhibit stronger state-dependent damping and more pronounced memristive behaviour.
Figures~\ref{fig:ashby_gamma_beta_alphaMH} and \ref{fig:ashby_alpha_beta_gammaMH} visualize this design space using Ashby-type material maps. Figure \ref{fig:ashby_gamma_beta_alphaMH} illustrates contours of constant \(M_H\) in the unit-transfer limit $|\mathcal H_\alpha|\to1$ in the \((\gamma,\beta)\)-plane, whereas Figure \ref{fig:ashby_alpha_beta_gammaMH} shows the relationship between dissipation and fractional memory characteristics. Together, these maps provide a compact representation of the mechanical-memristor landscape and reveal the principal trade-off between memory retention and energy dissipation. The values reported in Table~\ref{tab:materialmap} should be interpreted as first-order engineering estimates rather than intrinsic material constants. In particular, \(\beta\), \(\gamma\) and \(M_H\) depend upon specimen geometry, operating conditions, memory-state definition and identification methodology. Nevertheless, the resulting design maps provide a practical framework for comparing otherwise unrelated classes of mechanical memory systems and for identifying candidate materials suitable for future adaptive, neuromorphic and multifunctional mechanical architectures.

\section{Conclusion}
\label{sec:conclusion}
 In this work, mechanical memristance refers to a state-dependent dissipative response described within the memristive-systems framework, rather than to a direct mechanical realisation of the ideal memristor.
 This work has developed a scaling framework for describing and comparing mechanical memristive systems represented by the adopted state-dependent mem-dashpot model. Starting from a fractional-order mem-dashpot formulation, a set of dimensionless groups was identified that govern memory-dependent dissipation, hysteresis and dynamic response. The resulting framework links materials as diverse as shape-memory polymers, shape-memory alloys, hydrogels, lignocellulosic materials, piezoelectrics, electrorheological fluids, magnetorheological fluids and granular systems through a common state-space representation. The principal outcome of the study is the introduction of the effective mechanical memristance screening index $M_H$  which combines dissipation magnitude, state-scale ratio and memory transfer into a single measure of mechanical memristive behaviour. Although \(M_H\) is not an intrinsic material property, it provides a common basis for comparing memory-dependent damping systems operating under similar geometric and loading conditions. The framework shows that mechanical memristance is governed by three independent factors: the magnitude of dissipation, the relative scale of memory-state excursions and the ability of the memory state to retain information from previous loading. Dimensional scaling laws were derived for a range of candidate material classes and used to construct Ashby-type mechanical-memristor design maps. These maps reveal clear trade-offs between memory persistence, energy dissipation and state-dependent adaptability. Within the illustrative parameter scenarios considered here, hydrogels, nanocellulose and lignocellulosic materials are located in long-memory regions of the canonical design space, whereas shape-memory alloys, piezoelectric ceramics, granular systems and field-responsive fluids occupy regions associated with larger assigned values of the screening index $M_H$. More broadly, the present results suggest that a range of systems traditionally described as viscoelastic, hysteretic or adaptive may be examined within a common mechanical-memristive framework. By expressing memory-dependent dissipation through the dimensionless coordinates $(\alpha,\beta,\gamma)$ and the screening index $M_H$, the present framework provides a systematic basis for constitutive identification, comparison and future experimental assessment of mechanical memristive systems.
\vskip6pt

\appendix
\section{Detailed derivations}
\label{app:derivations}

\subsection{Dimensional force balance, power and passivity}\label{app:dime_force_balance}

In the single-oscillator schematic in Fig.~\ref{fig:canonical_oscillator}, the spring and mem-dashpot are connected in parallel, so their terminal displacement and velocity are identical and their resisting forces add. With the sign convention used in \eqref{eq:general_balance},
\begin{equation}
  F_{\mathrm s}=ku,
  \qquad
  F_{\mathrm d}=c_{\mathrm m}(q)\dot u,
  \qquad
  F_{\mathrm r}=F_{\mathrm s}+F_{\mathrm d}.
  \label{eq:appendix_parallel_forces}
\end{equation}
Substitution of \eqref{eq:appendix_parallel_forces} into Newton's law gives
\begin{equation}
  m\ddot u+ku+c_{\mathrm m}(q)\dot u=F_{\mathrm ext},
\end{equation}
which becomes \eqref{eq:common_mechanics_dimensional} after using \eqref{eq:forcing_dimensional}. This force--velocity representation is the mechanical counterpart of a state-dependent resistive port and is consistent with standard memristive and mechanical mem-element formulations \cite{ChuaKang1976,JeltsemaVanDerSchaft2010,PeiEtAl2015}.

Multiplication of the mechanical balance by $\dot u$ gives the instantaneous power identity
\begin{equation}
  \frac{\dd}{\dd t}
  \left(\frac{1}{2}m\dot u^2+\frac{1}{2}ku^2\right)
  =F_{\mathrm ext}\dot u-c_{\mathrm m}(q)\dot u^2.
  \label{eq:appendix_power_balance}
\end{equation}
The mem-dashpot power and the energy dissipated over $[t_0,t_1]$ are therefore
\begin{equation}
  P_{\mathrm d}=F_{\mathrm d}\dot u=c_{\mathrm m}(q)\dot u^2,
  \qquad
  E_{\mathrm d}[t_0,t_1]
  =\int_{t_0}^{t_1}c_{\mathrm m}(q(t))\dot u(t)^2\,\dd t.
  \label{eq:appendix_dissipated_energy}
\end{equation}
Thus $c_{\mathrm m}(q)\ge0$ is sufficient for passivity of the dissipative port: it cannot supply net energy through the mechanical terminals \cite{ChuaKang1976,JeltsemaVanDerSchaft2010}. Setting $m=0$ removes kinetic-energy storage but does not alter either the constitutive memory law or the non-negative dissipation condition.

\subsection{Positive regularisation and the meaning of \texorpdfstring{$\delta$}{delta}}
\label{app:delta_meaning}

To examine the smoothing effect of \(\delta\) on the sharp minimum, we first define \(a=\tanh z\) and consider the function replacing \(|a|\). Define
\begin{equation}
  a=\tanh z,
  \qquad
  r_{\delta}(a)=\sqrt{a^2+\delta^2}-\delta,
  \qquad
  r_{\delta}(a)=\frac{a^2}{\sqrt{a^2+\delta^2}+\delta}
  \quad\text{for } |a|+\delta>0,
  \label{eq:appendix_delta_regulariser}
\end{equation}
where $z=q/q_{\star}=\gamma y$ is the dimensionless state argument, $a$ is its hyperbolic-tangent transform and $r_{\delta}$ is the approximation to $|a|$ that vanishes at the origin and is smooth for $\delta>0$. With $r_0(0)=0$ defined by continuity, $\mathcal G(z)=\epsilon+r_{\delta}(\tanh z)$. The parameter $\delta$ is dimensionless and is an algebraic smoothing, or rounding, parameter.

For every $\delta\ge0$, $r_{\delta}$ is even and non-negative. For $\delta>0$ it is smooth for all $a$, with
\begin{equation}
  \frac{\dd r_{\delta}}{\dd a}
  =\frac{a}{\sqrt{a^2+\delta^2}},
  \qquad
  \frac{\dd^2 r_{\delta}}{\dd a^2}
  =\frac{\delta^2}{(a^2+\delta^2)^{3/2}}>0.
  \label{eq:appendix_delta_derivatives}
\end{equation}
The regularised function converges pointwise to the absolute value as $\delta\to0^+$, and the approximation error obeys the uniform bounds
\begin{equation}
  0\le r_{\delta}(a)\le |a|,
  \qquad
  0\le |a|-r_{\delta}(a)\le\delta.
  \label{eq:appendix_delta_bounds}
\end{equation}
Consequently,
\begin{equation}
  \mathcal G(z)
  \longrightarrow \epsilon+|\tanh z|
  \quad\text{as}\quad \delta\to0^+,
\end{equation}
while any finite $\delta>0$ removes the cusp, namely the jump between the left and right slopes at $z=0$. Since $|\tanh z|\le1$, the coefficient is bounded according to
\begin{equation}
  \epsilon
  \le \mathcal G(z)
  \le \epsilon+\sqrt{1+\delta^2}-\delta.
  \label{eq:appendix_delta_range}
\end{equation}
These inequalities also show that increasing $\delta$ reduces the available variation of the state-dependent contribution, for $\delta\gg1$, its maximum is approximately $1/(2\delta)$.

Two asymptotic regimes explain the shape change produced by $\delta$. In the rounded core, $|a|\ll\delta$,
\begin{equation}
  r_{\delta}(a)
  =\frac{a^2}{2\delta}
  +O\!\left(\frac{a^4}{\delta^3}\right),
  \label{eq:appendix_delta_inner}
\end{equation}
whereas outside that core, $\delta\ll|a|$,
\begin{equation}
  r_{\delta}(a)
  =|a|-\delta+\frac{\delta^2}{2|a|}
  +O\!\left(\frac{\delta^4}{|a|^3}\right).
  \label{eq:appendix_delta_outer}
\end{equation}
The crossover occurs when $|\tanh z|$ is of order $\delta$. For $0<\delta<1$, this corresponds approximately to
\begin{equation}
  |z|\sim\tanh^{-1}(\delta),
  \qquad
  |q|\sim q_{\star}\tanh^{-1}(\delta),
  \qquad
  |y|\sim\frac{\tanh^{-1}(\delta)}{\gamma}.
  \label{eq:appendix_delta_transition_scale}
\end{equation}
For small $\delta$, $\tanh^{-1}(\delta)\simeq\delta$, so $\delta$ can be interpreted as the approximate half-width of the smooth transition when the state is measured in units of $q_{\star}$. If $\delta\ge1$, the bounded quantity $|\tanh z|$ never substantially exceeds $\delta$, and the response remains in the rounded regime over essentially the entire state range.

The distinction between $\epsilon$ and $\delta$ is especially clear at the origin. For $\delta>0$,
\begin{equation}
  \mathcal G(0)=\epsilon,
  \qquad
  \left.\frac{\dd\mathcal G}{\dd z}\right|_{z=0}=0,
  \qquad
  \left.\frac{\dd^2\mathcal G}{\dd z^2}\right|_{z=0}=\frac{1}{\delta}.
  \label{eq:appendix_delta_origin}
\end{equation}
Therefore, for the nondimensional coefficient $C(y)=c_{\mathrm e}(y)$,
\begin{equation}
  C(0)=\beta\epsilon,
  \qquad
  C_y(0)=0,
  \qquad
  C_{yy}(0)=\frac{\beta\gamma^2}{\delta}.
  \label{eq:appendix_delta_curvature}
\end{equation}
The zero-state first-order tangent damping is consequently $\beta\epsilon$ and is independent of $\delta$. In the singular limit $\delta=0$, $C(y)=\beta[\epsilon+|\tanh(\gamma y)|]$ remains passive but is not differentiable at $y=0$, with left and right derivatives $-\beta\gamma$ and $+\beta\gamma$, respectively. A finite $\delta$ is therefore required whenever a classical derivative at the origin is needed for Newton iteration, gradient-based identification, or tangent linearisation.

Equation \eqref{eq:appendix_delta_curvature} also identifies a practical parameter-correlation issue. If measurements remain entirely inside the rounded core, the leading state-dependent term is
\begin{equation}
  C(y)\simeq\beta\epsilon+\frac{\beta\gamma^2}{2\delta}y^2,
  \label{eq:appendix_delta_local_identification}
\end{equation}
so local data identify the combination $\beta\gamma^2/\delta$ rather than $\gamma$ and $\delta$ separately. Reliable independent estimation of $\delta$ therefore requires data that span the crossover region in \eqref{eq:appendix_delta_transition_scale} and, preferably, part of the saturation region.

Finally, \eqref{eq:appendix_delta_regulariser} gives $r_{\delta}(a)\ge0$, and hence
\begin{equation}
  c_{\mathrm m}(q)
  =c_{\star}\mathcal G(q/q_{\star})
  \ge c_{\star}\epsilon>0.
\end{equation}
It follows from \eqref{eq:appendix_dissipated_energy} that
\begin{equation}
  E_{\mathrm d}[t_0,t_1]
  \ge c_{\star}\epsilon
  \int_{t_0}^{t_1}\dot u(t)^2\,\dd t\ge0.
  \label{eq:appendix_delta_passivity_bound}
\end{equation}
Thus $\delta$ changes smoothness and transition width without compromising passivity for any $\delta\ge0$; the non-negative memristance condition is the mechanical counterpart of the passivity condition for a memristive port \cite{ChuaKang1976,JeltsemaVanDerSchaft2010}.

\subsection{Nondimensionalisation of the mechanical equation}\label{app:nondimens_eom_details}

The scaling follows the usual dimensional-similarity construction \cite{Buckingham1914}. Use
\begin{equation}
  u=u_{\mathrm c}x,
  \quad q=u_{\mathrm c}y,
  \quad t=t_{\mathrm c}\tau,
  \quad F=F_{\mathrm c}f,
  \quad F_{\mathrm c}=ku_{\mathrm c}.
  \label{eq:appendix_scales}
\end{equation}
The time derivatives transform as
\begin{equation}
  \dot u=\frac{u_{\mathrm c}}{t_{\mathrm c}}x',
  \qquad
  \ddot u=\frac{u_{\mathrm c}}{t_{\mathrm c}^{2}}x'',
\end{equation}
and the state argument becomes
\begin{equation}
  \frac{q}{q_{\star}}
  =\frac{u_{\mathrm c}}{q_{\star}}y
  =\gamma y.
\end{equation}
Substitution into \eqref{eq:common_mechanics_dimensional} gives
\begin{align}
  m\frac{u_{\mathrm c}}{t_{\mathrm c}^{2}}x''
  +ku_{\mathrm c}x
  +c_{\star}\mathcal G(\gamma y)
   \frac{u_{\mathrm c}}{t_{\mathrm c}}x'
  =F_{\mathrm b}+F_{\mathrm a}\sin(\omega t_{\mathrm c}\tau).
\end{align}
Division by $F_{\mathrm c}=ku_{\mathrm c}$ yields
\begin{equation}
  \frac{m}{kt_{\mathrm c}^{2}}x''+x
  +\frac{c_{\star}}{kt_{\mathrm c}}
   \mathcal G(\gamma y)x'
  =\frac{F_{\mathrm b}}{F_{\mathrm c}}
   +\frac{F_{\mathrm a}}{F_{\mathrm c}}
    \sin(\omega t_{\mathrm c}\tau),
\end{equation}
which is \eqref{eq:first_order_nondim_mech} or \eqref{eq:fractional_nondim_mech}. The groups have direct time- and amplitude-scale interpretations:
\begin{equation}
  \mu=\left(\frac{\sqrt{m/k}}{t_{\mathrm c}}\right)^2,
  \qquad
  \beta=\frac{c_{\star}/k}{t_{\mathrm c}},
  \qquad
  \gamma=\frac{u_{\mathrm c}}{q_{\star}},
  \qquad
  \Omega=\omega t_{\mathrm c}.
  \label{eq:appendix_group_interpretation}
\end{equation}
Thus $\mu$ compares the undamped mechanical time $\sqrt{m/k}$ with the selected time scale, $\beta$ compares the viscous relaxation time $c_{\star}/k$ with that scale, $\gamma$ compares the imposed displacement scale with the memory-state scale, and $\Omega$ is the forcing frequency measured on the same clock. The parameters $\epsilon$ and $\delta$ require no additional dimensional factors because they shape the already dimensionless function $\mathcal G$.

\subsection{First-order memory scaling and harmonic transfer}\label{app:fo_scalng_memory_transfer}

Substitute $q=u_{\mathrm c}y$, $u=u_{\mathrm c}x$, and $t=t_{\mathrm c}\tau$ into \eqref{eq:first_order_memory_dimensional}:
\begin{equation}
  \tau_{\mathrm m}\frac{u_{\mathrm c}}{t_{\mathrm c}}y'
  +u_{\mathrm c}y=u_{\mathrm c}x.
\end{equation}
After division by $u_{\mathrm c}$,
\begin{equation}
  \lambda_1y'+y=x,
  \qquad
  \lambda_1=\frac{\tau_{\mathrm m}}{t_{\mathrm c}},
\end{equation}
which gives \eqref{eq:first_order_nondim_mem}. For a harmonic component $x=\widehat x\exp(\ii\Omega\tau)$ and $y=\widehat y\exp(\ii\Omega\tau)$,
\begin{equation}
  (1+\ii\lambda_1\Omega)\widehat y=\widehat x,
  \qquad
  \frac{\widehat y}{\widehat x}
  =\frac{1}{1+\ii\lambda_1\Omega}.
  \label{eq:appendix_first_order_harmonic}
\end{equation}
Hence the memory state follows displacement quasi-statically when $\lambda_1\Omega\ll1$, while its amplitude decreases and its phase lag approaches $\pi/2$ when $\lambda_1\Omega\gg1$.

\subsection{Scaling and transform properties of the Caputo derivative}\label{app:Caputo_derivative}

The Caputo derivative and its transform properties used here are standard in fractional relaxation and viscoelasticity \cite{Caputo1967,Podlubny1999,Mainardi2010}. Starting from \eqref{eq:caputo_definition}, let $t=t_{\mathrm c}\tau$, $\zeta=t_{\mathrm c}\sigma$, and $q=u_{\mathrm c}y$. Since
\begin{equation}
  \dot q(\zeta)=\frac{u_{\mathrm c}}{t_{\mathrm c}}y'(\sigma),
  \qquad
  \dd\zeta=t_{\mathrm c}\dd\sigma,
  \qquad
  (t-\zeta)^{-\alpha}
  =t_{\mathrm c}^{-\alpha}(\tau-\sigma)^{-\alpha},
\end{equation}
the following is obtained
\begin{align}
  \CD{t}{\alpha}q(t)
  &=\frac{u_{\mathrm c}t_{\mathrm c}^{-\alpha}}
          {\Gamma(1-\alpha)}
    \int_0^{\tau}
    \frac{y'(\sigma)}{(\tau-\sigma)^{\alpha}}\,\dd\sigma\\
  &=u_{\mathrm c}t_{\mathrm c}^{-\alpha}
    \CD{\tau}{\alpha}y(\tau).
\end{align}
Substitution into \eqref{eq:fractional_memory_dimensional}, followed by division by $u_{\mathrm c}$, gives
\begin{equation}
  \left(\frac{\tau_{\alpha}}{t_{\mathrm c}}\right)^{\alpha}
  \CD{\tau}{\alpha}y+y=x,
\end{equation}
which is \eqref{eq:fractional_nondim_mem} with $\lambda_{\alpha}=(\tau_{\alpha}/t_{\mathrm c})^{\alpha}$.

For $0<\alpha<1$, the one-sided Laplace transform is
\begin{equation}
  \mathcal L\left\{\CD{\tau}{\alpha}y\right\}
  =s^{\alpha}Y(s)-s^{\alpha-1}y(0^+).
  \label{eq:appendix_caputo_laplace_general}
\end{equation}
Consequently,
\begin{equation}
  \left(1+\lambda_{\alpha}s^{\alpha}\right)Y(s)
  =X(s)+\lambda_{\alpha}s^{\alpha-1}y(0^+).
  \label{eq:appendix_fractional_transform_initial}
\end{equation}
Under zero perturbation initial conditions, $Y/X=[1+\lambda_{\alpha}s^{\alpha}]^{-1}$. In a long-time periodic response, setting $s=\ii n\Omega$ yields \eqref{eq:fractional_harmonic_transfer_nondim}. On the principal branch and for $n\Omega>0$,
\begin{equation}
  (\ii n\Omega)^{\alpha}
  =(n\Omega)^{\alpha}
  \left[\cos\left(\frac{\pi\alpha}{2}\right)
  +\ii\sin\left(\frac{\pi\alpha}{2}\right)\right],
  \label{eq:appendix_fractional_branch}
\end{equation}
which makes explicit the simultaneous amplitude attenuation and fractional phase lag.

\subsection{Tangent linearisation}\label{app:Tangent_linearisation}

Tangent linearisation is the first Fr\'echet approximation of the nonlinear state equations about a reference solution \cite{khalil_nonlinear_2002}. Write $C(y)=c_{\mathrm e}(y)$ and use the perturbations in \eqref{eq:perturbations}. The nonlinear product expands as
\begin{align}
  C(y)x'
  &=\left[C(y_{\mathrm b})
    +\varepsilon C_y(y_{\mathrm b})\eta
    +O(\!\varepsilon^2)\right]
    \left[V_{\mathrm b}+\varepsilon\xi'\right]\\
  &=C_{\mathrm b}V_{\mathrm b}
    +\varepsilon\left(C_{\mathrm b}\xi'
    +C_y(y_{\mathrm b})V_{\mathrm b}\eta\right)
    +O(\!\varepsilon^2).
\end{align}
Subtracting the exact reference-state balance and retaining the coefficient of $\varepsilon$ gives
\begin{equation}
  \mu\xi''+\xi+C_{\mathrm b}\xi'+G_{\mathrm b}\eta=p,
  \qquad
  G_{\mathrm b}=C_y(y_{\mathrm b})V_{\mathrm b},
\end{equation}
which is \eqref{eq:linearised_mechanics}. Because the first-order and fractional memory equations are already linear in $x$ and $y$, their perturbations directly give \eqref{eq:linearised_first_order_memory} and \eqref{eq:linearised_fractional_memory}.

For the positive law, let $z=\gamma y$. Differentiating first with respect to $z$ and then applying the chain rule gives
\begin{equation}
  C_y(y)
  =\beta\gamma
  \frac{\tanh(\gamma y)\sech^2(\gamma y)}
  {\sqrt{\tanh^2(\gamma y)+\delta^2}},
\end{equation}
which is \eqref{eq:coefficient_derivative}. This expression is continuous for every $y$ when $\delta>0$. At $\delta=0$ it is valid only away from $y=0$, the origin has the cusp described in \eqref{eq:appendix_delta_curvature} and no two-sided classical tangent. At a static equilibrium $V_{\mathrm b}=0$, so $G_{\mathrm b}=0$ regardless of $C_y(y_{\mathrm b})$. At the zero state with $\delta>0$, $C_y(0)=0$ as well, and the first influence of state-dependent damping appears at quadratic order through $C_{yy}(0)$.

\subsection{Frequency-domain tangent models}\label{app:Frequency_domain_models}

Assume zero perturbation initial conditions. For the first-order memory law,
\begin{equation}
  (1+\lambda_1s)Y(s)=\Xi(s),
  \qquad
  Y(s)=\frac{\Xi(s)}{1+\lambda_1s}.
\end{equation}
The transformed mechanical equation is
\begin{equation}
  \left(\mu s^2+1+C_{\mathrm b}s\right)\Xi(s)
  +G_{\mathrm b}Y(s)=P(s).
\end{equation}
Eliminating $Y$ gives
\begin{equation}
  \frac{\Xi(s)}{P(s)}
  =\left[\mu s^2+1+C_{\mathrm b}s
  +\frac{G_{\mathrm b}}{1+\lambda_1s}\right]^{-1},
\end{equation}
which is \eqref{eq:first_order_linear_transfer}. For the fractional model, use \eqref{eq:appendix_caputo_laplace_general} with zero initial perturbation to obtain
\begin{equation}
  Y(s)=\frac{\Xi(s)}{1+\lambda_{\alpha}s^{\alpha}},
\end{equation}
and hence
\begin{equation}
  \frac{\Xi(s)}{P(s)}
  =\left[\mu s^2+1+C_{\mathrm b}s
  +\frac{G_{\mathrm b}}{1+\lambda_{\alpha}s^{\alpha}}\right]^{-1},
\end{equation}
which is \eqref{eq:fractional_linear_transfer}. Evaluating these expressions at $s=\ii\Omega$ produces the complex tangent stiffnesses in \eqref{eq:linear_complex_stiffnesses}; the compliance is their reciprocal. The use of the principal branch for $s^{\alpha}$ is the same convention as in \eqref{eq:appendix_fractional_branch} \cite{Podlubny1999,Mainardi2010}.

\subsection{Consistency limits}\label{app:Consistency}

Several limiting cases provide direct checks on the derivations. First, $\alpha=1$ and $\lambda_{\alpha}=\lambda_1$ reduce the fractional memory law and all of its transfer relations to the first-order model. Second, $\delta\to0^+$ recovers the passive but nonsmooth coefficient $\epsilon+|\tanh z|$, whereas any $\delta>0$ gives a smooth coefficient with the same minimum value $\epsilon$. Third, $\mu=0$ removes inertia and recovers the quasi-static balance without changing the memory dynamics. Finally, at a static operating point $V_{\mathrm b}=0$, the state-feedback term $G_{\mathrm b}\eta$ vanishes from the first-order mechanical tangent, so the distinction between first-order and fractional memory appears in the perturbation memory state but not in the static mechanical transfer function at first order.

\subsection{Numerical solution of the nonlinear equations} \label{app:numerical}
Equation \eqref{eq:first_order_nondim_attached} was used when $\alpha=1$, whereas the massless, unbiased and unit-force form of \eqref{eq:fractional_nondim_general}, with $\lambda_{\alpha}=1$, was used for $0<\alpha<1$. For the parameter maps, the phase $\phi=\Omega\tau$ was divided into $N=256$ equally spaced points over one forcing period. The response curves in Fig.~\ref{fig:periodic_response} use $N=1024$. For the fractional model, the memory state was evaluated directly from the harmonic transfer relation \eqref{eq:fractional_harmonic_transfer_nondim}, so that each displacement harmonic was multiplied by $[1+(\ii n\Omega)^{\alpha}]^{-1}$, in agreement with the standard transform representation of the Caputo derivative \cite{Podlubny1999,Mainardi2010}. The periodic derivative of $x$ was approximated from the current point and the two preceding points by the second-order backward difference formula \cite{HairerWanner1996}. This construction imposed periodicity over a single cycle and avoided selecting an arbitrary number of preliminary cycles.

For a trial displacement, the memory state and the coefficient $c_{\mathrm e}(y)$ in \eqref{eq:nondim_coefficient} were evaluated, the displacement balance was solved with that coefficient, and the process was repeated until the response was consistent with the state used to calculate it. Information from several preceding updates was combined to reach this agreement more rapidly \cite{WalkerNi2011}. If a point remained difficult, a Newton correction was applied and its linear equation was solved with a residual-minimising method \cite{SaadSchultz1986}. The nonlinearity or the change in parameters was then introduced through smaller steps, starting either from the constant-coefficient response or from a neighbouring converged solution \cite{AllgowerGeorg1990}. For difficult first-order cases, an additional calculation integrated the state over one period and corrected its initial value until the final state returned to the same value; the implicit integration followed the method of Shampine and Reichelt \cite{ShampineReichelt1997}. Fixed-point iterations used an update tolerance of $10^{-8}$. Final responses, including those obtained through Newton or periodic-shooting corrections, were accepted when the largest absolute residual in the governing displacement equation was no greater than $10^{-6}$.

Each two-parameter map contained $30\times30$ points. The parameters $\beta\in[10^{-2},10^{2}]$, $\gamma\in[10^{-2},10]$ and $\Omega\in[10^{-2},10]$ were spaced logarithmically, whereas $\alpha\in[0.30,1]$ was spaced linearly. The baseline values were $\beta=5$, $\gamma=0.1$ and $\Omega=0.2$, with $\alpha=0.6$ for the fractional maps. The coefficient in \eqref{eq:nondim_coefficient} was evaluated with $\epsilon=10^{-2}$ and $\delta=10^{-4}$. Consecutive rows were followed in opposite directions so that the response at one point provided the initial estimate for the next. If this estimate failed, a converged solution from the preceding row and then from the surrounding points was used.

The geometric area enclosed by the nondimensional $f$--$x$ loop was evaluated as
\begin{equation}
  \mathcal A
  =\left|\frac{1}{2}\sum_{j=1}^{N}
  \left(x_j f_{j+1}-x_{j+1}f_j\right)\right|,
  \qquad f_j=\sin\phi_j,
  \label{eq:numerical_loop_area}
\end{equation}
where the first point follows the last point in the sum. This geometric area and the force-work integral were compared with the cycle dissipation. The cycle-energy maps evaluate Eq.~\eqref{eq:exact_periodic_energy}, and the energy-equivalent damping coefficient was calculated as
\begin{equation}
  c_{\mathrm{eq}}
  =\frac{\sum_{j=1}^{N}c_{\mathrm e}(y_j)v_j^2}
  {\sum_{j=1}^{N}v_j^2},
  \qquad v_j=\Omega\,x_{\phi,j},
  \label{eq:numerical_equivalent_damping}
\end{equation}
The coefficient extrema were taken from the sampled periodic response. The tangent value was evaluated at the sampled positive displacement turning point as $C_{\mathrm b}=c_{\mathrm e}(y_{\mathrm b})$, consistently with \eqref{eq:linearised_mechanics}. Since $f$ and $x$ had already been nondimensionalised in Section~\ref{sec:nondimensional}, no additional normalisation of \eqref{eq:numerical_loop_area} was applied.

The contour lines join equal values of the calculated response measures. Convergence of the governing-equation residual does not establish phase-grid independence, uniqueness or stability of a periodic response. In particular, coefficient-based quantities at large $\gamma$ remain sensitive to the sampling of narrow velocity peaks; the $N=256$ maps should not be interpreted as a demonstration of convergence with respect to phase-grid refinement.

The MATLAB files used for the simulations can be downloaded at the following link: \href{https://github.com/Flagosw/Mechanical-memristor-oscillator}{Mem-dashpot model files}

\section{Illustrative parameter ranges and calculations}
\label{app:materialmap}

\subsection{Illustrative estimation of effective mechanical memristance for an epoxy shape-memory polymer} {\label{app:SMPexample}}
To illustrate the scaling procedure, representative parameter values are considered for an epoxy shape-memory polymer, drawing upon the thermomechanical experiments of Liu \emph{et al.} \cite{Liu2006} and the viscoelastic shape-memory characterisation reported by Chen \emph{et al.} \cite{Chen2014}. A possible coordinate scale may be defined in terms of the recoverable deformation of the polymer network:
\begin{equation} 
    q_\star \sim \varepsilon_r L , 
\end{equation}
where \(\varepsilon_r\) is the recoverable strain and \(L\) is a characteristic specimen length. The experiments reported in \cite{Liu2006,Chen2014} indicate recoverable strains of approximately $\varepsilon_r \approx 0.5$. Selecting a reference deformation of $u_{\mathrm c}/L \approx 0.05$, the state-scale ratio becomes:
\begin{equation} 
    \gamma = \frac{u_{\mathrm c}}{\varepsilon_r L} = \frac{0.05}{0.5} = 0.1 \end{equation}

Dynamic mechanical analysis of epoxy SMPs typically reports loss factors in the vicinity of the glass-transition temperature in the range $\tan\delta_{\mathrm{loss}} \approx 0.2-1$ consistent with moderate hysteretic energy dissipation \cite{Liu2006,Chen2014}, where $\delta_{\mathrm{loss}}$ denotes the mechanical loss angle. For the present numerical illustration, $\beta \approx 5$ and $\alpha \approx 0.7$ are assigned. The reported loss factors and fractional rheology \cite{Fang2015,Fang2016} provide guidance for these choices, but do not uniquely determine the corresponding parametres of the canonical force and state model. For loading frequencies comparable to the dominant memory timescale, the memory-transfer magnitude satisfies:

\begin{equation} 
    |\mathcal H_\alpha(\Omega)|=O(1). 
\end{equation}

Taking $|\mathcal H_\alpha(\Omega)|\approx0.5$ gives:

\begin{equation} 
    \Mh = \beta\gamma |\mathcal H_\alpha(\Omega)| = 5\times0.1\times0.5 = 0.25. 
\end{equation}

The resulting estimate $\Mh \approx 0.25$ lies within the range \begin{equation} \Mh = O(0.1-5) \end{equation} assigned to the illustrative SMP example in Table \ref{tab:materialmap}. This calculation serves only to illustrate the combined effect of the selected parameters on $\Mh$. Meaningful comparison of material performance additionally requires the constitutive identification procedure described in Section~\ref{sec:mapping_materials}. Material identification first requires that the force law and state evolution conform to the canonical model over the operating range, followed by joint calibration of the model parameters. Recoverable strain alone is insufficient to establish a state-dependent damping coefficient.

\subsection{Illustrative estimation of effective mechanical memristance for a NiTi shape-memory alloy}
\label{app:SMA_example}

The second calculation employs an illustrative set of NiTi parameters. The physical scales informing this designation are discussed in the experimental and modelling studies reviewed by Lagoudas \cite{Lagoudas2008}, together with representative thermomechanical characterisation data reported for NiTi actuators and springs \cite{PuenteCordova2023,Wang2023RateEffect}. For this dimensional illustration, a coordinate scale is defined from the transformation strain. This choice does not assign the transformation force to the mem-dashpot:

\begin{equation}
q_\star
\sim
\varepsilon_T L,
\end{equation}

where \(\varepsilon_T\) is the maximum transformation strain and \(L\)
is a characteristic specimen dimension.

For a typical pseudoelastic NiTi alloy,

\begin{equation}
\varepsilon_T
\approx
0.06.
\end{equation}

This value lies near the centre of the transformation-strain interval commonly reported for NiTi alloys (\(0.04 \lesssim \varepsilon_T \lesssim 0.08\)) \cite{Lagoudas2008,Wang2023RateEffect}. When selecting the following reference deformation:

\begin{equation}
\frac{u_{\mathrm c}}{L}
\approx
0.10,
\end{equation}

the state-scale ratio becomes:

\begin{equation}
\gamma
=
\frac{u_{\mathrm c}}{\varepsilon_T L}
=
\frac{0.10}{0.06}
\approx
1.7.
\end{equation}

Transformation-loop energy \cite{Lagoudas2008,Wang2023RateEffect} does not identify a bounded viscous coefficient. For a fixed displacement cycle traversed at progressively lower rates, Eq.~\eqref{eq:appendix_dissipated_energy}, together with the finite coefficient bound in Eq.~\eqref{eq:positive_law_bounds}, implies that the dissipated energy tends to zero. The present spring--mem-dashpot formulation therefore cannot sustain a finite quasistatic hysteresis loop. The value adopted below is therefore introduced solely for the illustrative scenario and does not constitute a calibration to the transformation loop:

\begin{equation}
\beta
\approx
10.
\end{equation}

Fractional descriptions of SMA-based systems motivate a range of state-law orders \cite{PuenteCordova2023,Wang2023RateEffect}. For the present illustrative scenario, the assigned range is:

\begin{equation}
0.75
\lesssim
\alpha
\lesssim
0.95,
\end{equation}

A representative value of:

\begin{equation}
    \alpha \approx 0.85
\end{equation}

is adopted for the present numerical illustration. The thermomechanical and spring-based studies reported in \cite{PuenteCordova2023,Guo2024FractionalSMA} do not uniquely determine this exponent of the canonical state-transfer law. For the present illustrative scenario, a harmonic transfer magnitude is assigned:

\begin{equation}
|\mathcal H_\alpha(\Omega)|
\approx
0.6.
\end{equation}

The effective mechanical memristance therefore becomes

\begin{equation}
\Mh
=
\beta\gamma|\mathcal H_\alpha(\Omega)|
=
10\times1.7\times0.6
\approx
10.
\end{equation}
The resulting estimate: 
\begin{equation}
    \Mh \approx 10 
\end{equation}

lies comfortably within the range \begin{equation} \Mh = O(2.5-100) \end{equation} assigned to the illustrative SMA scenario in Table~\ref{tab:materialmap}. The larger value relative to the SMP example follows directly from the adopted parametre assignments. It should not be interpreted as evidence that NiTi possesses greater mechanical memristance, since the specimens, loading conditions and constitutive state laws have not been identified under comparable conditions. The corresponding location on the design map should therefore be regarded as an illustrative scenario, rather than a material-specific characterisation.

\subsection{Illustrative estimation of the mechanical-memristor parametres for PVDF} \label{app:PVDF_example}
A third illustrative calculation employs a selected set of parameter values for a PVDF film. Plausible choices of state variable are informed by the electromechanical and damping measurements reported by Vinogradov \emph{et al.} \cite{Vinogradov2004PVDF}, together with the reviews of PVDF-based piezoelectric materials provided by Martins \emph{et al.}, Mohammadpourfazeli \emph{et al.}, and Ahbab \emph{et al.} \cite{Martins2014,Mohammadpourfazeli2023PVDFReview,Ahbab2025PVDFReview}. To define a representative internal coordinate, consider the evolving polarisation. A normalised memory variable is then introduced, \begin{equation} \theta = \frac{P}{P_r}, \end{equation} where $P_r$ denotes the remanent polarisation. The variable $\theta$ is dimensionless and remains close to 1. This normalisation defines the coordinate scale only 1, it does not determine either the evolution law or the coupling to the mechanical force. A calibrated mapping between $\theta$ and $q$, together with an appropriate coupled constitutive model, is therefore required. Typical values reported for poled PVDF films are \begin{equation} P_r \approx 0.05-0.10 \; \mathrm{C\,m^{-2}}, \end{equation} depending on crystallinity, processing history and degree of poling \cite{Martins2014,Furukawa1989,Ahbab2025PVDFReview}. Since the present framework employs normalised memory states, the state-scale ratio \(\gamma\) cannot be obtained directly from \(u_{\mathrm c}/P_r\). Its value would require a calibrated displacement--state relation. For the present numerical scenario, the following value is adopted solely for illustrative purposes: \begin{equation} \gamma \approx 1. \end{equation} The electromechanical losses reported by Vinogradov \emph{et al.} \cite{Vinogradov2004PVDF} motivate consideration of dissipative effects, but do not identify the canonical parameter \(\beta\). Dynamic-mechanical measurements of PVDF typically report \begin{equation} \tan\delta_{\mathrm{loss}} \approx 0.1-0.3, \end{equation} together with appreciable dielectric relaxation and hysteretic electromechanical losses. For the numerical illustration, the parameter \begin{equation} \beta \approx 5 \end{equation} is assigned. Although distributed dielectric and polymer relaxation motivate consideration of fractional-order behaviour, the corresponding fitted exponents do not, in general, coincide with those of the canonical state law. The illustrative scenario therefore adopts the range \begin{equation} 0.4 \lesssim \alpha \lesssim 0.8, \end{equation} with the representative value \begin{equation} \alpha \approx 0.6 \end{equation} taken from within this interval \cite{Martins2014,Mohammadpourfazeli2023PVDFReview}. For the numerical illustration, the memory-transfer magnitude is assigned as \begin{equation} |\mathcal H_\alpha(\Omega)| \approx 0.5. \end{equation} The resulting effective mechanical memristance is \begin{equation} \Mh = \beta\gamma |\mathcal H_\alpha(\Omega)| = 5 \times 1 \times 0.5 = 2.5. \end{equation} Hence, \begin{equation} \Mh \approx 2.5, \end{equation} which lies within the interval \begin{equation} \Mh = O(0.2-45) \end{equation} assigned to the illustrative PVDF scenario in Table~\ref{tab:materialmap}. This calculation serves only to illustrate the scaling procedure for a selected set of parameter assignments. Any mechanical application would require identification of both the state mapping and the force law under specified electrical boundary conditions.

\subsection{Remarks on parameter estimation} 
The preceding examples illustrate calculations based on assigned parameter values rather than experimental identification of $\Mh$. The reported values should therefore not be interpreted as measured material ranges. Material-specific application requires validation of the canonical force and state laws, followed by experimental identification of the memory state variable, the characteristic memory scale $q_\star$, the damping coefficient $c_\star$, and the fractional-memory parameter $\alpha$ for the material system and operating conditions of interest.

\subsection{Table with order-of-magnitude parameter ranges}
The assigned parameter ranges used to position the illustrative material scenarios in the canonical design space are collected in Table~\ref{tab:materialmap}.

\begin{table*}[ht]
\centering
\caption{Assigned parameter ranges for the illustrative material scenarios and the conditional index $M_H=\beta\gamma|H_\alpha(\Omega)|$. These ranges do not constitute material-specific calibrations, measured uncertainty intervals, confidence bounds, or experimentally validated limits. The reported intervals for $M_H$ are illustrative ranges selected for the assumed parameter scenarios.}
\label{tab:materialmap}
\begin{tabular}{lcccc}
\toprule
Material
&
$\alpha$
&
$\beta$
&
$\gamma$
&
$\Mh$
\\
\midrule

Hydrogels
&
0.2--0.8
&
0.1--10
&
0.2--2
&
0.1--10
\\

Nanocellulose
&
0.3--0.8
&
0.5--10
&
0.2--2
&
0.1--20
\\

Natural fibres
&
0.4--0.9
&
0.5--15
&
0.1--1
&
0.05--15
\\

Lignin-rich systems
&
0.2--0.7
&
1--20
&
0.5--3
&
0.5--60
\\

Shape-memory polymers
&
0.6--0.9
&
1--10
&
0.05--0.5
&
0.1--5
\\

PVDF / PVDF-TrFE
&
0.4--0.8
&
1--15
&
0.2--3
&
0.2--45
\\

Shape-memory alloys
&
0.75--0.95
&
5--20
&
0.5--5
&
2.5--100
\\

Granular dampers
&
0.3--0.9
&
5--50
&
0.5--5
&
2.5--250
\\

Piezoelectric ceramics
&
0.7--1.0
&
5--50
&
1--10
&
5--500
\\

Electrorheological fluids
&
0.4--0.8
&
1--20
&
0.5--5
&
0.5--100
\\

Magnetorheological fluids
&
0.5--0.9
&
20--100
&
1--10
&
20--1000
\\

\bottomrule
\end{tabular}
\end{table*}

The ranges reported in Table~\ref{tab:materialmap} represent illustrative scenario assignments, rather than measured material ranges. No comparative uncertainty is implied for the corresponding parametres, as a fractional rheological exponent, damping loss, or recoverable strain does not by itself uniquely determine the associated parameter of the present state-dependent dissipator.

\begin{acknowledgments}
This work has been supported by the ERC-2020-AdG 101020715 NEUROMETA project. Alhembar acknowledges the financial support from Khalifa University, United Arab Emirates, through its faculty-track postdoctoral program. JPKA acknowledges funding from a UKRI Future Leaders Fellowship (MR/V024965/1). The authors acknowledge the use of OpenAI's GPT-5.6 sol model as an assistive tool during code development. OpenAI's Codex was also used to prepare the manuscript layout and submission files. All numerical simulations, model verification, analysis and interpretation were conducted and validated by the authors.
\end{acknowledgments}

\bibliographystyle{apsrev4-2}
\bibliography{sample}

\begin{thebibliography}{94}%
\makeatletter
\providecommand \@ifxundefined [1]{%
 \@ifx{#1\undefined}
}%
\providecommand \@ifnum [1]{%
 \ifnum #1\expandafter \@firstoftwo
 \else \expandafter \@secondoftwo
 \fi
}%
\providecommand \@ifx [1]{%
 \ifx #1\expandafter \@firstoftwo
 \else \expandafter \@secondoftwo
 \fi
}%
\providecommand \natexlab [1]{#1}%
\providecommand \enquote  [1]{``#1''}%
\providecommand \bibnamefont  [1]{#1}%
\providecommand \bibfnamefont [1]{#1}%
\providecommand \citenamefont [1]{#1}%
\providecommand \href@noop [0]{\@secondoftwo}%
\providecommand \href [0]{\begingroup \@sanitize@url \@href}%
\providecommand \@href[1]{\@@startlink{#1}\@@href}%
\providecommand \@@href[1]{\endgroup#1\@@endlink}%
\providecommand \@sanitize@url [0]{\catcode `\\12\catcode `\$12\catcode
  `\&12\catcode `\#12\catcode `\^12\catcode `\_12\catcode `\%12\relax}%
\providecommand \@@startlink[1]{}%
\providecommand \@@endlink[0]{}%
\providecommand \url  [0]{\begingroup\@sanitize@url \@url }%
\providecommand \@url [1]{\endgroup\@href {#1}{\urlprefix }}%
\providecommand \urlprefix  [0]{URL }%
\providecommand \Eprint [0]{\href }%
\providecommand \doibase [0]{https://doi.org/}%
\providecommand \selectlanguage [0]{\@gobble}%
\providecommand \bibinfo  [0]{\@secondoftwo}%
\providecommand \bibfield  [0]{\@secondoftwo}%
\providecommand \translation [1]{[#1]}%
\providecommand \BibitemOpen [0]{}%
\providecommand \bibitemStop [0]{}%
\providecommand \bibitemNoStop [0]{.\EOS\space}%
\providecommand \EOS [0]{\spacefactor3000\relax}%
\providecommand \BibitemShut  [1]{\csname bibitem#1\endcsname}%
\let\auto@bib@innerbib\@empty
\bibitem [{\citenamefont {Chua}(1971)}]{1083337}%
  \BibitemOpen
  \bibfield  {author} {\bibinfo {author} {\bibfnamefont {L.}~\bibnamefont
  {Chua}},\ }\href {https://doi.org/10.1109/TCT.1971.1083337} {\bibfield
  {journal} {\bibinfo  {journal} {IEEE Transactions on Circuit Theory}\
  }\textbf {\bibinfo {volume} {18}},\ \bibinfo {pages} {507} (\bibinfo {year}
  {1971})}\BibitemShut {NoStop}%
\bibitem [{\citenamefont {Chua}\ and\ \citenamefont
  {Kang}(1976)}]{ChuaKang1976}%
  \BibitemOpen
  \bibfield  {author} {\bibinfo {author} {\bibfnamefont {L.~O.}\ \bibnamefont
  {Chua}}\ and\ \bibinfo {author} {\bibfnamefont {S.-M.}\ \bibnamefont
  {Kang}},\ }\href {https://doi.org/10.1109/PROC.1976.10092} {\bibfield
  {journal} {\bibinfo  {journal} {Proceedings of the IEEE}\ }\textbf {\bibinfo
  {volume} {64}},\ \bibinfo {pages} {209} (\bibinfo {year} {1976})}\BibitemShut
  {NoStop}%
\bibitem [{\citenamefont {Chen}\ \emph
  {et~al.}(2014{\natexlab{a}})\citenamefont {Chen}, \citenamefont {Liu},
  \citenamefont {Wang}, \citenamefont {Zhang}, \citenamefont {Li},\ and\
  \citenamefont {Wang}}]{WOS:000348204900002}%
  \BibitemOpen
  \bibfield  {author} {\bibinfo {author} {\bibfnamefont {Y.}~\bibnamefont
  {Chen}}, \bibinfo {author} {\bibfnamefont {G.}~\bibnamefont {Liu}}, \bibinfo
  {author} {\bibfnamefont {C.}~\bibnamefont {Wang}}, \bibinfo {author}
  {\bibfnamefont {W.}~\bibnamefont {Zhang}}, \bibinfo {author} {\bibfnamefont
  {R.-W.}\ \bibnamefont {Li}},\ and\ \bibinfo {author} {\bibfnamefont
  {L.}~\bibnamefont {Wang}},\ }\href@noop {} {\bibfield  {journal} {\bibinfo
  {journal} {MATERIALS HORIZONS}\ }\textbf {\bibinfo {volume} {1}},\ \bibinfo
  {pages} {489} (\bibinfo {year} {2014}{\natexlab{a}})}\BibitemShut {NoStop}%
\bibitem [{\citenamefont {Guo}\ \emph {et~al.}(2020)\citenamefont {Guo},
  \citenamefont {Han},\ and\ \citenamefont {Zhou}}]{WOS:000581738300050}%
  \BibitemOpen
  \bibfield  {author} {\bibinfo {author} {\bibfnamefont {L.}~\bibnamefont
  {Guo}}, \bibinfo {author} {\bibfnamefont {S.-T.}\ \bibnamefont {Han}},\ and\
  \bibinfo {author} {\bibfnamefont {Y.}~\bibnamefont {Zhou}},\ }\href@noop {}
  {\bibfield  {journal} {\bibinfo  {journal} {NANO ENERGY}\ }\textbf {\bibinfo
  {volume} {77}} (\bibinfo {year} {2020})}\BibitemShut {NoStop}%
\bibitem [{\citenamefont {Park}\ \emph {et~al.}(2024)\citenamefont {Park},
  \citenamefont {Naqi}, \citenamefont {Lee}, \citenamefont {Park},
  \citenamefont {Hong},\ and\ \citenamefont {Lee}}]{WOS:001384802000001}%
  \BibitemOpen
  \bibfield  {author} {\bibinfo {author} {\bibfnamefont {S.}~\bibnamefont
  {Park}}, \bibinfo {author} {\bibfnamefont {M.}~\bibnamefont {Naqi}}, \bibinfo
  {author} {\bibfnamefont {N.}~\bibnamefont {Lee}}, \bibinfo {author}
  {\bibfnamefont {S.}~\bibnamefont {Park}}, \bibinfo {author} {\bibfnamefont
  {S.}~\bibnamefont {Hong}},\ and\ \bibinfo {author} {\bibfnamefont {B.~H.}\
  \bibnamefont {Lee}},\ }\href@noop {} {\bibfield  {journal} {\bibinfo
  {journal} {MICROMACHINES}\ }\textbf {\bibinfo {volume} {15}} (\bibinfo {year}
  {2024})}\BibitemShut {NoStop}%
\bibitem [{\citenamefont {Hota}\ \emph {et~al.}(2025)\citenamefont {Hota},
  \citenamefont {Pazos}, \citenamefont {Lanza},\ and\ \citenamefont
  {Alshareef}}]{WOS:001460474400001}%
  \BibitemOpen
  \bibfield  {author} {\bibinfo {author} {\bibfnamefont {M.~K.}\ \bibnamefont
  {Hota}}, \bibinfo {author} {\bibfnamefont {S.}~\bibnamefont {Pazos}},
  \bibinfo {author} {\bibfnamefont {M.}~\bibnamefont {Lanza}},\ and\ \bibinfo
  {author} {\bibfnamefont {H.~N.}\ \bibnamefont {Alshareef}},\ }\href@noop {}
  {\bibfield  {journal} {\bibinfo  {journal} {MATERIALS SCIENCE \& ENGINEERING
  R-REPORTS}\ }\textbf {\bibinfo {volume} {164}} (\bibinfo {year}
  {2025})}\BibitemShut {NoStop}%
\bibitem [{\citenamefont {Li}\ \emph {et~al.}(2021)\citenamefont {Li},
  \citenamefont {Wang}, \citenamefont {Song}, \citenamefont {Zhao},
  \citenamefont {Ren}, \citenamefont {Wang}, \citenamefont {Liang},
  \citenamefont {Li}, \citenamefont {Ma}, \citenamefont {Zhu}, \citenamefont
  {Wang},\ and\ \citenamefont {Hao}}]{WOS:000711790600083}%
  \BibitemOpen
  \bibfield  {author} {\bibinfo {author} {\bibfnamefont {F.}~\bibnamefont
  {Li}}, \bibinfo {author} {\bibfnamefont {R.}~\bibnamefont {Wang}}, \bibinfo
  {author} {\bibfnamefont {C.}~\bibnamefont {Song}}, \bibinfo {author}
  {\bibfnamefont {M.}~\bibnamefont {Zhao}}, \bibinfo {author} {\bibfnamefont
  {H.}~\bibnamefont {Ren}}, \bibinfo {author} {\bibfnamefont {S.}~\bibnamefont
  {Wang}}, \bibinfo {author} {\bibfnamefont {K.}~\bibnamefont {Liang}},
  \bibinfo {author} {\bibfnamefont {D.}~\bibnamefont {Li}}, \bibinfo {author}
  {\bibfnamefont {X.}~\bibnamefont {Ma}}, \bibinfo {author} {\bibfnamefont
  {B.}~\bibnamefont {Zhu}}, \bibinfo {author} {\bibfnamefont {H.}~\bibnamefont
  {Wang}},\ and\ \bibinfo {author} {\bibfnamefont {Y.}~\bibnamefont {Hao}},\
  }\href@noop {} {\bibfield  {journal} {\bibinfo  {journal} {ACS NANO}\
  }\textbf {\bibinfo {volume} {15}},\ \bibinfo {pages} {16422} (\bibinfo {year}
  {2021})}\BibitemShut {NoStop}%
\bibitem [{\citenamefont {Yang}\ \emph {et~al.}(2024)\citenamefont {Yang},
  \citenamefont {Wang}, \citenamefont {Zhou}, \citenamefont {Zhao},
  \citenamefont {Hou}, \citenamefont {Zhu}, \citenamefont {Zhao},\ and\
  \citenamefont {Sun}}]{WOS:001278072600001}%
  \BibitemOpen
  \bibfield  {author} {\bibinfo {author} {\bibfnamefont {C.}~\bibnamefont
  {Yang}}, \bibinfo {author} {\bibfnamefont {H.}~\bibnamefont {Wang}}, \bibinfo
  {author} {\bibfnamefont {G.}~\bibnamefont {Zhou}}, \bibinfo {author}
  {\bibfnamefont {H.}~\bibnamefont {Zhao}}, \bibinfo {author} {\bibfnamefont
  {W.}~\bibnamefont {Hou}}, \bibinfo {author} {\bibfnamefont {S.}~\bibnamefont
  {Zhu}}, \bibinfo {author} {\bibfnamefont {Y.}~\bibnamefont {Zhao}},\ and\
  \bibinfo {author} {\bibfnamefont {B.}~\bibnamefont {Sun}},\ }\href@noop {}
  {\bibfield  {journal} {\bibinfo  {journal} {SMALL}\ }\textbf {\bibinfo
  {volume} {20}} (\bibinfo {year} {2024})}\BibitemShut {NoStop}%
\bibitem [{\citenamefont {Kim}\ \emph {et~al.}(2025)\citenamefont {Kim},
  \citenamefont {Lee}, \citenamefont {Yoon},\ and\ \citenamefont
  {Son}}]{WOS:001453720000001}%
  \BibitemOpen
  \bibfield  {author} {\bibinfo {author} {\bibfnamefont {J.}~\bibnamefont
  {Kim}}, \bibinfo {author} {\bibfnamefont {S.}~\bibnamefont {Lee}}, \bibinfo
  {author} {\bibfnamefont {J.}~\bibnamefont {Yoon}},\ and\ \bibinfo {author}
  {\bibfnamefont {D.}~\bibnamefont {Son}},\ }\href@noop {} {\bibfield
  {journal} {\bibinfo  {journal} {INTERNATIONAL JOURNAL OF EXTREME
  MANUFACTURING}\ }\textbf {\bibinfo {volume} {7}} (\bibinfo {year}
  {2025})}\BibitemShut {NoStop}%
\bibitem [{\citenamefont {Ghazanfar}\ \emph {et~al.}(2026)\citenamefont
  {Ghazanfar}, \citenamefont {Rabeel}, \citenamefont {Kim}, \citenamefont
  {Abbas}, \citenamefont {Anis-ur Rehman}, \citenamefont {Nisar}, \citenamefont
  {Tahir}, \citenamefont {Zulfiqar}, \citenamefont {Ali}, \citenamefont
  {Dastgeer},\ and\ \citenamefont {Kim}}]{WOS:001768750700001}%
  \BibitemOpen
  \bibfield  {author} {\bibinfo {author} {\bibfnamefont {H.}~\bibnamefont
  {Ghazanfar}}, \bibinfo {author} {\bibfnamefont {M.}~\bibnamefont {Rabeel}},
  \bibinfo {author} {\bibfnamefont {H.}~\bibnamefont {Kim}}, \bibinfo {author}
  {\bibfnamefont {H.}~\bibnamefont {Abbas}}, \bibinfo {author} {\bibfnamefont
  {M.}~\bibnamefont {Anis-ur Rehman}}, \bibinfo {author} {\bibfnamefont
  {S.}~\bibnamefont {Nisar}}, \bibinfo {author} {\bibfnamefont {M.~S.}\
  \bibnamefont {Tahir}}, \bibinfo {author} {\bibfnamefont {M.~W.}\ \bibnamefont
  {Zulfiqar}}, \bibinfo {author} {\bibfnamefont {R.~F.}\ \bibnamefont {Ali}},
  \bibinfo {author} {\bibfnamefont {G.}~\bibnamefont {Dastgeer}},\ and\
  \bibinfo {author} {\bibfnamefont {D.-k.}\ \bibnamefont {Kim}},\ }\href@noop
  {} {\bibfield  {journal} {\bibinfo  {journal} {NANO ENERGY}\ }\textbf
  {\bibinfo {volume} {154}} (\bibinfo {year} {2026})}\BibitemShut {NoStop}%
\bibitem [{\citenamefont {Pei}\ \emph {et~al.}(2015{\natexlab{a}})\citenamefont
  {Pei}, \citenamefont {Wright}, \citenamefont {Todd}, \citenamefont {Masri},\
  and\ \citenamefont {Gay-Balmaz}}]{WOS:000352695100037}%
  \BibitemOpen
  \bibfield  {author} {\bibinfo {author} {\bibfnamefont {J.-S.}\ \bibnamefont
  {Pei}}, \bibinfo {author} {\bibfnamefont {J.~P.}\ \bibnamefont {Wright}},
  \bibinfo {author} {\bibfnamefont {M.~D.}\ \bibnamefont {Todd}}, \bibinfo
  {author} {\bibfnamefont {S.~F.}\ \bibnamefont {Masri}},\ and\ \bibinfo
  {author} {\bibfnamefont {F.}~\bibnamefont {Gay-Balmaz}},\ }\href@noop {}
  {\bibfield  {journal} {\bibinfo  {journal} {NONLINEAR DYNAMICS}\ }\textbf
  {\bibinfo {volume} {80}},\ \bibinfo {pages} {457} (\bibinfo {year}
  {2015}{\natexlab{a}})}\BibitemShut {NoStop}%
\bibitem [{\citenamefont {Fouda}\ \emph {et~al.}(2015)\citenamefont {Fouda},
  \citenamefont {Radwan}, \citenamefont {Elwakil},\ and\ \citenamefont
  {Nawayseh}}]{WOS:000380571000049}%
  \BibitemOpen
  \bibfield  {author} {\bibinfo {author} {\bibfnamefont {M.~E.}\ \bibnamefont
  {Fouda}}, \bibinfo {author} {\bibfnamefont {A.~G.}\ \bibnamefont {Radwan}},
  \bibinfo {author} {\bibfnamefont {A.~S.}\ \bibnamefont {Elwakil}},\ and\
  \bibinfo {author} {\bibfnamefont {N.~K.}\ \bibnamefont {Nawayseh}},\ }in\
  \href@noop {} {\emph {\bibinfo {booktitle} {2015 IEEE CONFERENCE ON
  ELECTRONICS, CIRCUITS, AND SYSTEMS (ICECS)}}},\ \bibinfo {series and number}
  {IEEE International Conference on Electronics, Circuits and Systems}\
  (\bibinfo {year} {2015})\ pp.\ \bibinfo {pages} {201--204}\BibitemShut
  {NoStop}%
\bibitem [{\citenamefont {Xiong}\ \emph {et~al.}(2020)\citenamefont {Xiong},
  \citenamefont {Zhu}, \citenamefont {Ye}, \citenamefont {Ren}, \citenamefont
  {Yu}, \citenamefont {Xiao}, \citenamefont {Xu}, \citenamefont {Zhou},
  \citenamefont {Zhou},\ and\ \citenamefont {Lu}}]{WOS:000521624000001}%
  \BibitemOpen
  \bibfield  {author} {\bibinfo {author} {\bibfnamefont {W.}~\bibnamefont
  {Xiong}}, \bibinfo {author} {\bibfnamefont {L.~Q.}\ \bibnamefont {Zhu}},
  \bibinfo {author} {\bibfnamefont {C.}~\bibnamefont {Ye}}, \bibinfo {author}
  {\bibfnamefont {Z.~Y.}\ \bibnamefont {Ren}}, \bibinfo {author} {\bibfnamefont
  {F.}~\bibnamefont {Yu}}, \bibinfo {author} {\bibfnamefont {H.}~\bibnamefont
  {Xiao}}, \bibinfo {author} {\bibfnamefont {Z.}~\bibnamefont {Xu}}, \bibinfo
  {author} {\bibfnamefont {Y.}~\bibnamefont {Zhou}}, \bibinfo {author}
  {\bibfnamefont {H.}~\bibnamefont {Zhou}},\ and\ \bibinfo {author}
  {\bibfnamefont {H.-L.}\ \bibnamefont {Lu}},\ }\href@noop {} {\bibfield
  {journal} {\bibinfo  {journal} {ADVANCED ELECTRONIC MATERIALS}\ }\textbf
  {\bibinfo {volume} {6}} (\bibinfo {year} {2020})}\BibitemShut {NoStop}%
\bibitem [{\citenamefont {Uka}\ and\ \citenamefont
  {Zhao}(2025)}]{WOS:001518837600001}%
  \BibitemOpen
  \bibfield  {author} {\bibinfo {author} {\bibfnamefont {E.}~\bibnamefont
  {Uka}}\ and\ \bibinfo {author} {\bibfnamefont {C.}~\bibnamefont {Zhao}},\
  }\href@noop {} {\bibfield  {journal} {\bibinfo  {journal} {JOURNAL OF
  MICROELECTROMECHANICAL SYSTEMS}\ }\textbf {\bibinfo {volume} {34}},\ \bibinfo
  {pages} {503} (\bibinfo {year} {2025})}\BibitemShut {NoStop}%
\bibitem [{\citenamefont {Firestone}(1933)}]{10.1121/1.1915605}%
  \BibitemOpen
  \bibfield  {author} {\bibinfo {author} {\bibfnamefont {F.~A.}\ \bibnamefont
  {Firestone}},\ }\href {https://doi.org/10.1121/1.1915605} {\bibfield
  {journal} {\bibinfo  {journal} {The Journal of the Acoustical Society of
  America}\ }\textbf {\bibinfo {volume} {4}},\ \bibinfo {pages} {249} (\bibinfo
  {year} {1933})},\ \Eprint
  {https://arxiv.org/abs/https://pubs.aip.org/asa/jasa/article-pdf/4/3/249/18722594/249\_1\_online.pdf}
  {https://pubs.aip.org/asa/jasa/article-pdf/4/3/249/18722594/249\_1\_online.pdf}
  \BibitemShut {NoStop}%
\bibitem [{\citenamefont {Jeltsema}\ and\ \citenamefont {van~der
  Schaft}(2010{\natexlab{a}})}]{Jeltsema14052010}%
  \BibitemOpen
  \bibfield  {author} {\bibinfo {author} {\bibfnamefont {D.}~\bibnamefont
  {Jeltsema}}\ and\ \bibinfo {author} {\bibfnamefont {A.~J.}\ \bibnamefont
  {van~der Schaft}},\ }\href {https://doi.org/10.1080/13873951003690824}
  {\bibfield  {journal} {\bibinfo  {journal} {Mathematical and Computer
  Modelling of Dynamical Systems}\ }\textbf {\bibinfo {volume} {16}},\ \bibinfo
  {pages} {75} (\bibinfo {year} {2010}{\natexlab{a}})}\BibitemShut {NoStop}%
\bibitem [{\citenamefont {Liu}\ \emph {et~al.}(2016)\citenamefont {Liu},
  \citenamefont {Hua}, \citenamefont {Yu}, \citenamefont {Yang}, \citenamefont
  {Zhang}, \citenamefont {Zhang},\ and\ \citenamefont
  {Pan}}]{WOS:000382548000014}%
  \BibitemOpen
  \bibfield  {author} {\bibinfo {author} {\bibfnamefont {H.}~\bibnamefont
  {Liu}}, \bibinfo {author} {\bibfnamefont {Q.}~\bibnamefont {Hua}}, \bibinfo
  {author} {\bibfnamefont {R.}~\bibnamefont {Yu}}, \bibinfo {author}
  {\bibfnamefont {Y.}~\bibnamefont {Yang}}, \bibinfo {author} {\bibfnamefont
  {T.}~\bibnamefont {Zhang}}, \bibinfo {author} {\bibfnamefont
  {Y.}~\bibnamefont {Zhang}},\ and\ \bibinfo {author} {\bibfnamefont
  {C.}~\bibnamefont {Pan}},\ }\href@noop {} {\bibfield  {journal} {\bibinfo
  {journal} {ADVANCED FUNCTIONAL MATERIALS}\ }\textbf {\bibinfo {volume}
  {26}},\ \bibinfo {pages} {5307} (\bibinfo {year} {2016})}\BibitemShut
  {NoStop}%
\bibitem [{\citenamefont {Khan}\ \emph {et~al.}(2024)\citenamefont {Khan},
  \citenamefont {Abbas}, \citenamefont {Rezeq}, \citenamefont {Alazzam},\ and\
  \citenamefont {Mohammad}}]{WOS:001040804900001}%
  \BibitemOpen
  \bibfield  {author} {\bibinfo {author} {\bibfnamefont {M.~U.}\ \bibnamefont
  {Khan}}, \bibinfo {author} {\bibfnamefont {Y.}~\bibnamefont {Abbas}},
  \bibinfo {author} {\bibfnamefont {M.}~\bibnamefont {Rezeq}}, \bibinfo
  {author} {\bibfnamefont {A.}~\bibnamefont {Alazzam}},\ and\ \bibinfo {author}
  {\bibfnamefont {B.}~\bibnamefont {Mohammad}},\ }\href@noop {} {\bibfield
  {journal} {\bibinfo  {journal} {ADVANCED FUNCTIONAL MATERIALS}\ }\textbf
  {\bibinfo {volume} {34}} (\bibinfo {year} {2024})}\BibitemShut {NoStop}%
\bibitem [{\citenamefont {He}\ \emph {et~al.}(2025)\citenamefont {He},
  \citenamefont {Lv}, \citenamefont {Zhang}, \citenamefont {Si}, \citenamefont
  {Sha}, \citenamefont {Chen},\ and\ \citenamefont {Ma}}]{WOS:001562294200001}%
  \BibitemOpen
  \bibfield  {author} {\bibinfo {author} {\bibfnamefont {Y.}~\bibnamefont
  {He}}, \bibinfo {author} {\bibfnamefont {H.}~\bibnamefont {Lv}}, \bibinfo
  {author} {\bibfnamefont {Y.}~\bibnamefont {Zhang}}, \bibinfo {author}
  {\bibfnamefont {W.}~\bibnamefont {Si}}, \bibinfo {author} {\bibfnamefont
  {J.}~\bibnamefont {Sha}}, \bibinfo {author} {\bibfnamefont {Y.}~\bibnamefont
  {Chen}},\ and\ \bibinfo {author} {\bibfnamefont {J.}~\bibnamefont {Ma}},\
  }\href@noop {} {\bibfield  {journal} {\bibinfo  {journal} {ACS APPLIED
  MATERIALS \& INTERFACES}\ }\textbf {\bibinfo {volume} {17}},\ \bibinfo
  {pages} {50292} (\bibinfo {year} {2025})}\BibitemShut {NoStop}%
\bibitem [{\citenamefont {Galucio}\ \emph {et~al.}(2005)\citenamefont
  {Galucio}, \citenamefont {De{\"u}},\ and\ \citenamefont
  {Ohayon}}]{WOS:000226389800004}%
  \BibitemOpen
  \bibfield  {author} {\bibinfo {author} {\bibfnamefont {A.}~\bibnamefont
  {Galucio}}, \bibinfo {author} {\bibfnamefont {J.}~\bibnamefont {De{\"u}}},\
  and\ \bibinfo {author} {\bibfnamefont {R.}~\bibnamefont {Ohayon}},\
  }\href@noop {} {\bibfield  {journal} {\bibinfo  {journal} {JOURNAL OF
  INTELLIGENT MATERIAL SYSTEMS AND STRUCTURES}\ }\textbf {\bibinfo {volume}
  {16}},\ \bibinfo {pages} {33} (\bibinfo {year} {2005})}\BibitemShut {NoStop}%
\bibitem [{\citenamefont {de~Lima}\ \emph {et~al.}(2014)\citenamefont
  {de~Lima}, \citenamefont {Guaraldo-Neto}, \citenamefont {Sales},\ and\
  \citenamefont {Rade}}]{WOS:000335708500008}%
  \BibitemOpen
  \bibfield  {author} {\bibinfo {author} {\bibfnamefont {A.~M.~G.}\
  \bibnamefont {de~Lima}}, \bibinfo {author} {\bibfnamefont {B.}~\bibnamefont
  {Guaraldo-Neto}}, \bibinfo {author} {\bibfnamefont {T.~P.}\ \bibnamefont
  {Sales}},\ and\ \bibinfo {author} {\bibfnamefont {D.~A.}\ \bibnamefont
  {Rade}},\ }\href@noop {} {\bibfield  {journal} {\bibinfo  {journal}
  {ENGINEERING STRUCTURES}\ }\textbf {\bibinfo {volume} {68}},\ \bibinfo
  {pages} {85} (\bibinfo {year} {2014})}\BibitemShut {NoStop}%
\bibitem [{\citenamefont {Remillat}\ \emph {et~al.}(2006)\citenamefont
  {Remillat}, \citenamefont {Hassan},\ and\ \citenamefont
  {Scarpa}}]{WOS:000239158700003}%
  \BibitemOpen
  \bibfield  {author} {\bibinfo {author} {\bibfnamefont {C.}~\bibnamefont
  {Remillat}}, \bibinfo {author} {\bibfnamefont {M.~R.}\ \bibnamefont
  {Hassan}},\ and\ \bibinfo {author} {\bibfnamefont {F.}~\bibnamefont
  {Scarpa}},\ }\href@noop {} {\bibfield  {journal} {\bibinfo  {journal}
  {JOURNAL OF ENGINEERING MATERIALS AND TECHNOLOGY-TRANSACTIONS OF THE ASME}\
  }\textbf {\bibinfo {volume} {128}},\ \bibinfo {pages} {260} (\bibinfo {year}
  {2006})}\BibitemShut {NoStop}%
\bibitem [{\citenamefont {Reyes-Melo}\ \emph {et~al.}(2016)\citenamefont
  {Reyes-Melo}, \citenamefont {Renteria-Baltierrez}, \citenamefont
  {Lopez-Walle}, \citenamefont {Lopez-Cuellar},\ and\ \citenamefont
  {de~Araujo}}]{WOS:000385246400024}%
  \BibitemOpen
  \bibfield  {author} {\bibinfo {author} {\bibfnamefont {M.~E.}\ \bibnamefont
  {Reyes-Melo}}, \bibinfo {author} {\bibfnamefont {F.~Y.}\ \bibnamefont
  {Renteria-Baltierrez}}, \bibinfo {author} {\bibfnamefont {B.}~\bibnamefont
  {Lopez-Walle}}, \bibinfo {author} {\bibfnamefont {E.}~\bibnamefont
  {Lopez-Cuellar}},\ and\ \bibinfo {author} {\bibfnamefont {C.~J.}\
  \bibnamefont {de~Araujo}},\ }\href@noop {} {\bibfield  {journal} {\bibinfo
  {journal} {JOURNAL OF THERMAL ANALYSIS AND CALORIMETRY}\ }\textbf {\bibinfo
  {volume} {126}},\ \bibinfo {pages} {593} (\bibinfo {year}
  {2016})}\BibitemShut {NoStop}%
\bibitem [{\citenamefont {Changqing}\ \emph {et~al.}(2018)\citenamefont
  {Changqing}, \citenamefont {Jinsong}, \citenamefont {Huiyu},\ and\
  \citenamefont {Jianping}}]{WOS:000430031700004}%
  \BibitemOpen
  \bibfield  {author} {\bibinfo {author} {\bibfnamefont {F.}~\bibnamefont
  {Changqing}}, \bibinfo {author} {\bibfnamefont {L.}~\bibnamefont {Jinsong}},
  \bibinfo {author} {\bibfnamefont {S.}~\bibnamefont {Huiyu}},\ and\ \bibinfo
  {author} {\bibfnamefont {G.}~\bibnamefont {Jianping}},\ }\href@noop {}
  {\bibfield  {journal} {\bibinfo  {journal} {MECHANICS OF MATERIALS}\ }\textbf
  {\bibinfo {volume} {120}},\ \bibinfo {pages} {34} (\bibinfo {year}
  {2018})}\BibitemShut {NoStop}%
\bibitem [{\citenamefont {Zeng}\ \emph {et~al.}(2018)\citenamefont {Zeng},
  \citenamefont {Leng}, \citenamefont {Gu}, \citenamefont {Yin},\ and\
  \citenamefont {Sun}}]{WOS:000436101200005}%
  \BibitemOpen
  \bibfield  {author} {\bibinfo {author} {\bibfnamefont {H.}~\bibnamefont
  {Zeng}}, \bibinfo {author} {\bibfnamefont {J.}~\bibnamefont {Leng}}, \bibinfo
  {author} {\bibfnamefont {J.}~\bibnamefont {Gu}}, \bibinfo {author}
  {\bibfnamefont {C.}~\bibnamefont {Yin}},\ and\ \bibinfo {author}
  {\bibfnamefont {H.}~\bibnamefont {Sun}},\ }\href@noop {} {\bibfield
  {journal} {\bibinfo  {journal} {SMART MATERIALS AND STRUCTURES}\ }\textbf
  {\bibinfo {volume} {27}} (\bibinfo {year} {2018})}\BibitemShut {NoStop}%
\bibitem [{\citenamefont {Doria-Cerezo}\ \emph {et~al.}(2013)\citenamefont
  {Doria-Cerezo}, \citenamefont {van~der Heijden},\ and\ \citenamefont
  {Scherpen}}]{WOS:000329274500004}%
  \BibitemOpen
  \bibfield  {author} {\bibinfo {author} {\bibfnamefont {A.}~\bibnamefont
  {Doria-Cerezo}}, \bibinfo {author} {\bibfnamefont {L.}~\bibnamefont {van~der
  Heijden}},\ and\ \bibinfo {author} {\bibfnamefont {J.~M.~A.}\ \bibnamefont
  {Scherpen}},\ }\href@noop {} {\bibfield  {journal} {\bibinfo  {journal}
  {EUROPEAN JOURNAL OF CONTROL}\ }\textbf {\bibinfo {volume} {19}},\ \bibinfo
  {pages} {454} (\bibinfo {year} {2013})}\BibitemShut {NoStop}%
\bibitem [{\citenamefont {Zhang}\ \emph {et~al.}(2020)\citenamefont {Zhang},
  \citenamefont {Geng}, \citenamefont {Nie},\ and\ \citenamefont
  {Gao}}]{WOS:000554423300001}%
  \BibitemOpen
  \bibfield  {author} {\bibinfo {author} {\bibfnamefont {X.-L.}\ \bibnamefont
  {Zhang}}, \bibinfo {author} {\bibfnamefont {C.}~\bibnamefont {Geng}},
  \bibinfo {author} {\bibfnamefont {J.-M.}\ \bibnamefont {Nie}},\ and\ \bibinfo
  {author} {\bibfnamefont {Q.}~\bibnamefont {Gao}},\ }\href@noop {} {\bibfield
  {journal} {\bibinfo  {journal} {NONLINEAR DYNAMICS}\ }\textbf {\bibinfo
  {volume} {101}},\ \bibinfo {pages} {835} (\bibinfo {year}
  {2020})}\BibitemShut {NoStop}%
\bibitem [{\citenamefont {Pei}\ \emph {et~al.}(2015{\natexlab{b}})\citenamefont
  {Pei}, \citenamefont {Wright}, \citenamefont {Todd}, \citenamefont {Masri},\
  and\ \citenamefont {Gay-Balmaz}}]{PeiEtAl2015}%
  \BibitemOpen
  \bibfield  {author} {\bibinfo {author} {\bibfnamefont {J.-S.}\ \bibnamefont
  {Pei}}, \bibinfo {author} {\bibfnamefont {J.~P.}\ \bibnamefont {Wright}},
  \bibinfo {author} {\bibfnamefont {M.~D.}\ \bibnamefont {Todd}}, \bibinfo
  {author} {\bibfnamefont {S.~F.}\ \bibnamefont {Masri}},\ and\ \bibinfo
  {author} {\bibfnamefont {F.}~\bibnamefont {Gay-Balmaz}},\ }\href
  {https://doi.org/10.1007/s11071-014-1882-3} {\bibfield  {journal} {\bibinfo
  {journal} {Nonlinear Dynamics}\ }\textbf {\bibinfo {volume} {80}},\ \bibinfo
  {pages} {457} (\bibinfo {year} {2015}{\natexlab{b}})}\BibitemShut {NoStop}%
\bibitem [{\citenamefont {Liu}\ \emph {et~al.}(2006)\citenamefont {Liu},
  \citenamefont {Gall}, \citenamefont {Dunn}, \citenamefont {Greenberg},\ and\
  \citenamefont {Diani}}]{Liu2006}%
  \BibitemOpen
  \bibfield  {author} {\bibinfo {author} {\bibfnamefont {Y.}~\bibnamefont
  {Liu}}, \bibinfo {author} {\bibfnamefont {K.}~\bibnamefont {Gall}}, \bibinfo
  {author} {\bibfnamefont {M.~L.}\ \bibnamefont {Dunn}}, \bibinfo {author}
  {\bibfnamefont {A.~R.}\ \bibnamefont {Greenberg}},\ and\ \bibinfo {author}
  {\bibfnamefont {J.}~\bibnamefont {Diani}},\ }\href
  {https://doi.org/10.1016/j.ijplas.2005.03.004} {\bibfield  {journal}
  {\bibinfo  {journal} {International Journal of Plasticity}\ }\textbf
  {\bibinfo {volume} {22}},\ \bibinfo {pages} {279} (\bibinfo {year}
  {2006})}\BibitemShut {NoStop}%
\bibitem [{\citenamefont {Lagoudas}(2008)}]{Lagoudas2008}%
  \BibitemOpen
  \bibinfo {editor} {\bibfnamefont {D.~C.}\ \bibnamefont {Lagoudas}},\ ed.,\
  \href {https://doi.org/10.1007/978-0-387-47685-8} {\emph {\bibinfo {title}
  {Shape Memory Alloys: Modeling and Engineering Applications}}}\ (\bibinfo
  {publisher} {Springer},\ \bibinfo {address} {New York},\ \bibinfo {year}
  {2008})\BibitemShut {NoStop}%
\bibitem [{\citenamefont {Damjanovic}(2006)}]{Damjanovic2006}%
  \BibitemOpen
  \bibfield  {author} {\bibinfo {author} {\bibfnamefont {D.}~\bibnamefont
  {Damjanovic}},\ }in\ \href {https://doi.org/10.1016/B978-012480874-4/50022-1}
  {\emph {\bibinfo {booktitle} {The Science of Hysteresis}}},\ \bibinfo
  {series} {The Science of Hysteresis}, Vol.~\bibinfo {volume} {3},\ \bibinfo
  {editor} {edited by\ \bibinfo {editor} {\bibfnamefont {G.}~\bibnamefont
  {Bertotti}}\ and\ \bibinfo {editor} {\bibfnamefont {I.~D.}\ \bibnamefont
  {Mayergoyz}}}\ (\bibinfo  {publisher} {Elsevier},\ \bibinfo {address}
  {Amsterdam},\ \bibinfo {year} {2006})\ pp.\ \bibinfo {pages}
  {337--465}\BibitemShut {NoStop}%
\bibitem [{\citenamefont {Winslow}(1949)}]{Winslow1949}%
  \BibitemOpen
  \bibfield  {author} {\bibinfo {author} {\bibfnamefont {W.~M.}\ \bibnamefont
  {Winslow}},\ }\href {https://doi.org/10.1063/1.1698285} {\bibfield  {journal}
  {\bibinfo  {journal} {Journal of Applied Physics}\ }\textbf {\bibinfo
  {volume} {20}},\ \bibinfo {pages} {1137} (\bibinfo {year}
  {1949})}\BibitemShut {NoStop}%
\bibitem [{\citenamefont {Jaeger}\ \emph {et~al.}(1996)\citenamefont {Jaeger},
  \citenamefont {Nagel},\ and\ \citenamefont {Behringer}}]{Jaeger1996}%
  \BibitemOpen
  \bibfield  {author} {\bibinfo {author} {\bibfnamefont {H.~M.}\ \bibnamefont
  {Jaeger}}, \bibinfo {author} {\bibfnamefont {S.~R.}\ \bibnamefont {Nagel}},\
  and\ \bibinfo {author} {\bibfnamefont {R.~P.}\ \bibnamefont {Behringer}},\
  }\href {https://doi.org/10.1103/RevModPhys.68.1259} {\bibfield  {journal}
  {\bibinfo  {journal} {Reviews of Modern Physics}\ }\textbf {\bibinfo {volume}
  {68}},\ \bibinfo {pages} {1259} (\bibinfo {year} {1996})}\BibitemShut
  {NoStop}%
\bibitem [{\citenamefont {Jeltsema}\ and\ \citenamefont {van~der
  Schaft}(2010{\natexlab{b}})}]{JeltsemaVanDerSchaft2010}%
  \BibitemOpen
  \bibfield  {author} {\bibinfo {author} {\bibfnamefont {D.}~\bibnamefont
  {Jeltsema}}\ and\ \bibinfo {author} {\bibfnamefont {A.~J.}\ \bibnamefont
  {van~der Schaft}},\ }\href {https://doi.org/10.1080/13873951003690824}
  {\bibfield  {journal} {\bibinfo  {journal} {Mathematical and Computer
  Modelling of Dynamical Systems}\ }\textbf {\bibinfo {volume} {16}},\ \bibinfo
  {pages} {75} (\bibinfo {year} {2010}{\natexlab{b}})}\BibitemShut {NoStop}%
\bibitem [{\citenamefont {Mainardi}(2010)}]{Mainardi2010}%
  \BibitemOpen
  \bibfield  {author} {\bibinfo {author} {\bibfnamefont {F.}~\bibnamefont
  {Mainardi}},\ }\href {https://doi.org/10.1142/P614} {\emph {\bibinfo {title}
  {Fractional Calculus and Waves in Linear Viscoelasticity: An Introduction to
  Mathematical Models}}}\ (\bibinfo  {publisher} {Imperial College Press},\
  \bibinfo {address} {London, UK},\ \bibinfo {year} {2010})\BibitemShut
  {NoStop}%
\bibitem [{\citenamefont {Caputo}(1967)}]{Caputo1967}%
  \BibitemOpen
  \bibfield  {author} {\bibinfo {author} {\bibfnamefont {M.}~\bibnamefont
  {Caputo}},\ }\href {https://doi.org/10.1111/j.1365-246X.1967.tb02303.x}
  {\bibfield  {journal} {\bibinfo  {journal} {Geophysical Journal of the Royal
  Astronomical Society}\ }\textbf {\bibinfo {volume} {13}},\ \bibinfo {pages}
  {529} (\bibinfo {year} {1967})}\BibitemShut {NoStop}%
\bibitem [{\citenamefont {Podlubny}(1999)}]{Podlubny1999}%
  \BibitemOpen
  \bibfield  {author} {\bibinfo {author} {\bibfnamefont {I.}~\bibnamefont
  {Podlubny}},\ }\href@noop {} {\emph {\bibinfo {title} {Fractional
  Differential Equations: An Introduction to Fractional Derivatives, Fractional
  Differential Equations, to Methods of Their Solution and Some of Their
  Applications}}},\ \bibinfo {series} {Mathematics in Science and Engineering},
  Vol.\ \bibinfo {volume} {198}\ (\bibinfo  {publisher} {Academic Press},\
  \bibinfo {address} {San Diego, CA},\ \bibinfo {year} {1999})\BibitemShut
  {NoStop}%
\bibitem [{\citenamefont {Jacobsen}(1930)}]{Jacobsen1930Damping}%
  \BibitemOpen
  \bibfield  {author} {\bibinfo {author} {\bibfnamefont {L.~S.}\ \bibnamefont
  {Jacobsen}},\ }\href {https://doi.org/10.1115/1.4057368} {\bibfield
  {journal} {\bibinfo  {journal} {Transactions of the American Society of
  Mechanical Engineers}\ }\textbf {\bibinfo {volume} {52}},\ \bibinfo {pages}
  {169} (\bibinfo {year} {1930})}\BibitemShut {NoStop}%
\bibitem [{\citenamefont
  {Papagiannopoulos}(2018)}]{Papagiannopoulos2018Equivalent}%
  \BibitemOpen
  \bibfield  {author} {\bibinfo {author} {\bibfnamefont {G.~A.}\ \bibnamefont
  {Papagiannopoulos}},\ }\href {https://doi.org/10.1016/j.soildyn.2018.08.001}
  {\bibfield  {journal} {\bibinfo  {journal} {Soil Dynamics and Earthquake
  Engineering}\ }\textbf {\bibinfo {volume} {115}},\ \bibinfo {pages} {82}
  (\bibinfo {year} {2018})}\BibitemShut {NoStop}%
\bibitem [{\citenamefont {Christensen}(1982)}]{Christensen1982}%
  \BibitemOpen
  \bibfield  {author} {\bibinfo {author} {\bibfnamefont {R.~M.}\ \bibnamefont
  {Christensen}},\ }\href@noop {} {\emph {\bibinfo {title} {Theory of
  Viscoelasticity: An Introduction}}},\ \bibinfo {edition} {2nd}\ ed.\
  (\bibinfo  {publisher} {Academic Press},\ \bibinfo {address} {New York},\
  \bibinfo {year} {1982})\BibitemShut {NoStop}%
\bibitem [{\citenamefont {Oster}\ and\ \citenamefont
  {Auslander}(1972)}]{OsterAuslander1972}%
  \BibitemOpen
  \bibfield  {author} {\bibinfo {author} {\bibfnamefont {G.~F.}\ \bibnamefont
  {Oster}}\ and\ \bibinfo {author} {\bibfnamefont {D.~M.}\ \bibnamefont
  {Auslander}},\ }\href {https://doi.org/10.1115/1.3426595} {\bibfield
  {journal} {\bibinfo  {journal} {Journal of Dynamic Systems, Measurement, and
  Control}\ }\textbf {\bibinfo {volume} {94}},\ \bibinfo {pages} {249}
  (\bibinfo {year} {1972})}\BibitemShut {NoStop}%
\bibitem [{\citenamefont {Happel}\ and\ \citenamefont
  {Brenner}(1983)}]{HappelBrenner1983}%
  \BibitemOpen
  \bibfield  {author} {\bibinfo {author} {\bibfnamefont {J.}~\bibnamefont
  {Happel}}\ and\ \bibinfo {author} {\bibfnamefont {H.}~\bibnamefont
  {Brenner}},\ }\href {https://doi.org/10.1007/978-94-009-8352-6} {\emph
  {\bibinfo {title} {Low Reynolds Number Hydrodynamics: With Special
  Applications to Particulate Media}}}\ (\bibinfo  {publisher} {Martinus
  Nijhoff},\ \bibinfo {address} {Dordrecht},\ \bibinfo {year}
  {1983})\BibitemShut {NoStop}%
\bibitem [{\citenamefont {Bonfanti}\ \emph {et~al.}(2020)\citenamefont
  {Bonfanti}, \citenamefont {Kaplan}, \citenamefont {Charras},\ and\
  \citenamefont {Kabla}}]{Bonfanti2020}%
  \BibitemOpen
  \bibfield  {author} {\bibinfo {author} {\bibfnamefont {A.}~\bibnamefont
  {Bonfanti}}, \bibinfo {author} {\bibfnamefont {J.~L.}\ \bibnamefont
  {Kaplan}}, \bibinfo {author} {\bibfnamefont {G.}~\bibnamefont {Charras}},\
  and\ \bibinfo {author} {\bibfnamefont {A.}~\bibnamefont {Kabla}},\ }\href
  {https://doi.org/10.1039/D0SM00354A} {\bibfield  {journal} {\bibinfo
  {journal} {Soft Matter}\ }\textbf {\bibinfo {volume} {16}},\ \bibinfo {pages}
  {6002} (\bibinfo {year} {2020})}\BibitemShut {NoStop}%
\bibitem [{\citenamefont {Vinogradov}\ \emph
  {et~al.}(2004{\natexlab{a}})\citenamefont {Vinogradov}, \citenamefont
  {Schmidt}, \citenamefont {Tuthill},\ and\ \citenamefont
  {Bohannan}}]{Vinogradov2004}%
  \BibitemOpen
  \bibfield  {author} {\bibinfo {author} {\bibfnamefont {A.~M.}\ \bibnamefont
  {Vinogradov}}, \bibinfo {author} {\bibfnamefont {V.~H.}\ \bibnamefont
  {Schmidt}}, \bibinfo {author} {\bibfnamefont {G.~F.}\ \bibnamefont
  {Tuthill}},\ and\ \bibinfo {author} {\bibfnamefont {G.~W.}\ \bibnamefont
  {Bohannan}},\ }\href {https://doi.org/10.1016/j.mechmat.2003.04.002}
  {\bibfield  {journal} {\bibinfo  {journal} {Mechanics of Materials}\ }\textbf
  {\bibinfo {volume} {36}},\ \bibinfo {pages} {1007} (\bibinfo {year}
  {2004}{\natexlab{a}})}\BibitemShut {NoStop}%
\bibitem [{\citenamefont {Spencer}\ \emph {et~al.}(1997)\citenamefont
  {Spencer}, \citenamefont {Dyke}, \citenamefont {Sain},\ and\ \citenamefont
  {Carlson}}]{Spencer1997}%
  \BibitemOpen
  \bibfield  {author} {\bibinfo {author} {\bibfnamefont {B.~F.}\ \bibnamefont
  {Spencer}}, \bibinfo {author} {\bibfnamefont {S.~J.}\ \bibnamefont {Dyke}},
  \bibinfo {author} {\bibfnamefont {M.~K.}\ \bibnamefont {Sain}},\ and\
  \bibinfo {author} {\bibfnamefont {J.~D.}\ \bibnamefont {Carlson}},\ }\href
  {https://doi.org/10.1061/(ASCE)0733-9399(1997)123:3(230)} {\bibfield
  {journal} {\bibinfo  {journal} {Journal of Engineering Mechanics}\ }\textbf
  {\bibinfo {volume} {123}},\ \bibinfo {pages} {230} (\bibinfo {year}
  {1997})}\BibitemShut {NoStop}%
\bibitem [{\citenamefont {Josserand}\ \emph {et~al.}(2000)\citenamefont
  {Josserand}, \citenamefont {Tkachenko}, \citenamefont {Mueth},\ and\
  \citenamefont {Jaeger}}]{Josserand2000}%
  \BibitemOpen
  \bibfield  {author} {\bibinfo {author} {\bibfnamefont {C.}~\bibnamefont
  {Josserand}}, \bibinfo {author} {\bibfnamefont {A.~V.}\ \bibnamefont
  {Tkachenko}}, \bibinfo {author} {\bibfnamefont {D.~M.}\ \bibnamefont
  {Mueth}},\ and\ \bibinfo {author} {\bibfnamefont {H.~M.}\ \bibnamefont
  {Jaeger}},\ }\href {https://doi.org/10.1103/PhysRevLett.85.3632} {\bibfield
  {journal} {\bibinfo  {journal} {Physical Review Letters}\ }\textbf {\bibinfo
  {volume} {85}},\ \bibinfo {pages} {3632} (\bibinfo {year}
  {2000})}\BibitemShut {NoStop}%
\bibitem [{\citenamefont {Salm{\'e}n}(2004)}]{Salmen2004}%
  \BibitemOpen
  \bibfield  {author} {\bibinfo {author} {\bibfnamefont {L.}~\bibnamefont
  {Salm{\'e}n}},\ }\href {https://doi.org/10.1016/j.crvi.2004.03.010}
  {\bibfield  {journal} {\bibinfo  {journal} {Comptes Rendus Biologies}\
  }\textbf {\bibinfo {volume} {327}},\ \bibinfo {pages} {873} (\bibinfo {year}
  {2004})}\BibitemShut {NoStop}%
\bibitem [{\citenamefont {Thybring}\ \emph {et~al.}(2022)\citenamefont
  {Thybring}, \citenamefont {Fredriksson}, \citenamefont {Zelinka},\ and\
  \citenamefont {Glass}}]{Thybring2022}%
  \BibitemOpen
  \bibfield  {author} {\bibinfo {author} {\bibfnamefont {E.~E.}\ \bibnamefont
  {Thybring}}, \bibinfo {author} {\bibfnamefont {M.}~\bibnamefont
  {Fredriksson}}, \bibinfo {author} {\bibfnamefont {S.~L.}\ \bibnamefont
  {Zelinka}},\ and\ \bibinfo {author} {\bibfnamefont {S.~V.}\ \bibnamefont
  {Glass}},\ }\href {https://doi.org/10.3390/f13122051} {\bibfield  {journal}
  {\bibinfo  {journal} {Forests}\ }\textbf {\bibinfo {volume} {13}},\ \bibinfo
  {pages} {2051} (\bibinfo {year} {2022})}\BibitemShut {NoStop}%
\bibitem [{\citenamefont {Chen}\ \emph
  {et~al.}(2014{\natexlab{b}})\citenamefont {Chen}, \citenamefont {Liu},
  \citenamefont {Liu},\ and\ \citenamefont {Leng}}]{Chen2014}%
  \BibitemOpen
  \bibfield  {author} {\bibinfo {author} {\bibfnamefont {J.}~\bibnamefont
  {Chen}}, \bibinfo {author} {\bibfnamefont {L.}~\bibnamefont {Liu}}, \bibinfo
  {author} {\bibfnamefont {Y.}~\bibnamefont {Liu}},\ and\ \bibinfo {author}
  {\bibfnamefont {J.}~\bibnamefont {Leng}},\ }\href
  {https://doi.org/10.1088/0964-1726/23/5/055025} {\bibfield  {journal}
  {\bibinfo  {journal} {Smart Materials and Structures}\ }\textbf {\bibinfo
  {volume} {23}},\ \bibinfo {pages} {055025} (\bibinfo {year}
  {2014}{\natexlab{b}})}\BibitemShut {NoStop}%
\bibitem [{\citenamefont {Leng}\ \emph {et~al.}(2011)\citenamefont {Leng},
  \citenamefont {Lan}, \citenamefont {Liu},\ and\ \citenamefont
  {Du}}]{Leng2011}%
  \BibitemOpen
  \bibfield  {author} {\bibinfo {author} {\bibfnamefont {J.}~\bibnamefont
  {Leng}}, \bibinfo {author} {\bibfnamefont {X.}~\bibnamefont {Lan}}, \bibinfo
  {author} {\bibfnamefont {Y.}~\bibnamefont {Liu}},\ and\ \bibinfo {author}
  {\bibfnamefont {S.}~\bibnamefont {Du}},\ }\href
  {https://doi.org/10.1016/j.pmatsci.2011.03.001} {\bibfield  {journal}
  {\bibinfo  {journal} {Progress in Materials Science}\ }\textbf {\bibinfo
  {volume} {56}},\ \bibinfo {pages} {1077} (\bibinfo {year}
  {2011})}\BibitemShut {NoStop}%
\bibitem [{\citenamefont {Fang}\ \emph {et~al.}(2015)\citenamefont {Fang},
  \citenamefont {Sun},\ and\ \citenamefont {Gu}}]{Fang2015}%
  \BibitemOpen
  \bibfield  {author} {\bibinfo {author} {\bibfnamefont {C.-Q.}\ \bibnamefont
  {Fang}}, \bibinfo {author} {\bibfnamefont {H.-Y.}\ \bibnamefont {Sun}},\ and\
  \bibinfo {author} {\bibfnamefont {J.-P.}\ \bibnamefont {Gu}},\ }\href
  {https://doi.org/10.1017/jmech.2014.98} {\bibfield  {journal} {\bibinfo
  {journal} {Journal of Mechanics}\ }\textbf {\bibinfo {volume} {31}},\
  \bibinfo {pages} {427} (\bibinfo {year} {2015})}\BibitemShut {NoStop}%
\bibitem [{\citenamefont {Fang}\ \emph {et~al.}(2016)\citenamefont {Fang},
  \citenamefont {Sun},\ and\ \citenamefont {Gu}}]{Fang2016}%
  \BibitemOpen
  \bibfield  {author} {\bibinfo {author} {\bibfnamefont {C.-Q.}\ \bibnamefont
  {Fang}}, \bibinfo {author} {\bibfnamefont {H.-Y.}\ \bibnamefont {Sun}},\ and\
  \bibinfo {author} {\bibfnamefont {J.-P.}\ \bibnamefont {Gu}},\ }\href
  {https://doi.org/10.1017/jmech.2015.82} {\bibfield  {journal} {\bibinfo
  {journal} {Journal of Mechanics}\ }\textbf {\bibinfo {volume} {32}},\
  \bibinfo {pages} {11} (\bibinfo {year} {2016})}\BibitemShut {NoStop}%
\bibitem [{\citenamefont {Puente-C{\'o}rdova}\ \emph
  {et~al.}(2023)\citenamefont {Puente-C{\'o}rdova}, \citenamefont
  {Renter{\'i}a-Balti{\'e}rrez}, \citenamefont {Diabb-Zavala}, \citenamefont
  {Mohamed-Noriega}, \citenamefont {Bello-G{\'o}mez},\ and\ \citenamefont
  {Luna-Mart{\'i}nez}}]{PuenteCordova2023}%
  \BibitemOpen
  \bibfield  {author} {\bibinfo {author} {\bibfnamefont {J.~G.}\ \bibnamefont
  {Puente-C{\'o}rdova}}, \bibinfo {author} {\bibfnamefont {F.~Y.}\ \bibnamefont
  {Renter{\'i}a-Balti{\'e}rrez}}, \bibinfo {author} {\bibfnamefont {J.~M.}\
  \bibnamefont {Diabb-Zavala}}, \bibinfo {author} {\bibfnamefont
  {N.}~\bibnamefont {Mohamed-Noriega}}, \bibinfo {author} {\bibfnamefont
  {M.~A.}\ \bibnamefont {Bello-G{\'o}mez}},\ and\ \bibinfo {author}
  {\bibfnamefont {J.~F.}\ \bibnamefont {Luna-Mart{\'i}nez}},\ }\href
  {https://doi.org/10.3390/ma16103673} {\bibfield  {journal} {\bibinfo
  {journal} {Materials}\ }\textbf {\bibinfo {volume} {16}},\ \bibinfo {pages}
  {3673} (\bibinfo {year} {2023})}\BibitemShut {NoStop}%
\bibitem [{\citenamefont {Wang}\ \emph {et~al.}(2023)\citenamefont {Wang},
  \citenamefont {Luo}, \citenamefont {Kuang}, \citenamefont {Jin},
  \citenamefont {Liu}, \citenamefont {Jin},\ and\ \citenamefont
  {Shen}}]{Wang2023RateEffect}%
  \BibitemOpen
  \bibfield  {author} {\bibinfo {author} {\bibfnamefont {Z.}~\bibnamefont
  {Wang}}, \bibinfo {author} {\bibfnamefont {J.}~\bibnamefont {Luo}}, \bibinfo
  {author} {\bibfnamefont {W.}~\bibnamefont {Kuang}}, \bibinfo {author}
  {\bibfnamefont {M.}~\bibnamefont {Jin}}, \bibinfo {author} {\bibfnamefont
  {G.}~\bibnamefont {Liu}}, \bibinfo {author} {\bibfnamefont {X.}~\bibnamefont
  {Jin}},\ and\ \bibinfo {author} {\bibfnamefont {Y.}~\bibnamefont {Shen}},\
  }\href {https://doi.org/10.3390/met13010058} {\bibfield  {journal} {\bibinfo
  {journal} {Metals}\ }\textbf {\bibinfo {volume} {13}},\ \bibinfo {pages} {58}
  (\bibinfo {year} {2023})}\BibitemShut {NoStop}%
\bibitem [{\citenamefont {Guo}\ \emph {et~al.}(2024)\citenamefont {Guo} \emph
  {et~al.}}]{Guo2024FractionalSMA}%
  \BibitemOpen
  \bibfield  {author} {\bibinfo {author} {\bibfnamefont {R.}~\bibnamefont
  {Guo}} \emph {et~al.},\ }\href@noop {} {\bibfield  {journal} {\bibinfo
  {journal} {Axioms}\ }\textbf {\bibinfo {volume} {13}},\ \bibinfo {pages}
  {803} (\bibinfo {year} {2024})}\BibitemShut {NoStop}%
\bibitem [{\citenamefont {Raffaelli}\ and\ \citenamefont
  {Ellenbroek}(2021)}]{Raffaelli2021}%
  \BibitemOpen
  \bibfield  {author} {\bibinfo {author} {\bibfnamefont {C.}~\bibnamefont
  {Raffaelli}}\ and\ \bibinfo {author} {\bibfnamefont {W.~G.}\ \bibnamefont
  {Ellenbroek}},\ }\href {https://doi.org/10.1039/D1SM00091H} {\bibfield
  {journal} {\bibinfo  {journal} {Soft Matter}\ }\textbf {\bibinfo {volume}
  {17}},\ \bibinfo {pages} {10254} (\bibinfo {year} {2021})}\BibitemShut
  {NoStop}%
\bibitem [{\citenamefont {Lenoch}\ \emph {et~al.}(2022)\citenamefont {Lenoch},
  \citenamefont {Sch{\"o}nhoff},\ and\ \citenamefont {Cramer}}]{Lenoch2022}%
  \BibitemOpen
  \bibfield  {author} {\bibinfo {author} {\bibfnamefont {A.}~\bibnamefont
  {Lenoch}}, \bibinfo {author} {\bibfnamefont {M.}~\bibnamefont
  {Sch{\"o}nhoff}},\ and\ \bibinfo {author} {\bibfnamefont {C.}~\bibnamefont
  {Cramer}},\ }\href {https://doi.org/10.1039/D2SM01122K} {\bibfield  {journal}
  {\bibinfo  {journal} {Soft Matter}\ }\textbf {\bibinfo {volume} {18}},\
  \bibinfo {pages} {8467} (\bibinfo {year} {2022})}\BibitemShut {NoStop}%
\bibitem [{\citenamefont {Miranda-Valdez}\ \emph {et~al.}(2024)\citenamefont
  {Miranda-Valdez}, \citenamefont {Sourroubille}, \citenamefont {M\"akinen}
  \emph {et~al.}}]{MirandaValdez2024}%
  \BibitemOpen
  \bibfield  {author} {\bibinfo {author} {\bibfnamefont {I.~Y.}\ \bibnamefont
  {Miranda-Valdez}}, \bibinfo {author} {\bibfnamefont {M.}~\bibnamefont
  {Sourroubille}}, \bibinfo {author} {\bibfnamefont {T.}~\bibnamefont
  {M\"akinen}}, \emph {et~al.},\ }\bibfield  {journal} {\bibinfo  {journal}
  {Cellulose}\ }\href {https://doi.org/10.1007/s10570-023-05694-8}
  {10.1007/s10570-023-05694-8} (\bibinfo {year} {2024})\BibitemShut {NoStop}%
\bibitem [{\citenamefont {Puente-C\'ordova}\ \emph {et~al.}(2025)\citenamefont
  {Puente-C\'ordova}, \citenamefont {Renter\'ia-Balti\'errez},\ and\
  \citenamefont {Miranda-Valdez}}]{PuenteCordova2025}%
  \BibitemOpen
  \bibfield  {author} {\bibinfo {author} {\bibfnamefont {J.~G.}\ \bibnamefont
  {Puente-C\'ordova}}, \bibinfo {author} {\bibfnamefont {F.~Y.}\ \bibnamefont
  {Renter\'ia-Balti\'errez}},\ and\ \bibinfo {author} {\bibfnamefont {I.~Y.}\
  \bibnamefont {Miranda-Valdez}},\ }\href
  {https://doi.org/10.1007/s10570-025-06468-0} {\bibfield  {journal} {\bibinfo
  {journal} {Cellulose}\ }\textbf {\bibinfo {volume} {32}},\ \bibinfo {pages}
  {3619} (\bibinfo {year} {2025})}\BibitemShut {NoStop}%
\bibitem [{\citenamefont {Song}\ \emph {et~al.}(2018)\citenamefont {Song},
  \citenamefont {Chen}, \citenamefont {Zhu}, \citenamefont {Zhu}, \citenamefont
  {Dai}, \citenamefont {Ray}, \citenamefont {Li}, \citenamefont {Kuang},
  \citenamefont {Li}, \citenamefont {Quispe}, \citenamefont {Yao},
  \citenamefont {Gong}, \citenamefont {Leiste}, \citenamefont {Bruck},
  \citenamefont {Zhu}, \citenamefont {Vellore}, \citenamefont {Li},
  \citenamefont {Minus}, \citenamefont {Jia}, \citenamefont {Martini},
  \citenamefont {Li},\ and\ \citenamefont {Hu}}]{Song2018}%
  \BibitemOpen
  \bibfield  {author} {\bibinfo {author} {\bibfnamefont {J.}~\bibnamefont
  {Song}}, \bibinfo {author} {\bibfnamefont {C.}~\bibnamefont {Chen}}, \bibinfo
  {author} {\bibfnamefont {S.}~\bibnamefont {Zhu}}, \bibinfo {author}
  {\bibfnamefont {M.}~\bibnamefont {Zhu}}, \bibinfo {author} {\bibfnamefont
  {J.}~\bibnamefont {Dai}}, \bibinfo {author} {\bibfnamefont {U.}~\bibnamefont
  {Ray}}, \bibinfo {author} {\bibfnamefont {Y.}~\bibnamefont {Li}}, \bibinfo
  {author} {\bibfnamefont {Y.}~\bibnamefont {Kuang}}, \bibinfo {author}
  {\bibfnamefont {Y.}~\bibnamefont {Li}}, \bibinfo {author} {\bibfnamefont
  {N.}~\bibnamefont {Quispe}}, \bibinfo {author} {\bibfnamefont
  {Y.}~\bibnamefont {Yao}}, \bibinfo {author} {\bibfnamefont {A.}~\bibnamefont
  {Gong}}, \bibinfo {author} {\bibfnamefont {U.~H.}\ \bibnamefont {Leiste}},
  \bibinfo {author} {\bibfnamefont {H.~A.}\ \bibnamefont {Bruck}}, \bibinfo
  {author} {\bibfnamefont {J.~Y.}\ \bibnamefont {Zhu}}, \bibinfo {author}
  {\bibfnamefont {A.}~\bibnamefont {Vellore}}, \bibinfo {author} {\bibfnamefont
  {H.}~\bibnamefont {Li}}, \bibinfo {author} {\bibfnamefont {M.~L.}\
  \bibnamefont {Minus}}, \bibinfo {author} {\bibfnamefont {Z.}~\bibnamefont
  {Jia}}, \bibinfo {author} {\bibfnamefont {A.}~\bibnamefont {Martini}},
  \bibinfo {author} {\bibfnamefont {T.}~\bibnamefont {Li}},\ and\ \bibinfo
  {author} {\bibfnamefont {L.}~\bibnamefont {Hu}},\ }\href
  {https://doi.org/10.1038/nature25476} {\bibfield  {journal} {\bibinfo
  {journal} {Nature}\ }\textbf {\bibinfo {volume} {554}},\ \bibinfo {pages}
  {224} (\bibinfo {year} {2018})}\BibitemShut {NoStop}%
\bibitem [{\citenamefont {Bhattacharyya}\ \emph {et~al.}(2020)\citenamefont
  {Bhattacharyya}, \citenamefont {Matsakas}, \citenamefont {Rova},\ and\
  \citenamefont {Christakopoulos}}]{Bhattacharyya2020}%
  \BibitemOpen
  \bibfield  {author} {\bibinfo {author} {\bibfnamefont {S.}~\bibnamefont
  {Bhattacharyya}}, \bibinfo {author} {\bibfnamefont {L.}~\bibnamefont
  {Matsakas}}, \bibinfo {author} {\bibfnamefont {U.}~\bibnamefont {Rova}},\
  and\ \bibinfo {author} {\bibfnamefont {P.}~\bibnamefont {Christakopoulos}},\
  }\href {https://doi.org/10.3390/pr8091108} {\bibfield  {journal} {\bibinfo
  {journal} {Processes}\ }\textbf {\bibinfo {volume} {8}},\ \bibinfo {pages}
  {1108} (\bibinfo {year} {2020})}\BibitemShut {NoStop}%
\bibitem [{\citenamefont {Sternberg}\ and\ \citenamefont
  {Pilla}(2023)}]{Sternberg2023}%
  \BibitemOpen
  \bibfield  {author} {\bibinfo {author} {\bibfnamefont {J.}~\bibnamefont
  {Sternberg}}\ and\ \bibinfo {author} {\bibfnamefont {S.}~\bibnamefont
  {Pilla}},\ }\href {https://doi.org/10.1021/acsomega.3c01259} {\bibfield
  {journal} {\bibinfo  {journal} {ACS Omega}\ }\textbf {\bibinfo {volume}
  {8}},\ \bibinfo {pages} {40110} (\bibinfo {year} {2023})}\BibitemShut
  {NoStop}%
\bibitem [{\citenamefont {Nadányi}\ \emph {et~al.}(2025)\citenamefont
  {Nadányi}, \citenamefont {Džuganová}, \citenamefont {Vanovčanová} \emph
  {et~al.}}]{Nadanyi2025}%
  \BibitemOpen
  \bibfield  {author} {\bibinfo {author} {\bibfnamefont {R.}~\bibnamefont
  {Nadányi}}, \bibinfo {author} {\bibfnamefont {M.}~\bibnamefont
  {Džuganová}}, \bibinfo {author} {\bibfnamefont {Z.}~\bibnamefont
  {Vanovčanová}}, \emph {et~al.},\ }\bibfield  {journal} {\bibinfo  {journal}
  {MRS Bulletin}\ }\href {https://doi.org/10.1557/s43577-025-01037-z}
  {10.1557/s43577-025-01037-z} (\bibinfo {year} {2025})\BibitemShut {NoStop}%
\bibitem [{\citenamefont {Ruwoldt}\ \emph {et~al.}(2024)\citenamefont
  {Ruwoldt}, \citenamefont {Chinga-Carrasco},\ and\ \citenamefont
  {Tanase-Opedal}}]{Ruwoldt2024}%
  \BibitemOpen
  \bibfield  {author} {\bibinfo {author} {\bibfnamefont {J.}~\bibnamefont
  {Ruwoldt}}, \bibinfo {author} {\bibfnamefont {G.}~\bibnamefont
  {Chinga-Carrasco}},\ and\ \bibinfo {author} {\bibfnamefont {M.}~\bibnamefont
  {Tanase-Opedal}},\ }\href {https://doi.org/10.3390/polym16030377} {\bibfield
  {journal} {\bibinfo  {journal} {Polymers}\ }\textbf {\bibinfo {volume}
  {16}},\ \bibinfo {pages} {377} (\bibinfo {year} {2024})}\BibitemShut
  {NoStop}%
\bibitem [{\citenamefont {Placet}(2009)}]{Placet2009}%
  \BibitemOpen
  \bibfield  {author} {\bibinfo {author} {\bibfnamefont {V.}~\bibnamefont
  {Placet}},\ }\href {https://doi.org/10.1016/j.compositesa.2009.04.031}
  {\bibfield  {journal} {\bibinfo  {journal} {Composites Part A: Applied
  Science and Manufacturing}\ }\textbf {\bibinfo {volume} {40}},\ \bibinfo
  {pages} {1111} (\bibinfo {year} {2009})}\BibitemShut {NoStop}%
\bibitem [{\citenamefont {Shah}(2013)}]{Shah2013}%
  \BibitemOpen
  \bibfield  {author} {\bibinfo {author} {\bibfnamefont {D.~U.}\ \bibnamefont
  {Shah}},\ }\href {https://doi.org/10.1007/s10853-013-7458-7} {\bibfield
  {journal} {\bibinfo  {journal} {Journal of Materials Science}\ }\textbf
  {\bibinfo {volume} {48}},\ \bibinfo {pages} {6083} (\bibinfo {year}
  {2013})}\BibitemShut {NoStop}%
\bibitem [{\citenamefont {Lovinger}(1983)}]{Lovinger1983}%
  \BibitemOpen
  \bibfield  {author} {\bibinfo {author} {\bibfnamefont {A.~J.}\ \bibnamefont
  {Lovinger}},\ }\href {https://doi.org/10.1126/science.220.4602.1115}
  {\bibfield  {journal} {\bibinfo  {journal} {Science}\ }\textbf {\bibinfo
  {volume} {220}},\ \bibinfo {pages} {1115} (\bibinfo {year}
  {1983})}\BibitemShut {NoStop}%
\bibitem [{\citenamefont {Furukawa}(1989)}]{Furukawa1989}%
  \BibitemOpen
  \bibfield  {author} {\bibinfo {author} {\bibfnamefont {T.}~\bibnamefont
  {Furukawa}},\ }\href@noop {} {\bibfield  {journal} {\bibinfo  {journal}
  {Phase Transitions}\ }\textbf {\bibinfo {volume} {18}},\ \bibinfo {pages}
  {143} (\bibinfo {year} {1989})}\BibitemShut {NoStop}%
\bibitem [{\citenamefont {Martins}\ \emph {et~al.}(2014)\citenamefont
  {Martins}, \citenamefont {Lopes},\ and\ \citenamefont
  {Lanceros-Mendez}}]{Martins2014}%
  \BibitemOpen
  \bibfield  {author} {\bibinfo {author} {\bibfnamefont {P.}~\bibnamefont
  {Martins}}, \bibinfo {author} {\bibfnamefont {A.~C.}\ \bibnamefont {Lopes}},\
  and\ \bibinfo {author} {\bibfnamefont {S.}~\bibnamefont {Lanceros-Mendez}},\
  }\href {https://doi.org/10.1016/j.progpolymsci.2013.07.006} {\bibfield
  {journal} {\bibinfo  {journal} {Progress in Polymer Science}\ }\textbf
  {\bibinfo {volume} {39}},\ \bibinfo {pages} {683} (\bibinfo {year}
  {2014})}\BibitemShut {NoStop}%
\bibitem [{\citenamefont {Mohammadpourfazeli}\ \emph
  {et~al.}(2023)\citenamefont {Mohammadpourfazeli}, \citenamefont {Arash},
  \citenamefont {Ansari}, \citenamefont {Yang}, \citenamefont {Mallick},\ and\
  \citenamefont {Bagherzadeh}}]{Mohammadpourfazeli2023PVDFReview}%
  \BibitemOpen
  \bibfield  {author} {\bibinfo {author} {\bibfnamefont {S.}~\bibnamefont
  {Mohammadpourfazeli}}, \bibinfo {author} {\bibfnamefont {S.}~\bibnamefont
  {Arash}}, \bibinfo {author} {\bibfnamefont {A.}~\bibnamefont {Ansari}},
  \bibinfo {author} {\bibfnamefont {S.}~\bibnamefont {Yang}}, \bibinfo {author}
  {\bibfnamefont {K.}~\bibnamefont {Mallick}},\ and\ \bibinfo {author}
  {\bibfnamefont {R.}~\bibnamefont {Bagherzadeh}},\ }\href
  {https://doi.org/10.1039/D2RA06774A} {\bibfield  {journal} {\bibinfo
  {journal} {RSC Advances}\ }\textbf {\bibinfo {volume} {13}},\ \bibinfo
  {pages} {370} (\bibinfo {year} {2023})}\BibitemShut {NoStop}%
\bibitem [{\citenamefont {Ahbab}\ \emph {et~al.}(2025)\citenamefont {Ahbab},
  \citenamefont {Naz}, \citenamefont {Xu},\ and\ \citenamefont
  {Zhang}}]{Ahbab2025PVDFReview}%
  \BibitemOpen
  \bibfield  {author} {\bibinfo {author} {\bibfnamefont {N.}~\bibnamefont
  {Ahbab}}, \bibinfo {author} {\bibfnamefont {S.}~\bibnamefont {Naz}}, \bibinfo
  {author} {\bibfnamefont {T.-B.}\ \bibnamefont {Xu}},\ and\ \bibinfo {author}
  {\bibfnamefont {S.}~\bibnamefont {Zhang}},\ }\href
  {https://doi.org/10.3390/mi16040386} {\bibfield  {journal} {\bibinfo
  {journal} {Micromachines}\ }\textbf {\bibinfo {volume} {16}},\ \bibinfo
  {pages} {386} (\bibinfo {year} {2025})}\BibitemShut {NoStop}%
\bibitem [{\citenamefont {Jaffe}\ \emph {et~al.}(1971)\citenamefont {Jaffe},
  \citenamefont {Cook},\ and\ \citenamefont {Jaffe}}]{Jaffe1971}%
  \BibitemOpen
  \bibfield  {author} {\bibinfo {author} {\bibfnamefont {B.}~\bibnamefont
  {Jaffe}}, \bibinfo {author} {\bibfnamefont {W.~R.}\ \bibnamefont {Cook}},\
  and\ \bibinfo {author} {\bibfnamefont {H.}~\bibnamefont {Jaffe}},\
  }\href@noop {} {\emph {\bibinfo {title} {Piezoelectric Ceramics}}}\ (\bibinfo
   {publisher} {Academic Press},\ \bibinfo {year} {1971})\BibitemShut {NoStop}%
\bibitem [{\citenamefont {Lines}\ and\ \citenamefont
  {Glass}(1977)}]{LinesGlass1977}%
  \BibitemOpen
  \bibfield  {author} {\bibinfo {author} {\bibfnamefont {M.~E.}\ \bibnamefont
  {Lines}}\ and\ \bibinfo {author} {\bibfnamefont {A.~M.}\ \bibnamefont
  {Glass}},\ }\href@noop {} {\emph {\bibinfo {title} {Principles and
  Applications of Ferroelectrics and Related Materials}}}\ (\bibinfo
  {publisher} {Oxford University Press},\ \bibinfo {year} {1977})\BibitemShut
  {NoStop}%
\bibitem [{\citenamefont {Uchino}(2010)}]{Uchino2010}%
  \BibitemOpen
  \bibfield  {author} {\bibinfo {author} {\bibfnamefont {K.}~\bibnamefont
  {Uchino}},\ }\href@noop {} {\emph {\bibinfo {title} {Ferroelectric
  Devices}}},\ \bibinfo {edition} {2nd}\ ed.\ (\bibinfo  {publisher} {CRC
  Press},\ \bibinfo {year} {2010})\BibitemShut {NoStop}%
\bibitem [{\citenamefont {Damjanovic}(1998)}]{Damjanovic1998}%
  \BibitemOpen
  \bibfield  {author} {\bibinfo {author} {\bibfnamefont {D.}~\bibnamefont
  {Damjanovic}},\ }\href {https://doi.org/10.1088/0034-4885/61/9/002}
  {\bibfield  {journal} {\bibinfo  {journal} {Reports on Progress in Physics}\
  }\textbf {\bibinfo {volume} {61}},\ \bibinfo {pages} {1267} (\bibinfo {year}
  {1998})}\BibitemShut {NoStop}%
\bibitem [{\citenamefont {Conrad}\ and\ \citenamefont
  {Sprecher}(1991)}]{Conrad1991}%
  \BibitemOpen
  \bibfield  {author} {\bibinfo {author} {\bibfnamefont {H.}~\bibnamefont
  {Conrad}}\ and\ \bibinfo {author} {\bibfnamefont {A.~F.}\ \bibnamefont
  {Sprecher}},\ }\href {https://doi.org/10.1007/BF01048815} {\bibfield
  {journal} {\bibinfo  {journal} {Journal of Materials Science}\ }\textbf
  {\bibinfo {volume} {26}},\ \bibinfo {pages} {1073} (\bibinfo {year}
  {1991})}\BibitemShut {NoStop}%
\bibitem [{\citenamefont {Munteanu}\ \emph {et~al.}(2025)\citenamefont
  {Munteanu}, \citenamefont {Munteanu},\ and\ \citenamefont
  {Sedlacik}}]{Munteanu2025}%
  \BibitemOpen
  \bibfield  {author} {\bibinfo {author} {\bibfnamefont {L.}~\bibnamefont
  {Munteanu}}, \bibinfo {author} {\bibfnamefont {A.}~\bibnamefont {Munteanu}},\
  and\ \bibinfo {author} {\bibfnamefont {M.}~\bibnamefont {Sedlacik}},\
  }\href@noop {} {\bibfield  {journal} {\bibinfo  {journal} {Materials Today
  Communications}\ } (\bibinfo {year} {2025})}\BibitemShut {NoStop}%
\bibitem [{\citenamefont {Liu}\ \emph {et~al.}(2025)\citenamefont {Liu},
  \citenamefont {Liu}, \citenamefont {Wang}, \citenamefont {Tao},\ and\
  \citenamefont {Bi}}]{Liu2025}%
  \BibitemOpen
  \bibfield  {author} {\bibinfo {author} {\bibfnamefont {X.}~\bibnamefont
  {Liu}}, \bibinfo {author} {\bibfnamefont {M.}~\bibnamefont {Liu}}, \bibinfo
  {author} {\bibfnamefont {S.}~\bibnamefont {Wang}}, \bibinfo {author}
  {\bibfnamefont {B.}~\bibnamefont {Tao}},\ and\ \bibinfo {author}
  {\bibfnamefont {X.}~\bibnamefont {Bi}},\ }\href
  {https://doi.org/10.1088/2631-8695/ae19da} {\bibfield  {journal} {\bibinfo
  {journal} {Engineering Research Express}\ }\textbf {\bibinfo {volume} {7}},\
  \bibinfo {pages} {042002} (\bibinfo {year} {2025})}\BibitemShut {NoStop}%
\bibitem [{\citenamefont {Morillas}\ and\ \citenamefont
  {de~Vicente}(2020)}]{Morillas2020}%
  \BibitemOpen
  \bibfield  {author} {\bibinfo {author} {\bibfnamefont {J.~R.}\ \bibnamefont
  {Morillas}}\ and\ \bibinfo {author} {\bibfnamefont {J.}~\bibnamefont
  {de~Vicente}},\ }\href {https://doi.org/10.1039/D0SM01082K} {\bibfield
  {journal} {\bibinfo  {journal} {Soft Matter}\ }\textbf {\bibinfo {volume}
  {16}},\ \bibinfo {pages} {9614} (\bibinfo {year} {2020})}\BibitemShut
  {NoStop}%
\bibitem [{\citenamefont {Maurya}\ and\ \citenamefont
  {Sarkar}(2024)}]{Maurya2024}%
  \BibitemOpen
  \bibfield  {author} {\bibinfo {author} {\bibfnamefont {C.~S.}\ \bibnamefont
  {Maurya}}\ and\ \bibinfo {author} {\bibfnamefont {C.}~\bibnamefont
  {Sarkar}},\ }\bibfield  {journal} {\bibinfo  {journal} {Rheologica Acta}\
  }\href {https://doi.org/10.1007/s00397-024-01470-y}
  {10.1007/s00397-024-01470-y} (\bibinfo {year} {2024})\BibitemShut {NoStop}%
\bibitem [{\citenamefont {Osial}\ \emph {et~al.}(2023)\citenamefont {Osial},
  \citenamefont {Pregowska}, \citenamefont {Warczak},\ and\ \citenamefont
  {Giersig}}]{Osial2023}%
  \BibitemOpen
  \bibfield  {author} {\bibinfo {author} {\bibfnamefont {M.}~\bibnamefont
  {Osial}}, \bibinfo {author} {\bibfnamefont {A.}~\bibnamefont {Pregowska}},
  \bibinfo {author} {\bibfnamefont {M.}~\bibnamefont {Warczak}},\ and\ \bibinfo
  {author} {\bibfnamefont {M.}~\bibnamefont {Giersig}},\ }\bibfield  {journal}
  {\bibinfo  {journal} {Journal of Intelligent Material Systems and
  Structures}\ }\href {https://doi.org/10.1177/1045389X231157357}
  {10.1177/1045389X231157357} (\bibinfo {year} {2023})\BibitemShut {NoStop}%
\bibitem [{\citenamefont {Escalante-Martinez}(2020)}]{Escalante2020}%
  \BibitemOpen
  \bibfield  {author} {\bibinfo {author} {\bibfnamefont {J.~E.}\ \bibnamefont
  {Escalante-Martinez}},\ }\href
  {https://doi.org/10.1140/epjp/s13360-020-00802-0} {\bibfield  {journal}
  {\bibinfo  {journal} {European Physical Journal Plus}\ }\textbf {\bibinfo
  {volume} {135}},\ \bibinfo {pages} {860} (\bibinfo {year}
  {2020})}\BibitemShut {NoStop}%
\bibitem [{\citenamefont {Saeki}(2002)}]{Saeki2002}%
  \BibitemOpen
  \bibfield  {author} {\bibinfo {author} {\bibfnamefont {M.}~\bibnamefont
  {Saeki}},\ }\href {https://doi.org/10.1006/jsvi.2001.3980} {\bibfield
  {journal} {\bibinfo  {journal} {Journal of Sound and Vibration}\ }\textbf
  {\bibinfo {volume} {251}},\ \bibinfo {pages} {153} (\bibinfo {year}
  {2002})}\BibitemShut {NoStop}%
\bibitem [{\citenamefont {Marhadi}\ and\ \citenamefont
  {Kinra}(2005)}]{Marhadi2005}%
  \BibitemOpen
  \bibfield  {author} {\bibinfo {author} {\bibfnamefont {K.~S.}\ \bibnamefont
  {Marhadi}}\ and\ \bibinfo {author} {\bibfnamefont {V.~K.}\ \bibnamefont
  {Kinra}},\ }\href {https://doi.org/10.1016/j.jsv.2004.04.031} {\bibfield
  {journal} {\bibinfo  {journal} {Journal of Sound and Vibration}\ }\textbf
  {\bibinfo {volume} {283}},\ \bibinfo {pages} {433} (\bibinfo {year}
  {2005})}\BibitemShut {NoStop}%
\bibitem [{\citenamefont {Lu}\ \emph {et~al.}(2011)\citenamefont {Lu},
  \citenamefont {Wang}, \citenamefont {Masri},\ and\ \citenamefont
  {Lu}}]{Lu2011}%
  \BibitemOpen
  \bibfield  {author} {\bibinfo {author} {\bibfnamefont {Z.}~\bibnamefont
  {Lu}}, \bibinfo {author} {\bibfnamefont {Z.}~\bibnamefont {Wang}}, \bibinfo
  {author} {\bibfnamefont {S.~F.}\ \bibnamefont {Masri}},\ and\ \bibinfo
  {author} {\bibfnamefont {X.}~\bibnamefont {Lu}},\ }\href
  {https://doi.org/10.1002/stc.382} {\bibfield  {journal} {\bibinfo  {journal}
  {Structural Control and Health Monitoring}\ }\textbf {\bibinfo {volume}
  {18}},\ \bibinfo {pages} {79} (\bibinfo {year} {2011})}\BibitemShut {NoStop}%
\bibitem [{\citenamefont {DeGiuli}\ \emph {et~al.}(2016)\citenamefont
  {DeGiuli}, \citenamefont {McElwaine},\ and\ \citenamefont
  {Wyart}}]{DeGiuli2016}%
  \BibitemOpen
  \bibfield  {author} {\bibinfo {author} {\bibfnamefont {E.}~\bibnamefont
  {DeGiuli}}, \bibinfo {author} {\bibfnamefont {J.~N.}\ \bibnamefont
  {McElwaine}},\ and\ \bibinfo {author} {\bibfnamefont {M.}~\bibnamefont
  {Wyart}},\ }\href@noop {} {\bibfield  {journal} {\bibinfo  {journal}
  {Physical Review E}\ }\textbf {\bibinfo {volume} {94}},\ \bibinfo {pages}
  {012904} (\bibinfo {year} {2016})}\BibitemShut {NoStop}%
\bibitem [{\citenamefont {Buckingham}(1914)}]{Buckingham1914}%
  \BibitemOpen
  \bibfield  {author} {\bibinfo {author} {\bibfnamefont {E.}~\bibnamefont
  {Buckingham}},\ }\href {https://doi.org/10.1103/PhysRev.4.345} {\bibfield
  {journal} {\bibinfo  {journal} {Physical Review}\ }\textbf {\bibinfo {volume}
  {4}},\ \bibinfo {pages} {345} (\bibinfo {year} {1914})}\BibitemShut {NoStop}%
\bibitem [{\citenamefont {Khalil}(2002)}]{khalil_nonlinear_2002}%
  \BibitemOpen
  \bibfield  {author} {\bibinfo {author} {\bibfnamefont {H.~K.}\ \bibnamefont
  {Khalil}},\ }\href@noop {} {{\selectlanguage {English}\emph {\bibinfo {title}
  {Nonlinear systems}}}}\ (\bibinfo  {publisher} {Prentice Hall},\ \bibinfo
  {address} {Upper Saddle River, {N.J.}},\ \bibinfo {year} {2002})\BibitemShut
  {NoStop}%
\bibitem [{\citenamefont {Hairer}\ and\ \citenamefont
  {Wanner}(1996)}]{HairerWanner1996}%
  \BibitemOpen
  \bibfield  {author} {\bibinfo {author} {\bibfnamefont {E.}~\bibnamefont
  {Hairer}}\ and\ \bibinfo {author} {\bibfnamefont {G.}~\bibnamefont
  {Wanner}},\ }\href {https://doi.org/10.1007/978-3-642-05221-7} {\emph
  {\bibinfo {title} {Solving Ordinary Differential Equations II: Stiff and
  Differential-Algebraic Problems}}},\ \bibinfo {edition} {2nd}\ ed.,\ \bibinfo
  {series} {Springer Series in Computational Mathematics}, Vol.~\bibinfo
  {volume} {14}\ (\bibinfo  {publisher} {Springer},\ \bibinfo {address}
  {Berlin, Germany},\ \bibinfo {year} {1996})\BibitemShut {NoStop}%
\bibitem [{\citenamefont {Walker}\ and\ \citenamefont
  {Ni}(2011)}]{WalkerNi2011}%
  \BibitemOpen
  \bibfield  {author} {\bibinfo {author} {\bibfnamefont {H.~F.}\ \bibnamefont
  {Walker}}\ and\ \bibinfo {author} {\bibfnamefont {P.}~\bibnamefont {Ni}},\
  }\href {https://doi.org/10.1137/10078356X} {\bibfield  {journal} {\bibinfo
  {journal} {SIAM Journal on Numerical Analysis}\ }\textbf {\bibinfo {volume}
  {49}},\ \bibinfo {pages} {1715} (\bibinfo {year} {2011})}\BibitemShut
  {NoStop}%
\bibitem [{\citenamefont {Saad}\ and\ \citenamefont
  {Schultz}(1986)}]{SaadSchultz1986}%
  \BibitemOpen
  \bibfield  {author} {\bibinfo {author} {\bibfnamefont {Y.}~\bibnamefont
  {Saad}}\ and\ \bibinfo {author} {\bibfnamefont {M.~H.}\ \bibnamefont
  {Schultz}},\ }\href {https://doi.org/10.1137/0907058} {\bibfield  {journal}
  {\bibinfo  {journal} {SIAM Journal on Scientific and Statistical Computing}\
  }\textbf {\bibinfo {volume} {7}},\ \bibinfo {pages} {856} (\bibinfo {year}
  {1986})}\BibitemShut {NoStop}%
\bibitem [{\citenamefont {Allgower}\ and\ \citenamefont
  {Georg}(1990)}]{AllgowerGeorg1990}%
  \BibitemOpen
  \bibfield  {author} {\bibinfo {author} {\bibfnamefont {E.~L.}\ \bibnamefont
  {Allgower}}\ and\ \bibinfo {author} {\bibfnamefont {K.}~\bibnamefont
  {Georg}},\ }\href {https://doi.org/10.1007/978-3-642-61257-2} {\emph
  {\bibinfo {title} {Numerical Continuation Methods: An Introduction}}},\
  \bibinfo {series} {Springer Series in Computational Mathematics},
  Vol.~\bibinfo {volume} {13}\ (\bibinfo  {publisher} {Springer},\ \bibinfo
  {address} {Berlin, Germany},\ \bibinfo {year} {1990})\BibitemShut {NoStop}%
\bibitem [{\citenamefont {Shampine}\ and\ \citenamefont
  {Reichelt}(1997)}]{ShampineReichelt1997}%
  \BibitemOpen
  \bibfield  {author} {\bibinfo {author} {\bibfnamefont {L.~F.}\ \bibnamefont
  {Shampine}}\ and\ \bibinfo {author} {\bibfnamefont {M.~W.}\ \bibnamefont
  {Reichelt}},\ }\href {https://doi.org/10.1137/S1064827594276424} {\bibfield
  {journal} {\bibinfo  {journal} {SIAM Journal on Scientific Computing}\
  }\textbf {\bibinfo {volume} {18}},\ \bibinfo {pages} {1} (\bibinfo {year}
  {1997})}\BibitemShut {NoStop}%
\bibitem [{\citenamefont {Vinogradov}\ \emph
  {et~al.}(2004{\natexlab{b}})\citenamefont {Vinogradov}, \citenamefont
  {Schmidt}, \citenamefont {Tuthill},\ and\ \citenamefont
  {Bohannan}}]{Vinogradov2004PVDF}%
  \BibitemOpen
  \bibfield  {author} {\bibinfo {author} {\bibfnamefont {A.~M.}\ \bibnamefont
  {Vinogradov}}, \bibinfo {author} {\bibfnamefont {V.~H.}\ \bibnamefont
  {Schmidt}}, \bibinfo {author} {\bibfnamefont {G.~F.}\ \bibnamefont
  {Tuthill}},\ and\ \bibinfo {author} {\bibfnamefont {G.~W.}\ \bibnamefont
  {Bohannan}},\ }\href {https://doi.org/10.1016/j.mechmat.2003.04.002}
  {\bibfield  {journal} {\bibinfo  {journal} {Mechanics of Materials}\ }\textbf
  {\bibinfo {volume} {36}},\ \bibinfo {pages} {1007} (\bibinfo {year}
  {2004}{\natexlab{b}})}\BibitemShut {NoStop}%
\end{thebibliography}%

\end{document}